\pdfoutput=1
\documentclass[12pt,a4paper]{article}

\usepackage{ifthen} 
\newboolean{pdflatex}
\setboolean{pdflatex}{true} 

\newboolean{articletitles}
\setboolean{articletitles}{true} 

\newboolean{uprightparticles}
\setboolean{uprightparticles}{false} 

\def\paperauthors{LHCb collaboration} 
\def\paperasciititle{Commissioning and Performance of the LHCb Upstream Tracker} 
\def\papertitle{Commissioning and Performance of the LHCb Upstream Tracker} 
\def\paperkeywords{{High Energy Physics}, {LHCb}} 
\def\papercopyright{\the\year\ CERN for the benefit of the LHCb collaboration} 
\def\paperlicence{CC BY 4.0 licence}
\def\paperlicenceurl{https://creativecommons.org/licenses/by/4.0/}

\newif\ifEnableSectionTOCLinks
\EnableSectionTOCLinksfalse 

\usepackage[top=1in, bottom=1.25in, left=1in, right=1in]{geometry}

\usepackage{microtype}
\usepackage{lineno}  
\usepackage{xspace} 
\usepackage{caption} 

\usepackage{graphicx}  
\usepackage{color}
\usepackage{colortbl}
\graphicspath{{./figs/}} 

\usepackage{amsmath} 
\usepackage{amssymb}
\usepackage{amsfonts}
\usepackage{upgreek} 

\newcommand*\patchAmsMathEnvironmentForLineno[1]{%
\expandafter\let\csname old#1\expandafter\endcsname\csname #1\endcsname
\expandafter\let\csname oldend#1\expandafter\endcsname\csname
end#1\endcsname
 \renewenvironment{#1}%
   {\linenomath\csname old#1\endcsname}%
   {\csname oldend#1\endcsname\endlinenomath}%
}
\newcommand*\patchBothAmsMathEnvironmentsForLineno[1]{%
  \patchAmsMathEnvironmentForLineno{#1}%
  \patchAmsMathEnvironmentForLineno{#1*}%
}
\AtBeginDocument{%
\patchBothAmsMathEnvironmentsForLineno{equation}%
\patchBothAmsMathEnvironmentsForLineno{align}%
\patchBothAmsMathEnvironmentsForLineno{flalign}%
\patchBothAmsMathEnvironmentsForLineno{alignat}%
\patchBothAmsMathEnvironmentsForLineno{gather}%
\patchBothAmsMathEnvironmentsForLineno{multline}%
\patchBothAmsMathEnvironmentsForLineno{eqnarray}%
}

\usepackage[pdftex]{hyperref}

\usepackage{hyperxmp}

\hypersetup{
    pdfauthor={\paperauthors},
    pdftitle={\paperasciititle},
    pdfkeywords={\paperkeywords},
    pdfcopyright={Copyright (C) \papercopyright},
    pdflicenseurl={\paperlicenceurl}
}

\usepackage[colorinlistoftodos,textsize=scriptsize]{todonotes}

\usepackage[bottom,flushmargin,hang,multiple]{footmisc}

\usepackage[all]{hypcap} 

\usepackage{xspace} 
\usepackage{upgreek}

\def\zsth   {\mbox{ZS$_{\rm th}$}\xspace}
\def\nzsmax {\mbox{N$_{\rm max}^{\rm ZS}$}\xspace}

\def\lhcb   {\mbox{LHCb}\xspace}

\def\MagUp {\mbox{\em Mag\kern -0.05em Up}\xspace}

\ifthenelse{\boolean{uprightparticles}}%
{

 \def\Pmu         {\ensuremath{\upmu}\xspace}

 \def\Ppi         {\ensuremath{\uppi}\xspace}

 \def\Ppsi        {\ensuremath{\uppsi}\xspace}

 \def\PDelta      {\ensuremath{\Delta}\xspace}                 
 \def\PXi         {\ensuremath{\Xi}\xspace}                 
 \def\PLambda     {\ensuremath{\Lambda}\xspace}                 
 \def\PSigma      {\ensuremath{\Sigma}\xspace}                 
 \def\POmega      {\ensuremath{\Omega}\xspace}                 
 \def\PUpsilon    {\ensuremath{\Upsilon}\xspace}
 \let\oldPi\Pi
 \def\PPi         {\ensuremath{\oldPi}\xspace}

 \def\PB      {\ensuremath{\mathrm{B}}\xspace}                 
 \def\PD      {\ensuremath{\mathrm{D}}\xspace}                 
 \def\PJ      {\ensuremath{\mathrm{J}}\xspace}                 
 \def\PK      {\ensuremath{\mathrm{K}}\xspace}                 
 \def\Ps      {\ensuremath{\mathrm{s}}\xspace}

 \def\thebaroffset{0.0em}
}
{

 \def\Pmu         {\ensuremath{\mu}\xspace}

 \def\Ppi         {\ensuremath{\pi}\xspace}

 \def\Ppsi        {\ensuremath{\psi}\xspace}                 
                  
 \mathchardef\PDelta="7101
 \mathchardef\PXi="7104
 \mathchardef\PLambda="7103
 \mathchardef\PSigma="7106
 \mathchardef\POmega="710A
 \mathchardef\PUpsilon="7107
 \mathchardef\PPi="7105
 \def\PB      {\ensuremath{B}\xspace}                 
 \def\PD      {\ensuremath{D}\xspace}                 
 \def\PJ      {\ensuremath{J}\xspace}                 
 \def\PK      {\ensuremath{K}\xspace}                 
 \def\Ps      {\ensuremath{s}\xspace}

 \def\thebaroffset{0.18em}
}
\newcommand{\offsetoverline}[2][\thebaroffset]{\kern #1\overline{\kern -#1 #2}}%

\makeatletter
\ifcase \@ptsize \relax
  \newcommand{\miniscule}{\@setfontsize\miniscule{4}{5}}
\or
  \newcommand{\miniscule}{\@setfontsize\miniscule{5}{6}}
\or
  \newcommand{\miniscule}{\@setfontsize\miniscule{5}{6}}
\fi
\makeatother

\DeclareRobustCommand{\optbar}[1]{\shortstack{{\miniscule (\rule[.5ex]{1.25em}{.18mm})}
  \\ [-.7ex] $#1$}}

\def\mup        {{\ensuremath{\Pmu^+}}\xspace}
\def\mun        {{\ensuremath{\Pmu^-}}\xspace} 

\def\squark    {{\ensuremath{\Ps}}\xspace}

\def\pion   {{\ensuremath{\Ppi}}\xspace}

\def\pip    {{\ensuremath{\pion^+}}\xspace}
\def\pim    {{\ensuremath{\pion^-}}\xspace}

\def\kaon    {{\ensuremath{\PK}}\xspace}

\def\KorKbar {\kern \thebaroffset\optbar{\kern -\thebaroffset \PK}{}\xspace}

\def\Kp      {{\ensuremath{\kaon^+}}\xspace}
\def\Km      {{\ensuremath{\kaon^-}}\xspace}

\def\KS      {{\ensuremath{\kaon^0_{\mathrm{S}}}}\xspace}

\def\D       {{\ensuremath{\PD}}\xspace}

\def\DorDbar {\kern \thebaroffset\optbar{\kern -\thebaroffset \PD}\xspace}
\def\Dz      {{\ensuremath{\D^0}}\xspace}

\def\Dp      {{\ensuremath{\D^+}}\xspace}
\def\Dm      {{\ensuremath{\D^-}}\xspace}

\def\DpDm    {\ensuremath{\Dp {\kern -0.16em \Dm}}\xspace}

\def\B       {{\ensuremath{\PB}}\xspace}

\def\BorBbar {\kern \thebaroffset\optbar{\kern -\thebaroffset \PB}\xspace}
\def\Bz      {{\ensuremath{\B^0}}\xspace}

\def\Bd      {{\ensuremath{\B^0}}\xspace}

\def\BdorBdbar {\kern \thebaroffset\optbar{\kern -\thebaroffset \Bd}\xspace}

\def\Bs      {{\ensuremath{\B^0_\squark}}\xspace}

\def\BsorBsbar {\kern \thebaroffset\optbar{\kern -\thebaroffset \Bs}\xspace}

\def\jpsi     {{\ensuremath{{\PJ\mskip -3mu/\mskip -2mu\Ppsi}}}\xspace}

\def\Y#1S{\ensuremath{\PUpsilon{(#1S)}}\xspace}

\def\Lz          {{\ensuremath{\PLambda}}\xspace}

\def\LorLbar     {\kern \thebaroffset\optbar{\kern -\thebaroffset \PLambda}\xspace}

\def\Xires       {{\ensuremath{\PXi}}\xspace}

\def\Xim         {{\ensuremath{\Xires^-}}\xspace}

\def\Omegares    {{\ensuremath{\POmega}}\xspace}

\def\Omegam      {{\ensuremath{\Omegares^-}}\xspace}

\def\to                 {\ensuremath{\rightarrow}\xspace}

\def\AT#1     {\ensuremath{A_{\mathrm{T}}^{#1}}\xspace}           

\def\C#1      {\ensuremath{\mathcal{C}_{#1}}\xspace}                       
\def\Cp#1     {\ensuremath{\mathcal{C}_{#1}^{'}}\xspace}                    
\def\Ceff#1   {\ensuremath{\mathcal{C}_{#1}^{\mathrm{(eff)}}}\xspace}        
\def\Cpeff#1  {\ensuremath{\mathcal{C}_{#1}^{'\mathrm{(eff)}}}\xspace}       
\def\Ope#1    {\ensuremath{\mathcal{O}_{#1}}\xspace}                       
\def\Opep#1   {\ensuremath{\mathcal{O}_{#1}^{'}}\xspace}                    

\newcommand{\nospaceunit}[1]{\ensuremath{\text{#1}}}       
\newcommand{\aunit}[1]{\ensuremath{\text{\,#1}}}       

\newcommand{\tev}{\aunit{Te\kern -0.1em V}\xspace}
\newcommand{\gev}{\aunit{Ge\kern -0.1em V}\xspace}
\newcommand{\mev}{\aunit{Me\kern -0.1em V}\xspace}
\newcommand{\kev}{\aunit{ke\kern -0.1em V}\xspace}
\newcommand{\ev}{\aunit{e\kern -0.1em V}\xspace}
 
\newcommand{\mevc}{\ensuremath{\aunit{Me\kern -0.1em V\!/}c}\xspace}
\newcommand{\gevc}{\ensuremath{\aunit{Ge\kern -0.1em V\!/}c}\xspace}
\newcommand{\mevcc}{\ensuremath{\aunit{Me\kern -0.1em V\!/}c^2}\xspace}
\newcommand{\gevcc}{\ensuremath{\aunit{Ge\kern -0.1em V\!/}c^2}\xspace}

\def\mum  {\ensuremath{\,\upmu\nospaceunit{m}}\xspace}

\newcommand{\chisq}{\ensuremath{\chi^2}\xspace}

\def\gsim{{~\raise.15em\hbox{$>$}\kern-.85em
          \lower.35em\hbox{$\sim$}~}\xspace}
\def\lsim{{~\raise.15em\hbox{$<$}\kern-.85em
          \lower.35em\hbox{$\sim$}~}\xspace}

\def\sPlot{\mbox{\em sPlot}\xspace}

\def\pt         {\ensuremath{p_{\mathrm{T}}}\xspace}

\def\fluka      {\mbox{\textsc{Fluka}}\xspace}

\def\tell1  {TELL1\xspace}
\def\ukl1   {UKL1\xspace}

\newcommand{\lhcborcid}[1]{\href{https://orcid.org/#1}{\hspace*{0.1em}\raisebox{-0.45ex}{\includegraphics[width=1em]{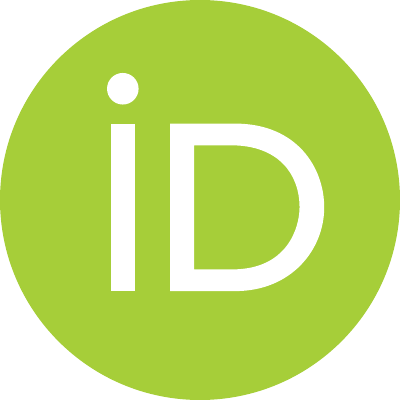}}}}

\hypersetup{
  colorlinks   = true, 
  urlcolor     = blue, 
  linkcolor    = blue, 
  citecolor    = red   
}

\usepackage{cite} 
\usepackage{mciteplus}

\usepackage{longtable} 

\begin{document}

\renewcommand{\thefootnote}{\fnsymbol{footnote}}
\setcounter{footnote}{1}


\begin{titlepage}
\pagenumbering{roman}

\vspace*{-1.5cm}
\centerline{\large EUROPEAN ORGANIZATION FOR NUCLEAR RESEARCH (CERN)}
\vspace*{1.5cm}
\noindent
\begin{tabular*}{\linewidth}{lc@{\extracolsep{\fill}}r@{\extracolsep{0pt}}}
\ifthenelse{\boolean{pdflatex}}
{\vspace*{-1.5cm}\mbox{\!\!\!\includegraphics[width=.14\textwidth]{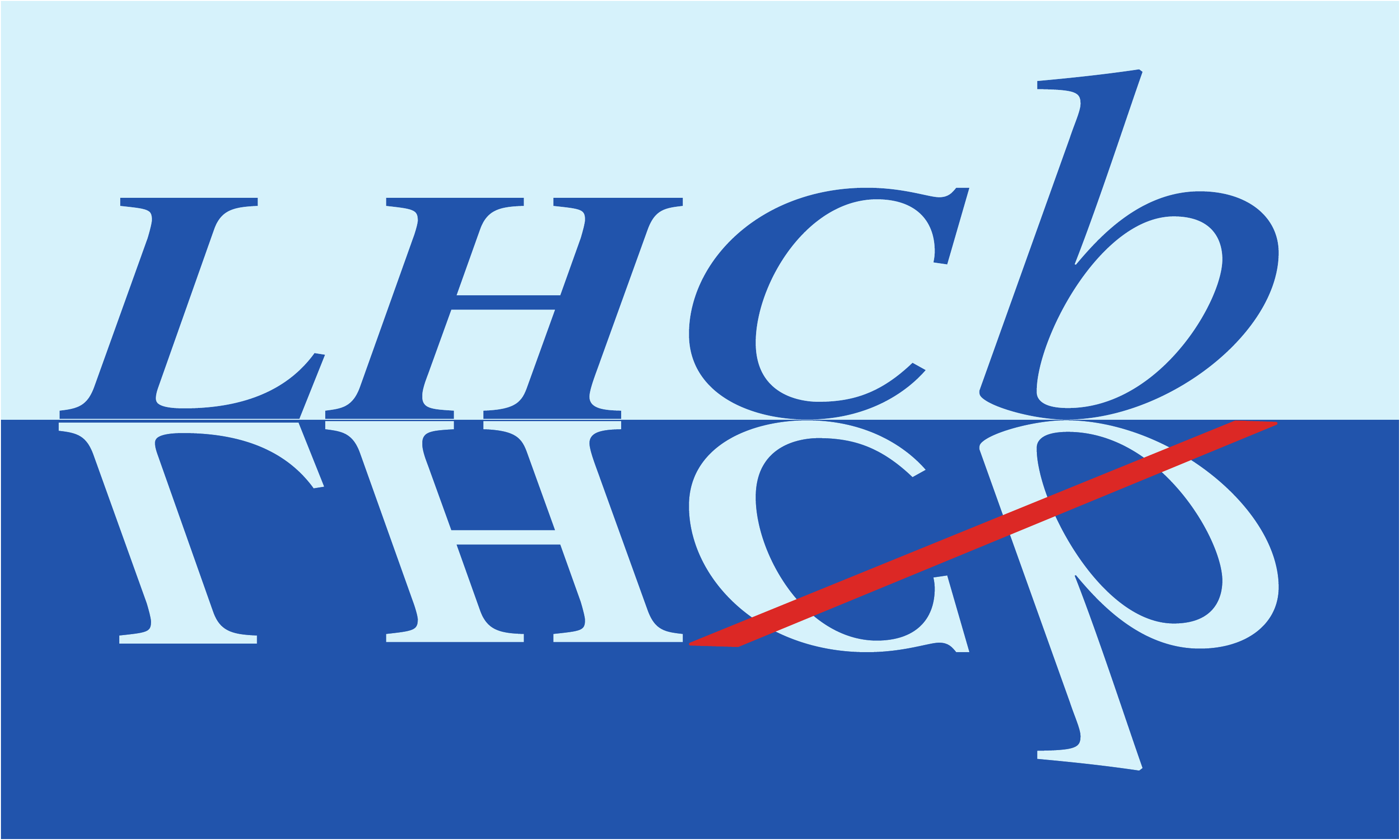}} & &}%
{\vspace*{-1.2cm}\mbox{\!\!\!\includegraphics[width=.12\textwidth]{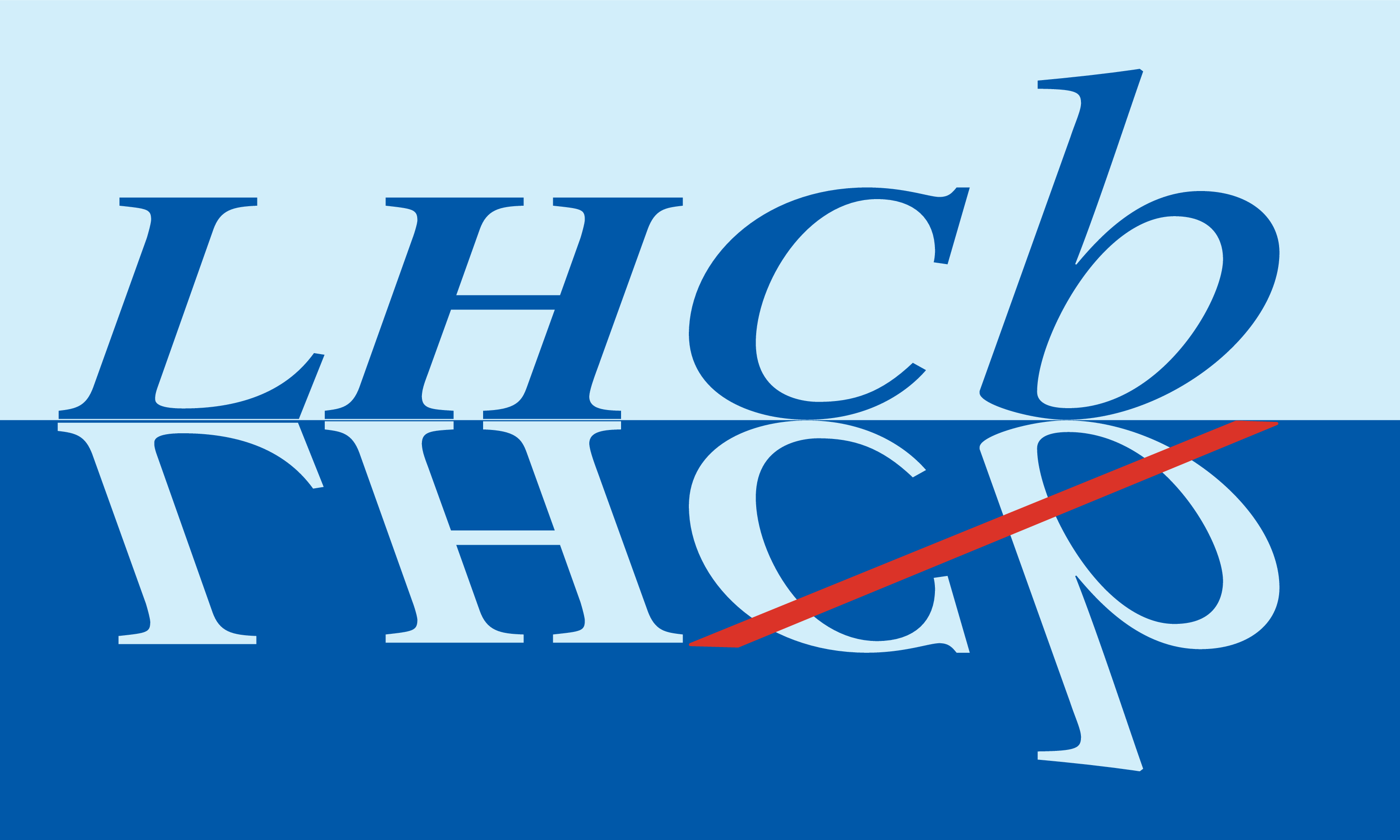}} & &}%
\\
 & & CERN-LHCb-DP-2026-003 \\  
 & & \today \\ 
 & & \\
\end{tabular*}

\vspace*{1.0cm}

{\normalfont\bfseries\boldmath\huge
\begin{center}
  \papertitle 
\end{center}
}

\vspace*{1.0cm}
\begin{flushleft}
\small

D.~Andreou$^{1}$\lhcborcid{0000-0001-6288-0558},
M.~Artuso$^{1}$\lhcborcid{0000-0002-5991-7273}, 
J.~Bartz$^{1}$\lhcborcid{0000-0002-2646-4124}, 
B.~Batsukh$^{1,d}$\lhcborcid{0000-0003-1020-2549}, 
A.~Beiter$^{1}$\lhcborcid{0000-0002-8315-1511}, 
H.~Bernstein$^{1}$, 
S.~Blusk$^{1,*}$\lhcborcid{0000-0001-9170-684X}, 
S.~Ding$^{1}$\lhcborcid{0009-0007-1406-3343}, 
S.~Ely$^{1,b}$\lhcborcid{0000-0003-1618-3617}, 
J.~Freiler$^{1}$
% ...

\end{flushleft}

\medskip
\begin{center}
{\footnotesize \it

$^{1}$Syracuse University, Syracuse, NY, United States\\
$^{2}$University of Maryland, College Park, MD, United States\\
$^{3}$INFN Sezione di Milano, Milano, Italy\\
$^{4}$Institute Of High Energy Physics (IHEP), Beijing, China\\
$^{5}$AGH - University of Krakow, Faculty of Physics and Applied Computer Science, Krak{\'o}w, Poland\\
$^{6}$Physik-Institut, Universit{\"a}t Z{\"u}rich, Z{\"u}rich, Switzerland\\
$^{7}$University of Michigan, Ann Arbor, MI, United States\\
$^{8}$European Organization for Nuclear Research (CERN), Geneva, Switzerland\\
$^{9}$University of Cincinnati, Cincinnati, Cincinnati, OH, United States\\
$^{10}$Department of Engineering Physics, Tsinghua University, Beijing, China\\
$^{11}$Universit{\'e} Savoie Mont Blanc, CNRS, IN2P3-LAPP, Annecy, France\\
$^{12}$Institute of Particle Physics, Central China Normal University, Wuhan, Hubei, China\\
$^{13}$State Key Laboratory of Nuclear Physics and Technology, South China Normal University, Guangzhou, China\\
$^{14}$University of Chinese Academy of Sciences, Beijing, China\\
$^{15}$INFN Sezione di Pisa, Pisa, Italy\\
$^{16}$INFN Sezione di Genova, Genova, Italy\\
$^{17}$School of Physics and Electronics, Hunan University, Changsha City, China\\
$^{18}$Universiteit Maastricht, Maastricht, Netherlands\\
$^{19}$Instituto de Fisica Corpuscular, Centro Mixto Universidad de Valencia - CSIC, Valencia, Spain\\
$^{20}$Department of Physics and Astronomy, University of Manchester, Manchester, United Kingdom\\
$^{21}$Institute of Physics, Ecole Polytechnique  F{\'e}d{\'e}rale de Lausanne (EPFL), Lausanne, Switzerland\\
$^{22}$Physikalisches Institut, Ruprecht-Karls-Universit{\"a}t Heidelberg, Heidelberg, Germany\\
$^{23}$Laboratoire de Physique Nucl{\'e}aire et de Hautes {\'E}nergies (LPNHE), Sorbonne Universit{\'e}, CNRS/IN2P3, Paris, France\\

\medskip

$^{a}$Present address: see affiliation $^{16}$\\
$^{b}$Present address: Imperial College London, London, United Kingdom\\
$^{c}$Present address: INFN Laboratori Nazionali di Frascati, Frascati, Italy\\
$^{d}$Present address: Institute of Physics and Technology, Mongolian Academy of Sciences, Ulan Bator, Mongolia\\
$^{e}$Present address: see affiliation $^{21}$\\
$^{f}$Present address: see affiliation $^{20}$\\
$^{g}$Present address: University of Oklahoma, Norman, OK, United States\\
$^{h}$Present address: Lanzhou University, Lanzhou, China\\
$^{i}$Present address: see affiliation $^{8}$\\
$^{j}$Present address: INFN Sezione di Padova, Padova, Italy\\
$^{k}$Present address: Central South University, Changsha, China\\
$^{l}$Present address: see affiliation $^{14}$\\
$^{m}$Present address: see affiliation $^{9}$\\

}

\end{center}
\medskip\flushleft{
$ ^*$ Corresponding author: sblusk@syr.edu 

$ ^{\dagger}$ Deceased

}


\vspace{\fill}
\begin{abstract}
  \noindent
  The Upstream Tracker is a new, high-precision silicon strip detector installed in the LHCb experiment as part of the overall LHCb Upgrade I project. The detector is an integral part of the tracking system, and plays a crucial role in the  reconstruction of charged particles in the new fully software-based trigger. An overview of the Upstream Tracker detector design, its commissioning, and performance are presented.
\end{abstract}

\vspace*{1.0cm}

\begin{center}
  Submitted to JINST
\end{center}

\vspace{\fill}

{\footnotesize 
\centerline{\copyright~\papercopyright. \href{\paperlicenceurl}{\paperlicence}.}}
\vspace*{2mm}

\end{titlepage}


\newpage
\setcounter{page}{2}
\mbox{~}

\renewcommand{\thefootnote}{\arabic{footnote}}
\setcounter{footnote}{0}

\tableofcontents

\cleardoublepage


\pagestyle{plain} 
\setcounter{page}{1}
\pagenumbering{arabic}


\section{Introduction}
\label{sec:Introduction}

\begin{figure}[b]
\centering
\includegraphics[width=0.88\textwidth]{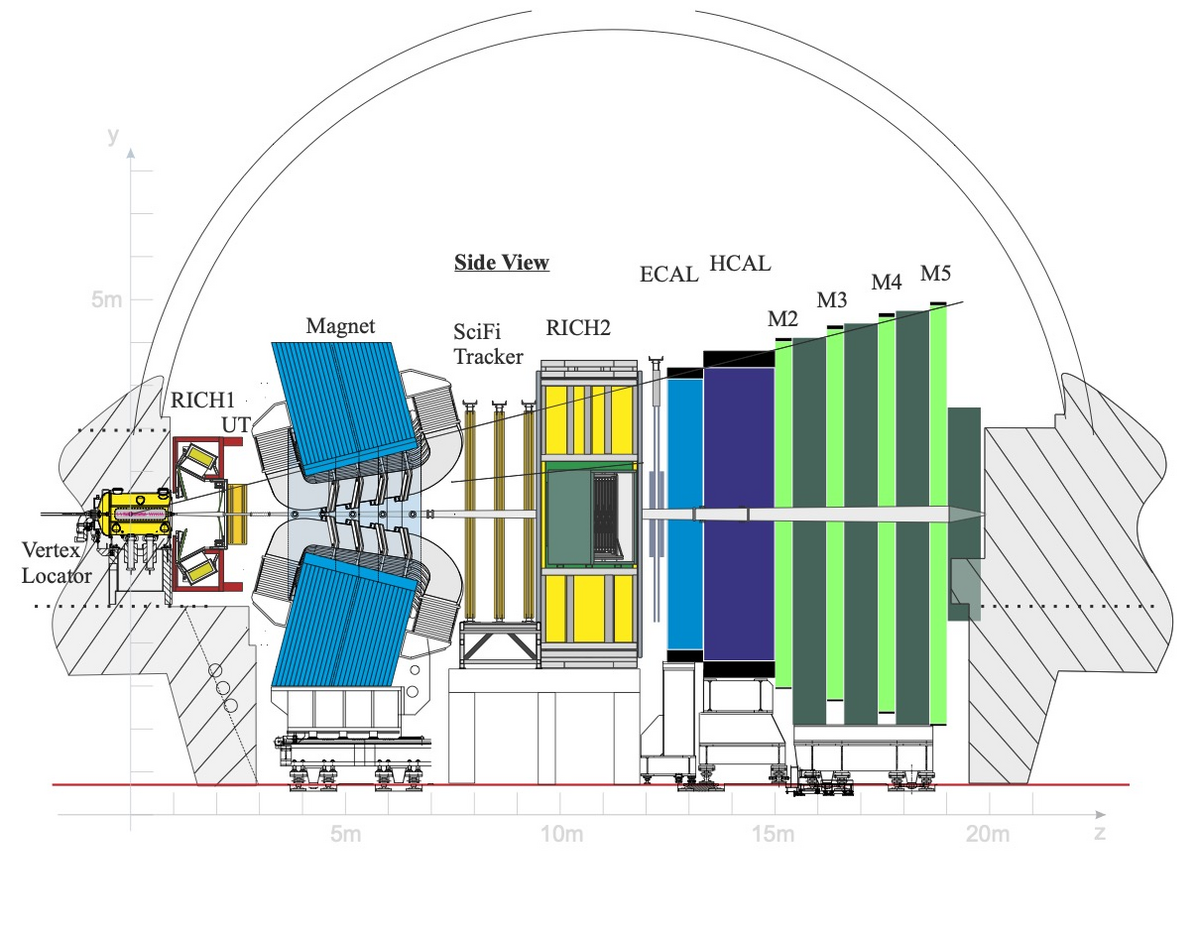}
\caption{\small{Layout of the Upgrade 1 LHCb detector}}
\label{fig:LHCbDetector}
\end{figure}

The LHCb detector~\cite{LHCb-DP-2008-001} is a single-arm spectrometer designed to study heavy flavour physics at the Large Hadron Collider (LHC) at CERN. from 2019--2022, the detector underwent a major upgrade~\cite{LHCb-DP-2022-002} in order to increase the data sample size by about a factor of five relative to the samples collected during Run 1 \& 2. This increase has been achieved by running at about five times larger instantaneous luminosity, up to $2\times10^{33}$~cm$^{-2}$\,s$^{-1}$, which results in around 5.3 visible $pp$ collisions (mu), on average, per LHC bunch crossing. A schematic view of the Upgrade 1 LHCb detector is shown in Fig.~\ref{fig:LHCbDetector}.

To cope with this higher instantaneous luminosity much of the original LHCb detector has been replaced with more granular detectors. One of the novel aspects of the LHCb upgrade is the new data acquisition and trigger~\cite{LHCb-TDR-016} systems, which analyze the collisions in every LHC crossing using custom reconstruction software running on GPU (first level trigger, HLT1) and CPU (second level trigger, HLT2) farms. In brief, HLT1 analyzes all visible events from the LHC ($\sim30$\,MHz), and using a limited number of interesting signatures, reduces the rate to $\sim$1\,MHz. The HLT2 then does a full event reconstruction and searches for a wide range of interesting decays, mostly, but not limited to flavor physics. To enable the fully software trigger, all sub-detectors developed new front-end electronics capable of shipping data out at the full LHC operating frequency of 40\,MHz. The removal of the L0 hardware trigger, which was present in Run 1 \& 2, more than doubles the efficiency for triggering on hadronic decays of $b$ and $c$ hadrons in the upgraded detector.

The upgraded LHCb tracking system~\cite{LHCb-TDR-015} consists of three main sub-detector components. Surrounding the interaction point, and spanning about 1 m in length along the beam ($z$) axis is the silicon pixel Vertex Locator (VELO)~\cite{LHCb-TDR-013}. Downstream of the large LHCb dipole magnet, from $z=7.9$ to $9.3$ m, are three scintillating fiber tracking stations (SciFi). Situated in front of the dipole magnet in between the VELO and SciFi at $z\simeq2.3$ to 2.6~m is the Upstream Tracker (UT). Due to its location, the UT is immersed in the fringe field of the LHCb magnet, and charged particles experience a bending power of approximately 0.12 T\,m from $z$=0 to the last layer of the UT. This feature is exploited in the track reconstruction to provide a fast estimate of the momentum of charged particles using VELO and UT hits, as described later.

One of the key features of the UT is its fine segmentation, resulting in low strip-hit occupancy, about 1\% on average and with a maximum of about 2.5\% in $pp$ collisions at the nominal operating luminosity. For tracks within the UT acceptance, the absence of UT hits is a powerful discriminating feature against fake tracks, which generally originate from incorrect combinations of hits in the VELO and SciFi detectors. Such tracks will hereafter be referred to as ghost tracks. The UT is also critical for reconstructing strange particles ($\KS$, $\Lambda$, $\Xim$, $\Omegam$). About 2/3 of these strange particles decay outside of the acceptance of the VELO, and are reconstructed using only the hits in the UT and SciFi detectors. Lastly, the addition of UT hits improves the overall momentum resolution by about 10\% relative to tracks formed without UT hits.

This paper reports on the commissioning and performance of the UT detector. All results are based on data collected in early 2025, where LHCb was running with $pp$ collisions at centre-of-mass energy 13.6\tev with the nominal operating luminosity. During normal $pp$ running, LHCb features a "fixed target" operational mode where protons collide with specific atomic nuclei that are injected into the beam pipe at approximately $z=-400$~mm~\cite{LHCb-TDR-020}. These $p$-nucleus interactions are sufficiently far away from the main $pp$ collisions, and have significantly lower rate than that of the primary $pp$ collisions. In addition, the LHC is capable to provide accelerate and bring into collision pairs of heavy nuclei to conduct unique studies of heavy-ion collisions.

\section{UT detector overview}
\begin{figure}[t]
\centering
\includegraphics[width=0.60\textwidth]{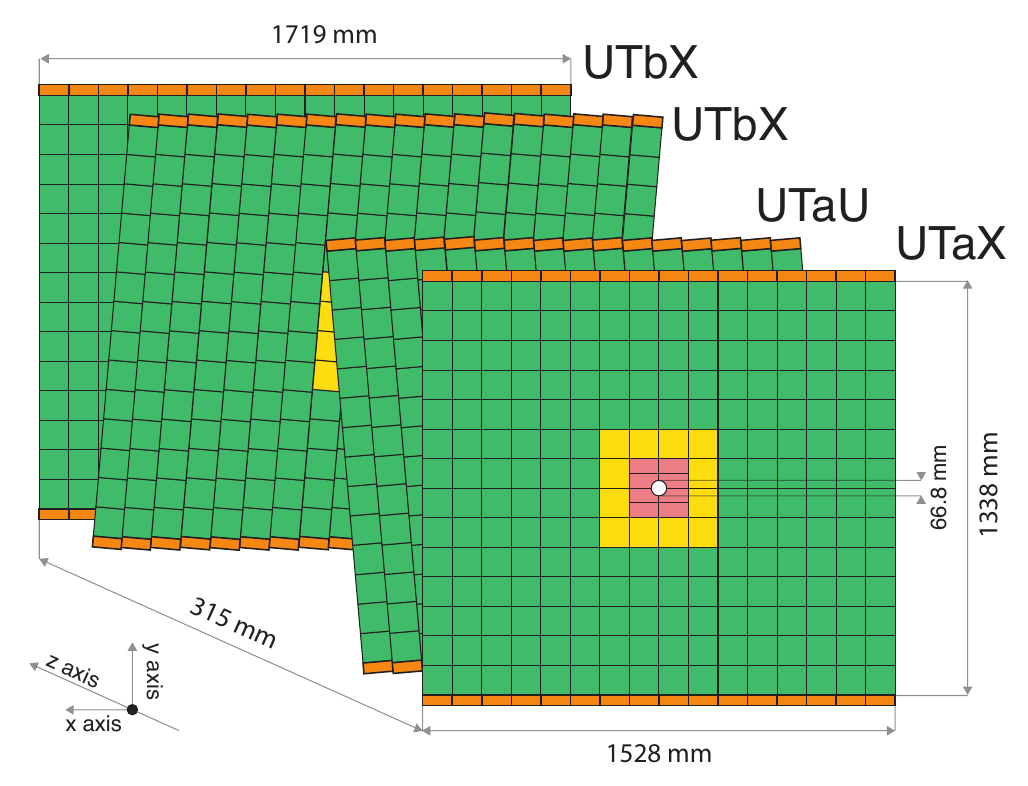}
\includegraphics[width=0.38\textwidth]{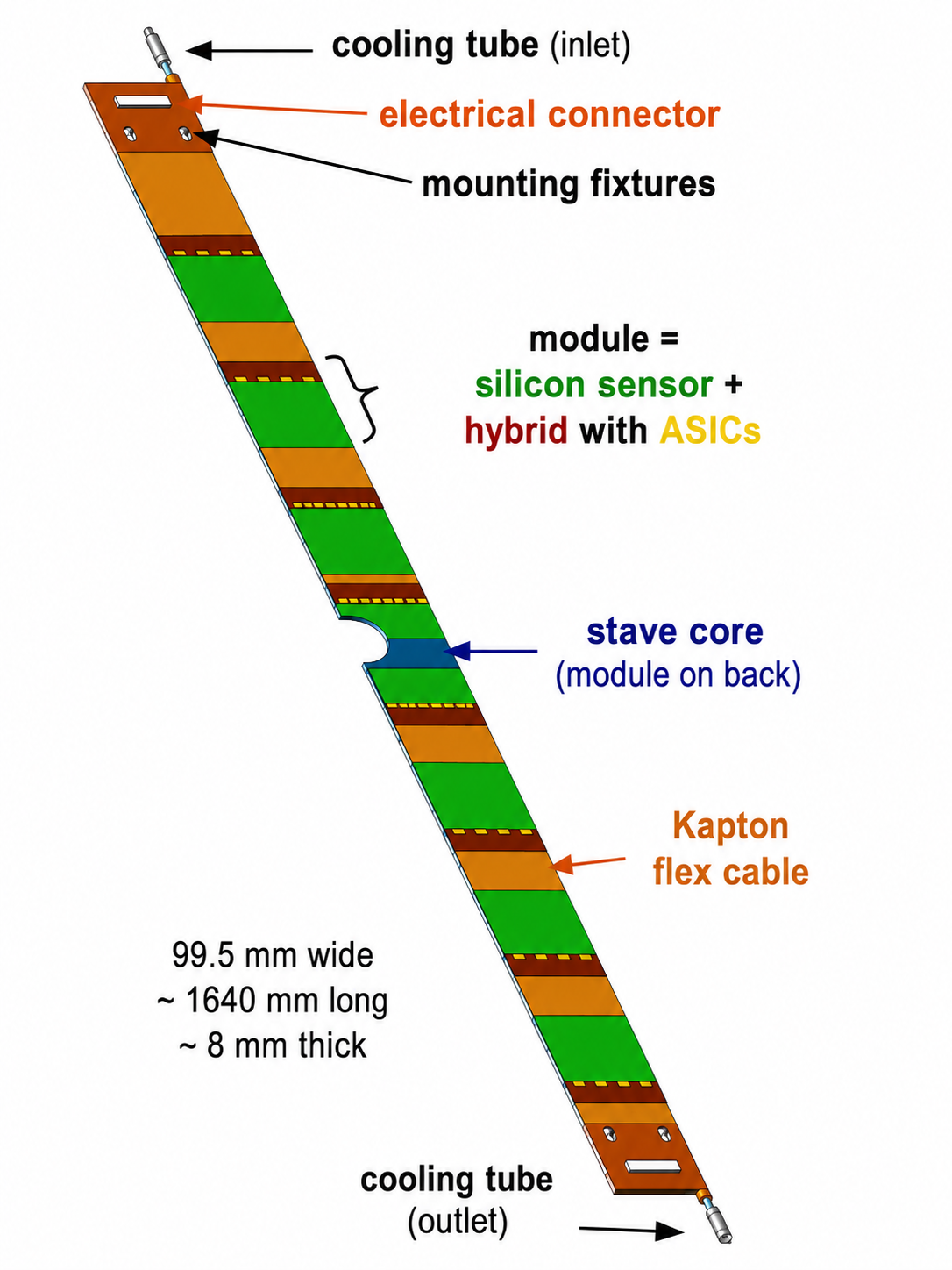}
\caption{\small{(Left) Diagram of the UT detector as seen from the interaction point in LHCb. Different sensor segmentations are shown by the different colours. The UTaX layer is closest to the interaction point, and UTbX the furthest. Each box corresponds to a unique module, with Type A (green), Type B (yellow), and Types C and D (pink), as described in the text. (Right) Drawing of a UT stave that is closest to the beam pipe.}}
\label{fig:UTDetector}
\end{figure}

The UT detector~\cite{LHCb-TDR-015} consists of four planes of silicon-strip detectors, as shown in Fig.~\ref{fig:UTDetector}\,(left). With respect to the vertical ($y$) direction, the strips in the four planes are oriented at $0^{\circ}$(UTaX), $-5^{\circ}$(UTaU), $+5^{\circ}$(UTbV) and $0^{\circ}$(UTbX), respectively, in order to optimize the position resolution in the bending ($xz$) plane of the magnet, while providing sufficient precision in the $y$ direction. Since the UTaU and UTbV hits contain both $x$ and $y$ information, the misassociation of UT hits to tracks is greatly reduced. The four planes are built from vertically-mounted double-sided staves (see Fig.~\ref{fig:UTDetector}\,(right)), which provide  mechanical support, cooling and routing of data, control signals, low voltage and sensor bias voltages needed to operate the detector. The double-sided design ensures that there are no gaps in the UT acceptance. Each stave hosts 7 or 8 modules on each side (face) depending on the location in the detector. The voltages and signals are routed to/from each module via a copper-Kapton flex cable. The UT detector is enclosed in a light-tight insulating box, that also serves as a Faraday cage. The heat generated by the electronics in the UT box, about 3.5 kW, is removed using an evaporative CO$_2$ cooling system~\cite{Coelli_2017} capable of cooling down to about -35$^o$C. The UT cooling system was set to -10$^{\circ}$C in 2024 and to $-15^{\circ}$C during the 2025 run.

\subsection{Silicon modules}
Each module is comprised of a silicon sensor, an electronics hybrid, and a supporting boron-nitride stiffener\footnote{Boron nitride was chosen for its high thermal conductivity, and high electrical resistivity.}. Four types of modules are integrated into the UT in order to provide low occupancy, excellent position resolution, and good signal-to-noise, which are important features for good performance in track pattern recognition. 
The main features of the four types of modules are shown in Table~\ref{tab:modules}. Since the particle flux falls exponentially with the distance from the center of the beam pipe, most of the UT active area is covered by type A modules with p-in-n (or n-type) sensors. A key consideration for the inner region of the UT is the high track density and, correspondingly, the large fluence expected to be accumulated over the course of Upgrade 1 (up to $\sim5\times10^{14}$ n$_{\rm eq}$\, cm$^{-2}$). Therefore, the modules close to the beam pipe, types B, C, and D, all have finer strip pitch, and are n-in-p (or p-type) sensors, which provide better charge collection properties under high irradiation. Types C and D are half as large in the vertical direction, and the four type D modules in each plane surrounding the beam pipe have a 34 mm radius quarter-circle cutout to maximize the acceptance close to the beam pipe. The C\&D type sensors are thinned to 250\,$\mu$m thickness, leveraging the fact that the shorter strips have lower noise. In total, the UT detector uses about 9 m$^2$ of silicon detectors and has 536,576 silicon strips.

\begin{table}[tb]
    \centering
    \begin{tabular}{c|c c c c c c}
        \hline
        Module/sensor &  Si type & pitch     & Number  & Number & Size  & thickness  \\
          type          &         & ($\mum$) & of strips & of sensors & (cm$^2$) &  ($\mum$) \\
        \hline
         A  & n-type & 187.5  & 512  & 888 & ~9.95  $\times$ 9.735  & 320\\
         B  & p-type & 93.5 & 1024 & 48  & ~9.95  $\times$ 9.735  & 320\\
         C  & p-type & 93.5 & 1024 & 16  & 5.145 $\times$ 9.735  & 250\\
         D  & p-type & 93.5 & 1024 & 16  & 5.145 $\times$ 9.735  & 250 \\
         \hline
    \end{tabular}
    \caption{\small{Sensor characteristics for the four types of UT modules/sensors. The type D modules/sensors correspond to those closest to the beam pipe.}}
    \label{tab:modules}
\end{table}

\subsection{Front-end ASIC for the LHCb Upstream Tracker}
\label{sec:SaltChip}

\begin{figure}[th]
\centering
\includegraphics[width=0.96\textwidth]{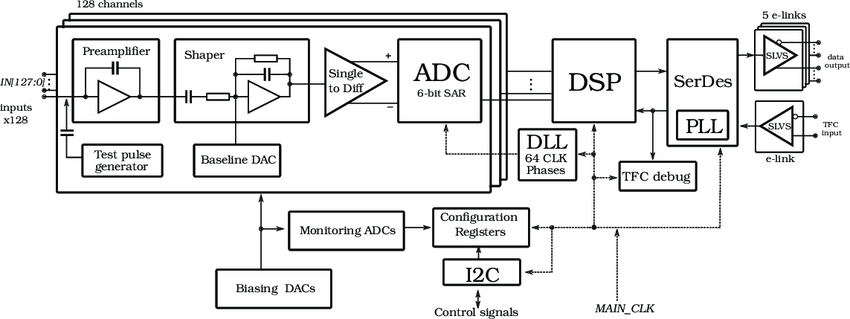}
\caption{\small{Block diagram of the SALT ASIC.}}
\label{fig:SaltAsic}
\end{figure}

The UT sensors are read out by  128-channel ASICs\, implemented in 130 nm CMOS technology\footnote{Application Specific Integrated Circuit.}  (SALT)~\cite{SaltAsic}. Each channel includes an analogue processor and an ADC. A digital signal processor block provides coherent noise subtraction and zero-suppression. The digitized signals are serialized and shipped out to the peripheral electronics via the copper-Kapton flex cables glued to the staves. A block diagram of the functional sections of the SALT ASIC is shown in Fig.~\ref{fig:SaltAsic}. The shaper circuit provides fast return to the baseline through a complex filtering network comprising real and complex poles and one zero~\cite{SaltAsic} , resulting in just a few percent spillover in consecutive beam crossings. A differential low-power 6-bit Successive Approximation Register (SAR) ADC digitizes signals at 40\,MHz. The device has the same dynamic range for  both positive and negative polarity, although it can be configured to have a higher dynamic range for one polarity. The baseline, namely the digitized value when no signal is present, can be equalized among channels either via an analogue current that shifts the DC working point of the shaper (baseline DAC) or by subtracting the mean ADC value of the channel in absence of a signal (pedestal). These parameters are determined via calibration scans and are stored in configuration registers inside the ASIC.

The digital back-end of the SALT ASIC incudes a digital signal processing (DSP) for pedestal subtraction, common-mode subtraction (CMS), and zero suppression. The CMS removes event-by-event common mode shifts of the baseline by subtracting the average ADC value of all the channels that have a digitized value within a certain window. With a nominal occupancy below 3\%, including all channels in the baseline estimation introduces a negligible bias on the mean. The window is therefore be set to the full dynamic range of the ASIC. The event-by-event baseline fluctuations have a standard deviation of approximately 1 ADC~\cite{Sowa_2025}.
Individual channels that are dead or noisy channels are removed via a mask register. 
Data produced in the DSP block are serialised and transmitted via SLVS\footnote{Scalable Low
Voltage Signalling (SLVS) standard.} e-links~\cite{Bonacini:1235849} to the peripheral electronics at 320\,Mb/s per link. The SALT ASIC  makes use of a Delay-Lock-Loop (DLL) circuit to make fine time adjustments to the sampling time in order to digitize the shaped pulse from a signal at the optimal time. The I2C provides the interface for signals from the slow control system, and is used for operations such as downloading and reading back the ASIC configuration, setting various chip registers and monitoring.

Each SALT ASIC can be configured to run in either non-zero suppressed (NZS) mode, where all channel data are saved, or in zero-suppressed (ZS) mode, where only channels with signals larger than a zero-suppression threshold, \zsth are transmitted. The value of \zsth is configurable, but applies to all channels in an ASIC. The ZS mode is used for normal operation, whereas NZS operation is used for  detector calibrations and other studies. During physics running, the detector operated in ZS mode with a nominal setting of \zsth$\ge4$.   

Another important feature of the SALT ASIC is the ability to configure the maximum number of strips, \nzsmax, to be transmitted. No hits are sent out if the limit is exceeded. The maximum of \nzsmax is 63 (50\% occupancy). To remove rare bursts with high hit counts that could cause readout problems, \nzsmax is set to 32, corresponding to about 25\% of the ASIC's channels. This loose requirement has a negligible impact on the UT hit efficiency. The performance of the SALT ASIC is optimized by tuning a variety of configuration parameters stored in the register section. The calibration steps pursued to choose the optimal value of the operationg conditions are discussed in Section~\ref{sec:asic_config}.

\subsubsection{Radiation hardness}
\label{sec:radhardasic}
Radiation hardness of the SALT ASIC is a critical requirement for the UT. In particular, a critical feature is the minimization of single event upset (SEU) in the configuration registers. Of particular importance are the two sets of registers storing the baseline equalization parameters. A loss of configuration in this case introduces spurious signals that adversely affect the data rates from the affected devices. 

Although the initial SALT versions did have triple modular redundancy on critical registers, a modest rate of SEU events was observed in proton irradiation studies~\cite{SHENG2024169356}. Specifically, bit-flips were observed more frequently in the baseline ADC registers than in the pedestal registers, and the flips were predominantly from "1" to "0". Mitigation strategies were implemented in a revised ASIC, however, due to a tight production schedule, SALT ASICs including this improvement arrived later, and were used primarily in the region closest to the beam pipe, where the radiation environment is harshest. In total, 896 of the 4192 ASICs deployed in the UT incorporate the improved baseline and pedestal register implementation.

Based on the SEU cross-section measurements from the test beam studies, and \fluka-based simulations~\cite{10.3389/fphy.2021.788253, Fluka_battistoni_2015} of the radiation environment in the \lhcb experiment, the estimated rate of SEU events from either the pedestal or baseline registers was estimated to be 3.5\,mHz at the nominal \lhcb instantaneous luminosity of $2\times10^{33}$~cm$^{-2}$\,s$^{-1}$. This level was deemed to be manageable for the full UT system with appropriate mitigation strategies. 

\subsection{Detector infrastructure}
\begin{figure}[tb]
\centering
\includegraphics[width=0.96\textwidth]{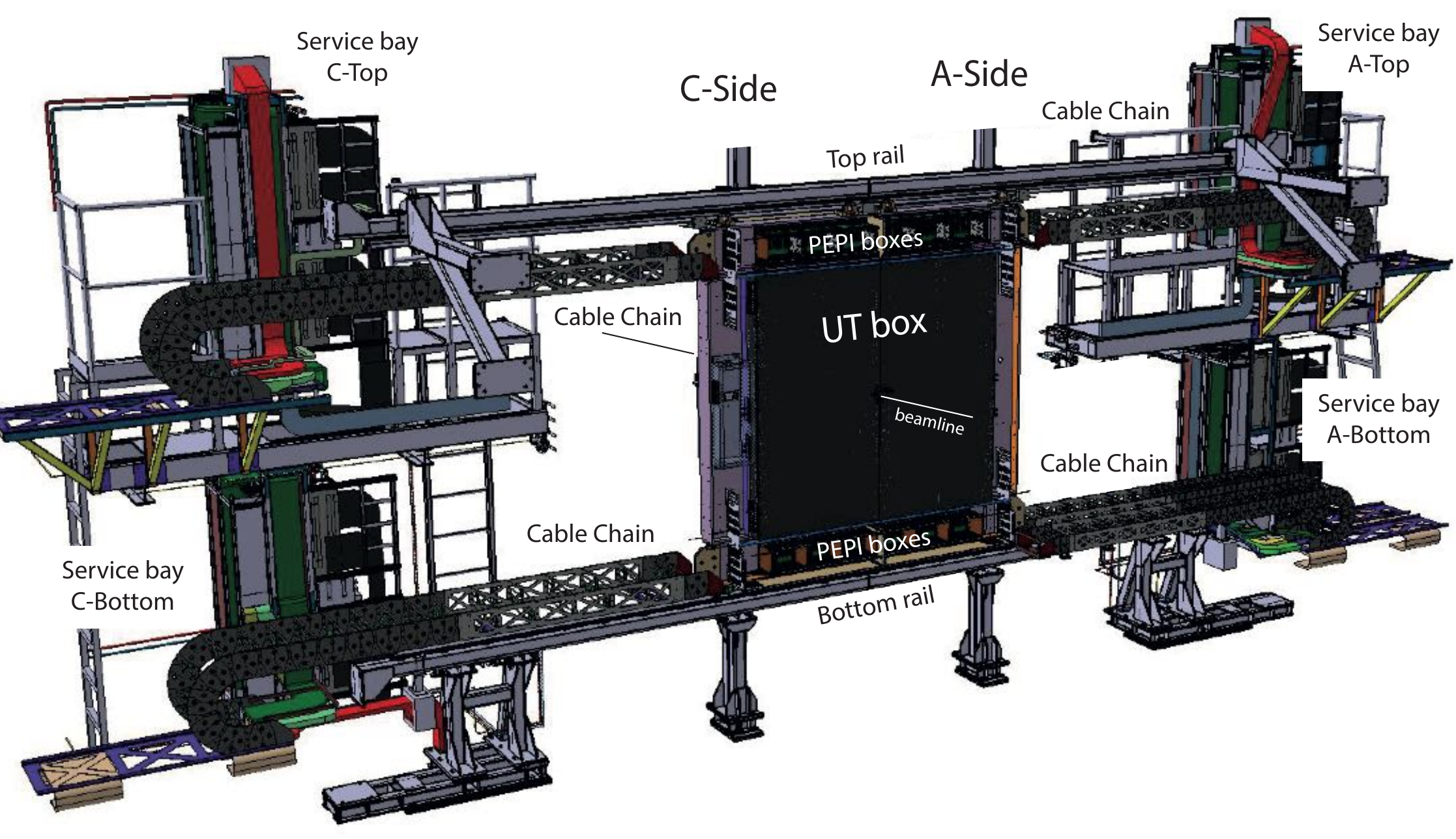}
\caption{\small{Three-dimensional rendering of the UT system.}}
\label{fig:UTFullSystem}
\end{figure}

\begin{figure}[tbh]
\centering
\includegraphics[width=0.85\textwidth]{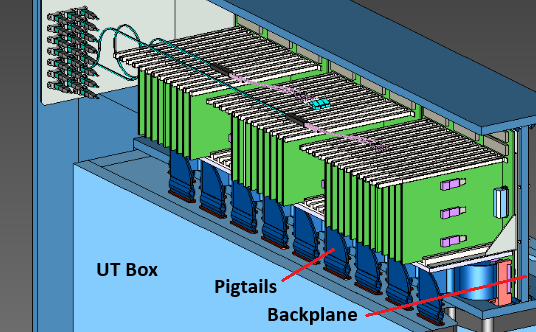}
\caption{\small{Schematic of the pigtail routing from the UT box to the PEPI backplane. The pigtails are shown in blue, and the data control boards (DCBs) are shown in green. The PEPI crates are mounted above and below the UT detector.}}
\label{fig:PEPI_DCBs}
\end{figure}

The layout of the UT system in the LHCb cavern is shown in Fig.~\ref{fig:UTFullSystem}~\cite{LHCb-TDR-015}. The UT box forms a Faraday cage around the detector and provides a dark, low-humidity environment for the sensors. Its top and bottom plates provide precision mounting points for the staves and incorporate slots for routing copper--Kapton flex cables, referred to as pigtails because of their curved shape, from the staves to the Peripheral Electronics Processing Interface (PEPI) crates mounted above and below the box, as shown in Fig.~\ref{fig:PEPI_DCBs}.
The pigtails carry data and control signals between the staves and the PEPI crates and route low-voltage power to all detector hybrids. The PEPI backplanes distribute data from the different stave types among modular data control boards (DCBs) to balance their data rates. The DCBs also provide an interface between the experiment control system and the SALT ASICs, including slow-control communication using the I$^2$C protocol. Further details on the DCBs are given in Section~\ref{sec:daq}.

All electrical and optical cables are routed through four cable chains to the service bays. These bays house low-voltage regulator boards with full differential remote sensing, together with patch panels for high-voltage channels, optical fibers, and other detector services. The two detector halves are mounted on a precision rail system that allows the detector to be opened when necessary and closed for normal data taking.

Operation of the UT detector relies on several subsystems: a high-voltage system to bias the sensors, a low-voltage system to power the DCBs and SALT ASICs, the experiment control and monitoring systems, and services for CO$_2$ cooling distribution, water cooling, dry air, and N$_2$ gas.

\section{Track reconstruction}
\label{sec:tracking}
\begin{figure}[tb]
\centering
\includegraphics[width=0.95\textwidth]{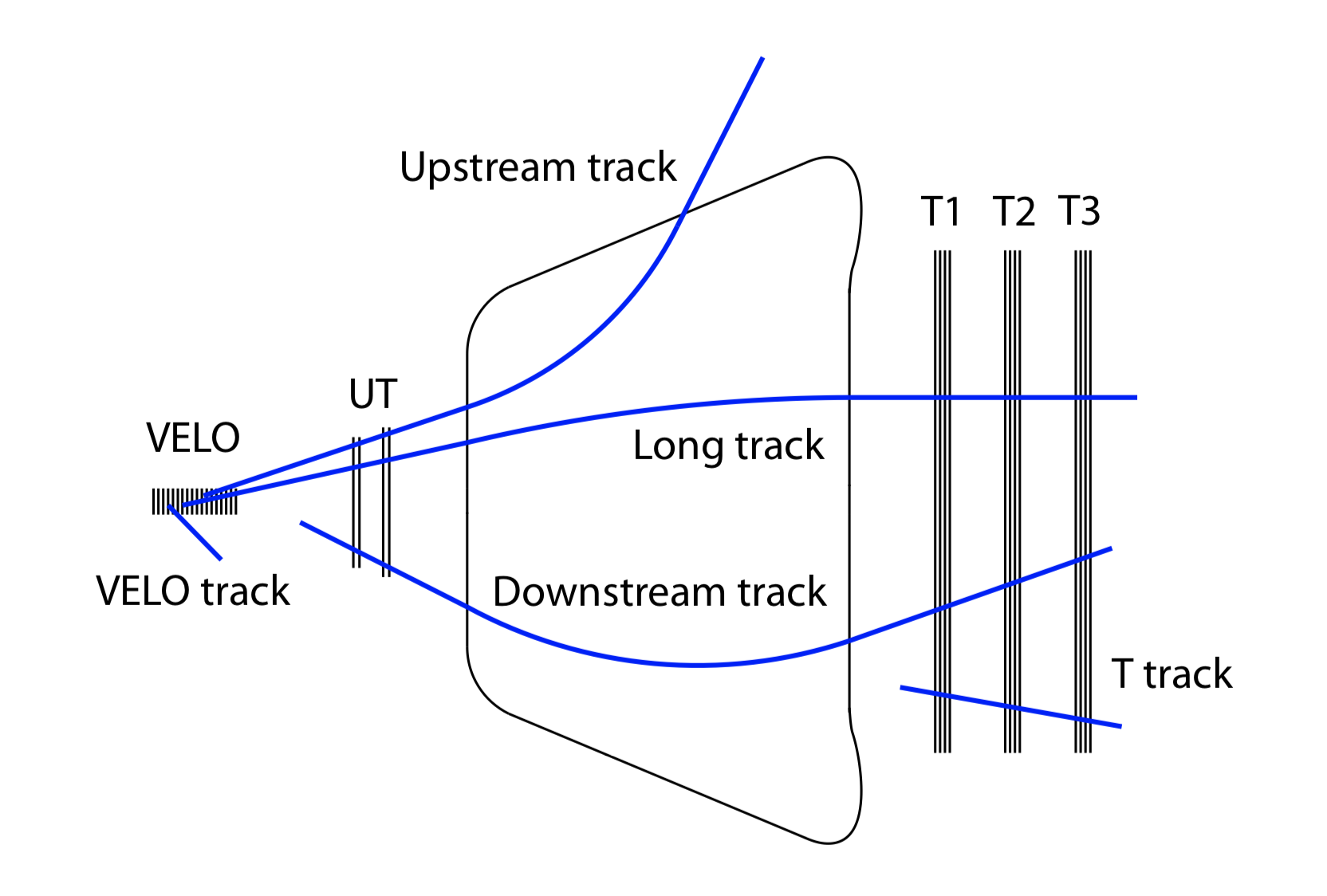}
\caption{\small{Overview of the different types of tracks reconstructed in the LHCb detector. The detectors labeled T1 - T3 are SciFi tracking stations.}}
\label{fig:TrackTypes}
\end{figure}

High track reconstruction efficiency and a low rate of ghost tracks are important to maximize the signal-to-background ratio in various $b$- and $c$-hadron decay channels. At design luminosity in triggered events there are, on average, approximately 130 forward-going track segments reconstructed in the VELO, of which about 100 (on average) are reconstructed as long tracks. An additional 65 VELO tracks, on average, are reconstructed in the backward direction, and are used in the reconstruction of primary vertices. The reconstruction of these tracks in real (or quasi-real) time in HLT1 and HLT2 present both combinatoric and computational challenges. 

The LHCb tracking system reconstructs several classes of tracks according to the sub-detectors contributing to the reconstruction, as illustrated in Fig.~\ref{fig:TrackTypes}. The three principal track types containing UT hits are long, downstream, and upstream tracks~\cite{LHCb-DP-2014-002}. Long tracks traverse the full tracking system, and generally have hits in the VELO, UT and SciFi detectors. Long tracks can also be reconstructed without the use of UT, in order to 
allow for long tracks outside the UT acceptance or UT inefficiencies. The vast majority of physics analyses use long tracks, since they provide excellent impact parameter (IP) resolution of $\sim$15\mum at high $\pt$, and an accurate momentum ($p$) determination, with $\Delta p/p\sim 0.5\%$ to 1.0\% from 3 to 100 \gevc. Downstream tracks are reconstructed from SciFi and UT hits when insufficient VELO information is available, as is typically the case for the decay products of long-lived strange particles. Upstream tracks include only VELO and UT hits, but have worse momentum resolution ($\Delta p/p\sim10\%)$, and are typically not used in physics analyses, except for flavour tagging.

\subsection{Long tracking}
Two algorithms are used to form long tracks, forward and matching. The forward algorithm forms VeloUT track segments from hits in the VELO and UT detectors~\cite{Bowen:2014oqx}, and with knowledge of the magnetic field integral, a measurement of the charge-to-momentum ratio ($q/p$) with about 10\% precision is obtained. With each VeloUT track, a momentum-dependent search window in the SciFi detectors can be formed to search for compatible SciFi hits. The limited search windows in the SciFi detector reduce the hit combinatorics, allowing the algorithm to run faster. The matching algorithm forms SciFi track segments from SciFi hits, and those segments are matched to VELO track segments, accounting for the effects of the LHCb magnetic field. After the track match is made, compatible UT hits, if present, are added to form the final long track.

For HLT1, the forward algorithm is run first, and the matching algorithm follows, only using hits that are not used on long tracks formed in the forward algorithm. For HLT2, there is much more time available due to about a factor of 30 reduction in rate by HLT1, and both algorithms are run on the full complement of hits. An algorithm then compares the tracks reconstructed by the two algorithms, and saves only those that are unique. Long tracks are fit using a Kalman filter fitter~\cite{KalmanFilter} to give the best estimate of their momenta. 

The detector resolution, the long extrapolation distance between the VELO and SciFi detectors, and the high track density result in a non-negligible fraction of ghost long tracks. These result from the incorrect association of VELO and SciFi segments belonging to different charged particles.
In most cases, but not all, these ghosts are formed from VELO and SciFi segments from two different charged particles. The presence of 2 or more UT hits on a long track provides a strong discriminant against ghost tracks, since the presence of compatible UT hits essentially confirms the VELO-SciFi match. This suppression is particularly important at large track multiplicity or in regions of high track density.

\subsection{Downstream tracking}
The downstream tracking algorithm is primarily used to reconstruct the decay products of long-lived strange particles, such as $\KS$, $\Lz$, $\Xim$ and $\Omegam$, which decay outside the VELO acceptance and upstream of the UT detector. Hence, the downstream algorithm is designed to reconstruct tracks that do not have hits in the VELO detector.

Downstream tracks are reconstructed in both HLT1 and HLT2 by extrapolating track segments reconstructed in the SciFi detector (SciFi seeds) to the UT, taking into account the magnetic field in the bending plane.

In HLT1, the downstream tracking is performed by an algorithm~\cite{Kholoimov:2025cqe} optimised for speed and performance. The SciFi seeds are reconstructed~\cite{Aiola:2020ydy} and propagated backwards to the last UT layer to find compatible hits. A new trajectory is computed for each compatible hit attached to the track and then it is extrapolated to the remaining UT layers.
The hit closest to the predicted trajectory in each layer is selected to form downstream track candidates.
A final downstream track candidate is retained for each SciFi seed, after removing duplicate tracks. The track momentum resolution achieved after the HLT1 reconstruction of downstream tracks is better than 2\% for momenta above $5\gevc$. The average reconstruction efficiency is about 75\% for both $\KS$ and $\Lz$ particles.

The downstream tracking in HLT2 follows a similar procedure but employs a different algorithm, which was migrated from the Run2 algorithm~\cite{Davis:2017ccl}. Contrary to HLT1, downstream track candidates in HLT2 only require hits in three out of the four UT layers.
The selected candidates are then passed to a Kalman filter algorithm to fit for the track parameters, yielding a momentum resolution comparable to that achieved in Run 2 (approximately 1\%).

\section{Data acquisition system}
\label{sec:daq}
\begin{figure}[tb]
\centering
\includegraphics[width=0.98\textwidth]{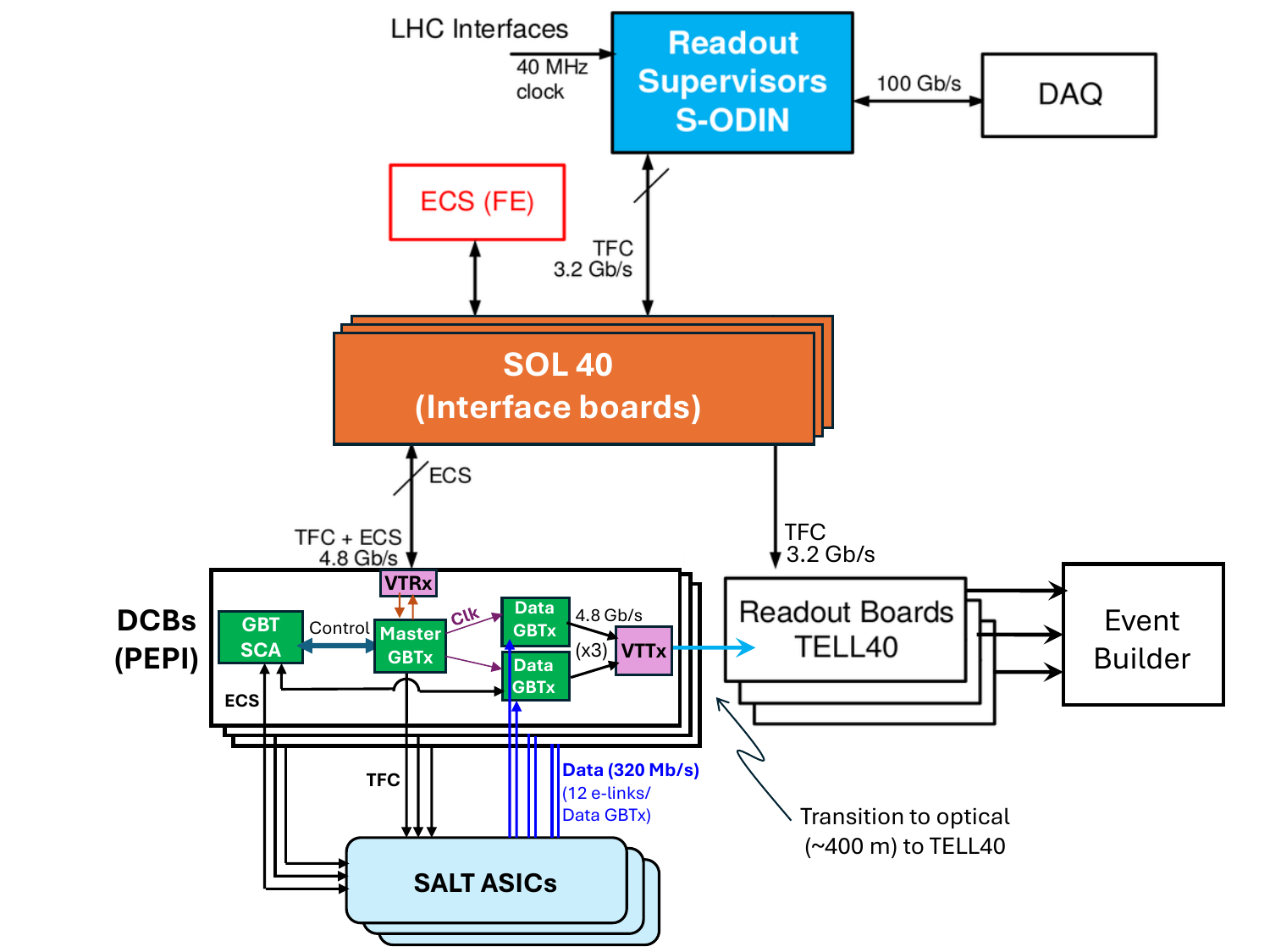}
\caption{\small{Simplified diagram of the LHCb readout and DAQ system. For more details, see Ref.~\cite{Wyllie_Alessio}.
}}
\label{fig:LHCb_DAQ}
\end{figure}
The LHCb data acquisition system (DAQ) is responsible for distributing the slow and fast control signals that manage the operation of the detectors during both normal data taking operations, as well as  in the various calibration modes necessary to tune the detector's operating parameters. The DAQ also handles the processing and organization of the information from the front-end devices to be transmitted to the real time analysis processors that form the HLT. A simplified diagram of the system is shown in Fig.~\ref{fig:LHCb_DAQ}. 

The data control boards (DCBs) receive data from the front end ASICs and provide the communications between the UT hybrids and the DAQ. The experimental control and data transmission between front-end devices and DAQ boards are distributed by Gigabit Transceiver (GBTx) ASICs\cite{Moreira:2009pem}, which operate together with Versatile Twin Transmitter (VTTx) ASICs\cite{CMS:2009chw} \cite{CMS:2009chw} to form  bi-directional serial bitstreams with a serial rate of 4.8 GB/s. Each DCB includes a master GBTx that distributes the timing and fast control (TFC) signals from the online system to the SALT ASICs, one GBT-SCA (GigaBit Transceiver Slow Control Adapter) providing a bi-directional flow of slow control signals, and six data GBTxs, which coordinate the data transfer from the front end to the readout boards (TELL40). 

The DAQ implements three back-end boards based on the same physical hardware, a custom-made PCIe40 board\cite{Cachemiche_2016,Colombo_2018,refId0}.
Each PCIe40 board contains a large FPGA that can be programmed with one of three different flavours of firmware that provides its functionality as either a S-Odin, SOL40 or TELL40 board.

The S-Odin board receives the LHC clock signal and generates the TFC signals that phase-align the different clocks distributed across the detector planes with fixed latencies. The SOL40 boards provide the main interface to the front-end electronics and the TELL40 readout boards. They fan out the timing and trigger signals to the TELL40 boards via an optical network and relay the same information to the front-end electronics. In addition, all Experiment Control System (ECS) commands are sent via bi-directional links from the SOL40s to the SALT chips via the GBT-SCA ASICs on the DCBs. The read-out system and the clock distribution are described in more detail in Ref.~\cite{TFC_10115510}.

The TELL40 boards are hosted in the event builder (EB) servers in the LHCb data centre on the surface, located approximately 400\,m from the LHCb experimental area. Each TELL40 boards hosts an FPGA (Intel$^{\rm TM}$ Arria 10), which is organized into lanes that feature several firmware processing blocks~\cite{abellan2025data}. Most of the blocks are designed centrally for all of the subdetectors, however a few detector-specific blocks are designed by developers working in the corresponding subdetector teams. One of the main challenges to the firmware code is to assemble fragments of the same event reaching the TELL40 board at very different times due to the different latencies and steep radial dependence of the number of hits. Smooth data processing relies on buffers that retain hit information until all hits associated with an event are available. Two blocks are particularly important: the Alignment block, which time-aligns asynchronously arriving data from different ASICs, and the Data Processing (DP) block, which combines the hit information from the ASICs to produce the event output. 
Periodically the buffers associated with the Alignment and DP blocks can exceed a safe occupancy for smooth data processing. In particular, if the buffer occupancy exceeds 80\% (40\%) in the Alignment (DP) block, the firmware algorithm temporarily truncates the incoming data in that TELL40 while the buffered data continues to be processed. Once the occupancy of the buffer in the Alignment (DP) block drops below 30\% (20\%), normal data processing is resumed. The tuning of the SALT ASIC parameters, \zsth and \nzsmax, as discussed in Section~\ref{sec:SaltChip} mitigates the impact of noisy channels and events with exceptionally high occupancy. Because the TELL40 protection mechanism is rarely activated, and the \nzsmax threshold is high, the loss in hit efficiency due to these protection mechanisms is negligible. 

The TELL40 boards assemble and deliver multi-event packets (MEPs) to the host EB server through the PCI interface, where data from all sub-detectors in a given event are combined. The event is then passed to the HLT1 software reconstruction, which runs in the same EB server.

In order to achieve optimum performance, there are several calibration procedures
that can be aggregated into two broad categories, timing alignments, including phase alignment between the different components, and baseline equalization. These calibrations are discussed in the next section. 

\subsection{Data and control boards}
\label{sec:DCBs}

The initial configuration parameters for the master GBTx in the DCBs were determined during the R\&D phase of the project. They were fused to the GBTx via separate E-fuse arrays for each configuration register during electronics assembly, allowing the GBTx to self-configure to an operational state after power-up. The current state of the master GBTx can be read from an internal register to check that it is fully operational. The SCA and data GBTx can then be programmed via their slow control interface, and their states can also be read back from internal registers.  

To validate the data flow architecture from the front-ends to the TELL40, both the GBTx and the SALT ASICs can be programmed to generate either fixed patterns or pseudorandom bit sequences (PRBS). The GBTx receiver on the TELL40 can be configured to count errors in the data and to estimate the bit error rate (BER) in the known data sequence. 
The data transmission from the DCB to the TELL40 was tested by configuring each data GBTx in turn to transmit PRBS data and checking that data was received on the expected link on the TELL40. The signal integrity was then measured by running BER tests with all electronics powered.  All control boards were tested to have $\mathrm{BER < 10^{-12}\ @\ 99.9\%\ c.l.}$ corresponding to the errorless transmission of approximately $1.5\times10^{13}$ bits.

\subsection{Calibration and front-end configuration}

The overall performance of the UT is critically dependent on the calibration of the front-end electronics. This section describes the procedures followed to optimize the configurable parameters in the SALT ASICs and the calibration procedures to monitor the optimal working point and the electronic noise.

\subsubsection{Configuration of SALT}
\label{sec:asic_config}

The SALT ASIC receives an external 40\,MHz clock that is used in the digital signal processing block (see DSP in Fig.~\ref{fig:SaltAsic}).  The phase of this clock relative to the ADC sampling clock is aligned using an internal DLL, while an ultra-low-power phase-locked loop (PLL) generates a 160\,MHz clock for data serialisation and transmission, as well as for the deserialisation of TFC commands. The initial configuration of the DLL and PLL was performed as described in Ref.~\cite{SaltManual}.  After the initialisation of these blocks, the serialiser and deserialiser circuits were tuned to ensure reliable communication both to and from the ASICs and the off-detector electronics.  

Reliable communication between the SALT ASICs and the off-detector electronics requires that the output data stream from each e-link on the SALT ASIC to be synchronized when they arrive at the TELL40 boards. During normal data-taking operations, the data from the DSP is serialised and transmitted to the data GBTx on the DCBs using up to five e-links (see Fig.~\ref{fig:SaltAsic}).  However, the {\it SerDes} circuit can also process data from alternative sources, including a fixed bit pattern, which can be selected using a control register in the SALT (see Fig.~30 in Ref.~\cite{SaltManual}). 

\begin{figure}[tb]
\centering
\includegraphics[width=0.9\textwidth]{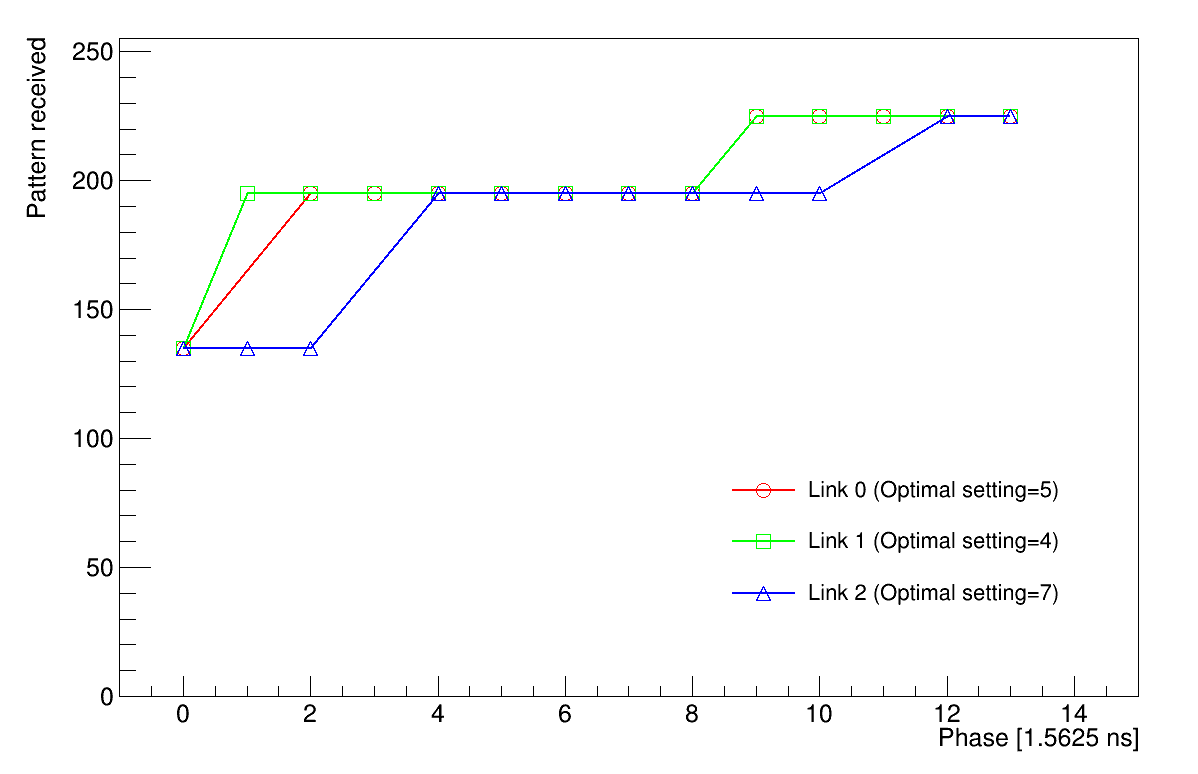}
\caption{\small{Decimal value of the bit pattern received on the TELL40 versus the phase on the data GBTx receiver (in units of 1/16 $\times$ 25 ns) for each e-link in one SALT ASIC. In this example, three e-links are used to serialise the data from the ASIC, and the optimal phase setting for each e-link was chosen as the midpoint of the region where all of the e-links receive the same pattern.
}}
\label{fig:GBTx_Tuning}
\end{figure}

To synchronize the data from each of the e-links on a SALT ASIC, a fixed pattern is generated on the SALT ASIC, selected as input to the {\it SerDes} circuit as described above, and decoded at the TELL40. The phase adjustment for each e-link is handled by the data GBTx on the DCBs, where any one of 16 values, each 1.5625 ns apart, can be selected. 
The optimal phase was chosen to be the value where all e-links from a given ASIC had the same decoded bit pattern, as shown in Fig.~\ref{fig:GBTx_Tuning}. At this stage, the pattern decoded in the TELL40 does not necessarily match the pattern generated in the SALT.  A common bit-shift is applied to each serialised byte in the serializer circuit to ensure that the bit-stream decoded in the TELL40 is exactly the same as the bit-stream generated by the SALT ASICs. Finally, the integrity of the data link was verified by running a BER test using the PRBS generator implemented in the SALT.

The SALT ASICs receive TFC commands from the SOL40 synchronously to the main 40\,MHz clock.  A single deserialiser circuit in the SALT is used to decode the incoming TFC commands, contained in a single byte received every 25\,ns, which are then used to control the DSP and back-end data processing.  The deserialised data stream must be properly bit-aligned with the 40\,MHz clock so that the correct TFC commands are decoded.  Similarly to the serialiser block, the first bit of a byte in the deserialised data stream can be defined using a configuration register.  The calibration of this register was performed by sending a known TFC pattern from the SOL40 and then transmitting the pattern decoded in the deserialiser block to the TELL40s using a dedicated loopback mode in the SALT.  The optimal phase was selected as the one where the transmitted and decoded command patterns were identical.

\subsubsection{Signal processing and noise performance} 
\label{sec:signal_proc}

In the absence of any signal, the measured ADC values for all the channel outputs have Gaussian distributions around a fixed number (ADC offset or baseline), with a standard deviation determined by the total electronic noise. The SALT chip has a symmetric dynamic range for positive and negative charge injections. We choose to operate it with a nominal baseline equal to the mid-range (nominal 0). The width of the baseline distribution is affected by two terms: incoherent noise, due to the SALT electronic noise, and coherent noise, where the outputs of all channels within a chip fluctuate coherently. The latter noise is subtracted with a CMS suppression algorithm," built in the SALT DSP, thus the only noise that affects the threshold, ideally, is the electronic noise. In principle, each channel has its own baseline value; the SALT ASIC provides two mechanisms for baseline equalization, one in the analogue section of the ASIC, and the second in the digital back-end.

\begin{figure}[tb]
\centering
\includegraphics[width=0.98\textwidth]{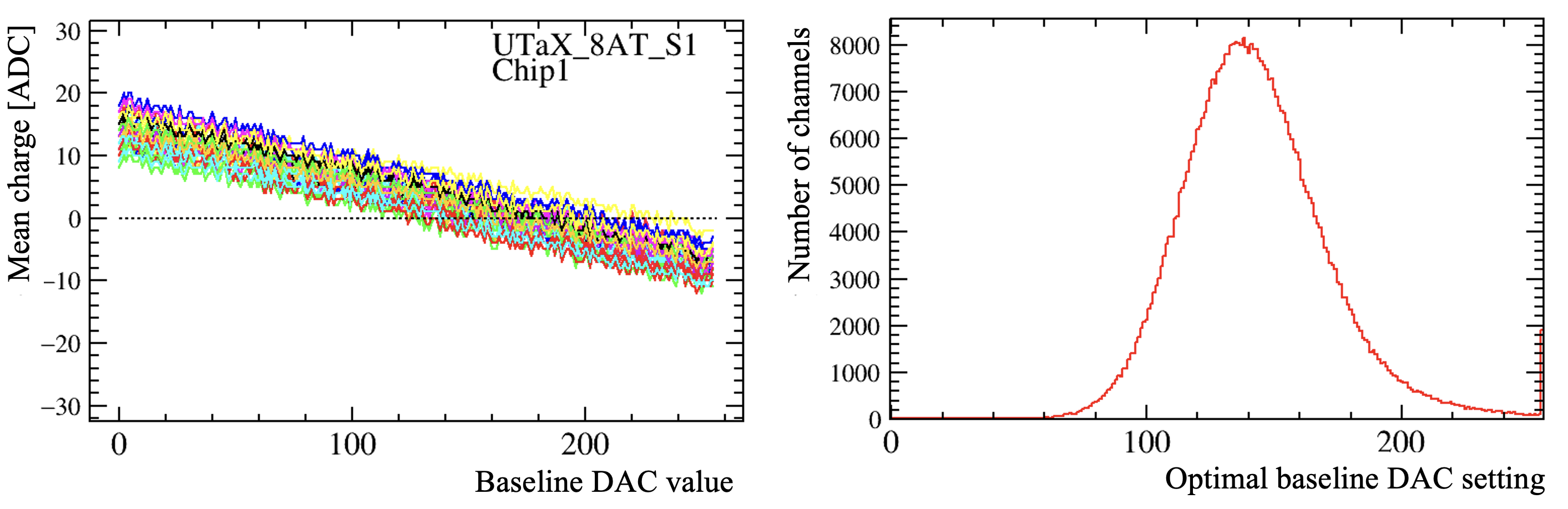}
\caption{\small{(Left) Example of a baseline DAC scan for 128 channels of a SALT ASIC, where each colour represents one of the 128 channels. (Right) Distribution of the optimal baseline DAC value for all channels in the UT. }}
\label{fig:TrimDAC}
\end{figure}

The primary method uses an 8-bit baseline DAC register (see Fig.~\ref{fig:SaltAsic}), allowing fine adjustments of the baseline voltages. A typical calibration scan is illustrated in Fig.~\ref{fig:TrimDAC}. The DAC values that determine the analogue settings of the channel baselines are systematically varied, and the corresponding mean ADC values are measured. The optimal baseline DAC settings aims to equalize all the baseline values to a mean value of zero using a linear fit to the scan data for each channel. 

This method provides baseline equalisation for most channels. However, in approximately 0.4\% of cases, the baseline DAC tuning is not adequate to bring the baseline value to 0. For these outlier channels, the pedestal calibration, described below, is employed to correct any residual offset not accounted for by the baseline corrections.

With the optimal baseline DAC set, NZS data is collected from which the mean value for each channel (the pedestal) and the standard deviation (total noise) are computed.  The pedestal values are rounded to the nearest integer and the 128 values are stored in the corresponding 6-bit registers of the SALT ASIC. Since the pedestal value can be negative, the two’s complement representation is used to encode both positive and negative values. Baseline DAC tuning and pedestal calibration are performed during periods without beam-beam collisions in LHCb to prevent collision-induced bias in the measured baseline and pedestal values. Together, they equalise the baselines of all readout channels and maximise the efficiency of hit processing. The nominal baseline DAC and pedestal values are stored in a configuration database for when the SALT ASICs require initialization. 

Decoding and analysis of the NZS data acquired during the baseline DAC and pedestal calibration scans are performed using the Vetra software package~\cite{Vetra}. The calibration algorithms are implemented within the Gaudi framework~\cite{Gaudi}, which serves as the standard event-processing architecture in LHCb. As mentioned previously, event-by-event common-mode (CM) shifts of the baseline are also removed. The SALT ASIC also allows for the masking of dead or noisy channels.

\begin{figure}[tb]
\centering
\includegraphics[width=0.55\textwidth]{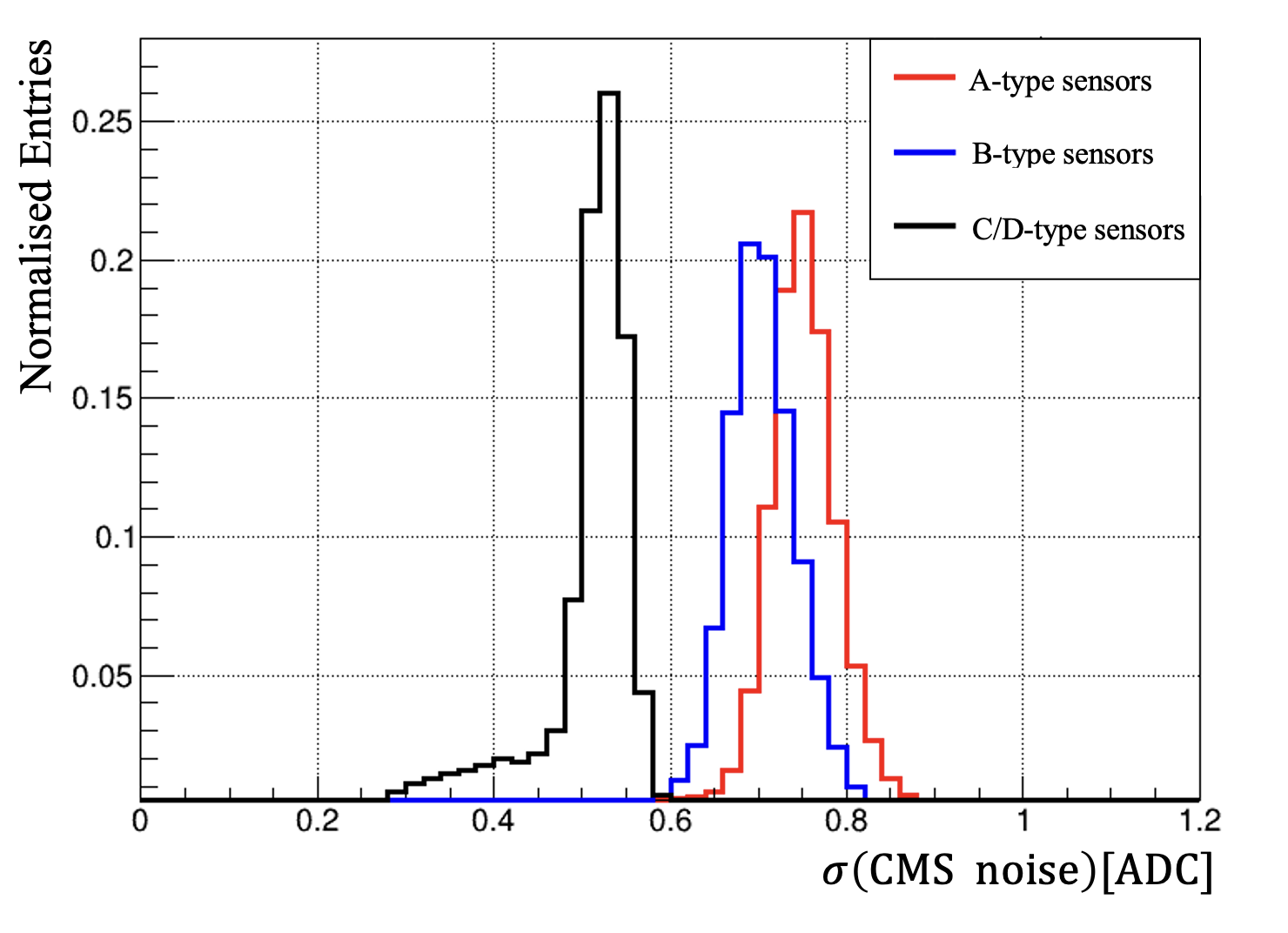}
\caption{\small{Distributions of RMS ($\sigma)$ of ADC values after pedestal and common-mode noise subtraction for all UT channels in NZS data.}}
\label{fig:RMSNoisePerASIC2D}
\end{figure}

Figure~\ref{fig:RMSNoisePerASIC2D} shows the root-mean-square (RMS) values of the ADC distributions for all UT channels, after subtraction of both pedestal and common-mode noise. The results are normalised to account for the different numbers of channels of each sensor type. As previously noted, Type A and B sensors exhibit higher noise levels compared to Types C and D due to their longer strip lengths. Type D sensors are located near the beam pipe, where the strip lengths decrease from approximately 50 mm to about 16 mm. Their shorter strip lengths account for the low-noise tail observed in  Fig.~\ref{fig:RMSNoisePerASIC2D}.

Pedestals are monitored, and variations over time of about 0.7 ADC counts are typical. Pedestal updates are performed routinely whenever operating conditions change or SALT ASIC settings are adjusted. The behaviour of both the baseline DAC and pedestal values has been systematically studied under various conditions, including the impact of temperature variations, as discussed in the following.

\subsubsection{{Temperature dependence of noise and calibration parameters}}

\begin{figure}[tb]
\centering
\includegraphics[width=0.95\textwidth]{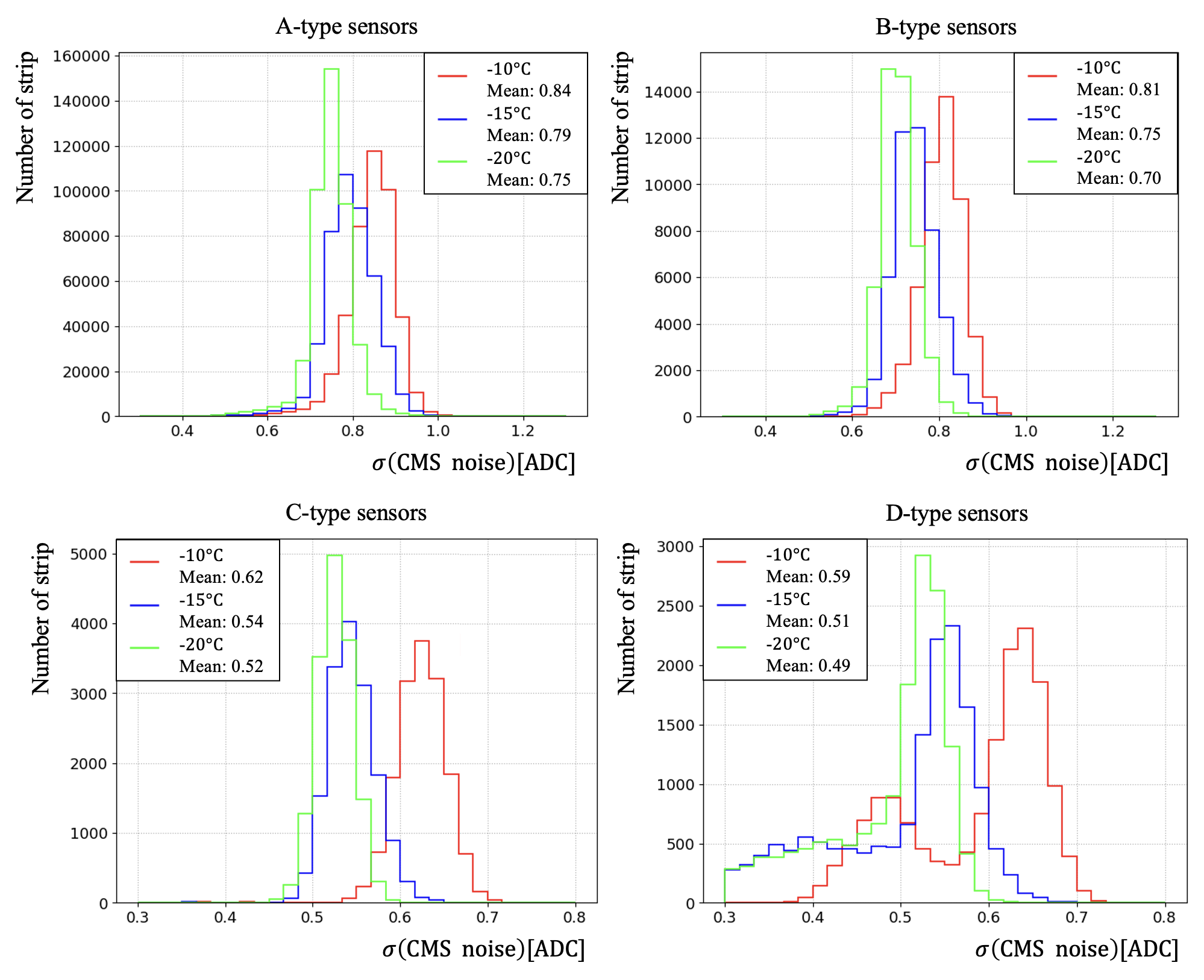}
\caption{\small{Comparison of distributions of RMS of CMS ADC for each sensor type at 3 different temperatures. At each temperature, the optimal UT parameters are recalibrated, as described in the text.}}
\label{fig:noise_temp}
\end{figure}

During the 2024 data-taking cycle, the UT operated at about $-10\,^{\circ}$C. Prior to the start of data collection in 2025, studies were conducted to determine the optimal running temperature, taking into account electrical and thermo-mechanical considerations. Data samples were collected at three different temperatures, $-10\,^{\circ}$C, $-15\,^{\circ}$C and $-20\,^{\circ}$C. The DC working point of the SALT ASICs exhibits a small temperature dependence. In particular, the optimal baseline DAC settings, pedestals, fine time alignment (c.f. Sec.~\ref{sec:fine_timing}), and ZS thresholds were re-evaluated, as the optimal values change with temperature. After recalibration at each temperature, the RMS of the CM-subtracted noise per strip was computed, and the results (by sensor type) are shown in Fig.~\ref{fig:noise_temp}. The noise decreases by (10--15)\% as the temperature is lowered from $–10^{\circ}$C to $-20^{\circ}$C. While $-20\,^{\circ}$C was achievable, it was chosen to operate at $-15\,^{\circ}$C to avoid thermo-mechanical stress and provides a better homogeneity in the ASIC setpoints.

\subsubsection{Time alignment}
\label{sec:fine_timing}
The synchronization of the SALT ASIC e-links discussed in Section~\ref{sec:asic_config} only accounts for the timing from the {\it SerDes} circuit to the TELL40s, and therefore it does not account for the time synchronization between the signal produced in the UT due to a given $pp$ beam crossing and the decoded data at the TELL40s. In order to ensure that the maximum detectable charge is collected at the front-end, it must be sampled (i) in the correct bunch crossing, and (ii) at precisely the correct time within the correct bunch crossing. These two steps are handled by first performing a coarse time alignment followed by a fine time alignment.

The coarse time alignment is performed using Time Alignment Events (TAE), in which data are collected from consecutive bunch crossings centred around well-isolated single bunch crossings in each LHC orbit. In these events, the correct bunch crossing is unambiguous as they are isolated from any other $pp$ collisions.  The LHCb data acquisition system can be configured to record the signals from several consecutive bunch crossings around the trigger signal. The coarse time alignment analysis is set at the level of the SOL40 link, corresponding to one DCB, which delays the TFC commands sent to the front-ends.  The algorithm finds the optimal offsets such that all ASICs within the DCB have the correct bunch crossing within $\pm1$ of the expected one.

\begin{figure}[tb]
\centering
\includegraphics[width=\textwidth]{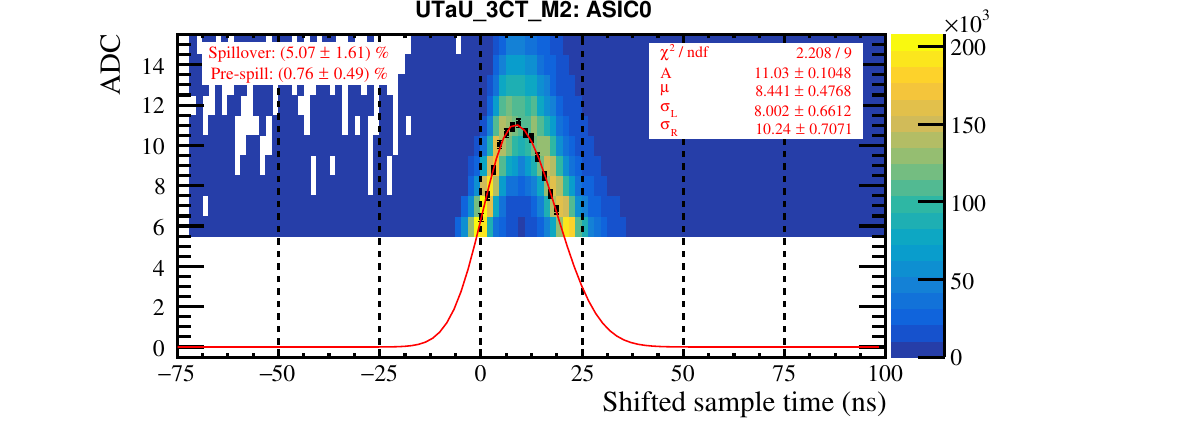}\\
\vspace{0.2cm}
\includegraphics[width=\textwidth]{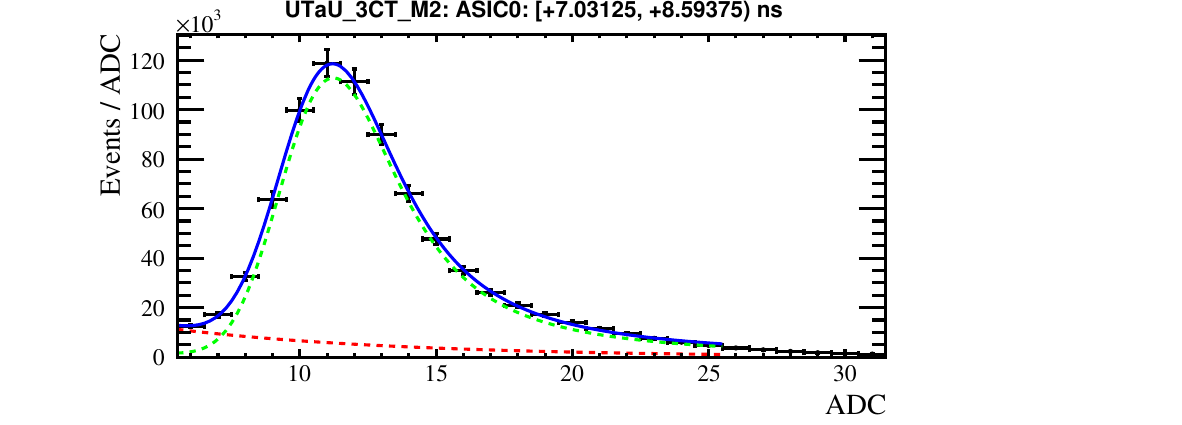}
\caption{\small{(Top) ADC vs. shifted sample time distribution and reconstructed pulse-shape obtained for a specific ASIC. Bottom: ADC distribution of the same ASIC at the optimal sampling phase, along with the fit as described in the text.}}
\label{fig:PulseShapeFTA}
\end{figure}

The fine time alignment is performed after the coarse time alignment using TAE data with single-strip hits in each sensor. Separate data samples are collected in fine time steps of 1.5625\,ns using the adjustable ADC sampling phase within the SALT ASIC. For each time step, NZS data is collected, in order to apply accurate pedestal corrections for the full scan. At each phase, the ADC distribution is fit to a Landau distribution convoluted with a Gaussian function on top of an exponential background to obtain the most probable value (MPV) of the Landau function. The distribution of MPV versus time is then fit to a bifurcated Gaussian function to obtain the optimal sampling time based on the peak value, as shown in Fig.~\ref{fig:PulseShapeFTA}(top) for a typical ASIC. The ADC distribution at the optimal sampling time along with the fit is shown in the bottom figure. The optimal sampling time is subsequently applied to the relevant registers of the SALT ASICs each time they are configured. The fitted function also predicts the amount of pre-spill (25 ns before the sampling point) and the spillover (25 ns after), which are both at the few percent level.

\section{Experiment control system and monitoring}

The UT detector was commissioned over the first two months of the 2024 LHC run. Thereafter, continuous monitoring was used to maintain stable operation and ensure high data quality.

\subsection{Experiment control system}

The LHCb ECS is responsible for the configuration, monitoring, and control of all hardware components of the experiment.  It is based on a commercial SCADA\footnote{Supervisory control and data acquisition.} system, WinCC-OA\footnote{http://www.etm.at/.}, with CERN extensions as described in Refs.~\cite{LHCb_ECS, LHCb-DP-2022-002, JCOP}.  This common framework was used to implement the control systems for all of the different LHCb sub-detectors.

The ECS allows users to power the detector on or off, to configure the electronics, and to start the DAQ, using pre-defined commands that read the relevant parameters from a configuration database and then apply them to the hardware.  The system includes a high level of automatic actions that minimise the need for human intervention when problems are detected.  The ECS also includes tools for the monitoring of temperatures, voltages, currents of the various hardware components plus other environmental conditions, such as the humidity and dew point inside the detector box.  In total, over 11,000 parameters are monitored.

\subsubsection{{Low voltage}}
\newcommand{\Vreg}{$V_{reg}$}
\newcommand{\Vrs}{$V_{rs}$}
\newcommand{\Vis}{$I_{s}$}
The Low Voltage (LV) system supplies regulated low-voltage power to the hybrids (1.2\,V) and the DCBs (1.5\,V and 2.5\,V), serving a total of 1,408 channels via 268 identical 8-channel low voltage regulator boards (LVRs), each configured to output one of the three required voltages. The voltage regulation is achieved using the LHC4913 low-dropout voltage regulator, specifically designed to be radiation-tolerant, and equipped with a precision remote-sense circuit, implemented with a commercial operational amplifier qualified for the expected radiation dose at their location (approximately 20 KRad). 

The LVRs are housed in the water-cooled service-bay racks outside the UT box, and are connected to the UT through the cable chains (see Fig.~\ref{fig:UTFullSystem}). Dedicated Telemetry Control Boards (TCBs), each serving 16 or 17 LVRs, provide the ECS interface to the LVRs and collect telemetry data from the regulators. Among the many internal readings, three key parameters are closely monitored: the regulated voltage \Vreg , the sense voltage \Vrs, which should be close to the target voltage, and the current \Vis.  Each of these parameters is monitored and individual alarms are set to flag anomalous operating conditions.

The LVRs and TCBs are powered by regulated 12-channel 6\,V MARATON power supplies located on the LHCb balcony and connected to the LVRs via 33–35\,A lines. For each MARATON channel, the monitored values include the regulated voltage on the LVR inputs, the current and status flags that signal anomalous conditions and trigger DCS alarms. The LV system featured  stable operating conditions, with only a few interventions needed.

\subsubsection{High voltage}

\newcommand{\Vbias}{$V_{\rm bias}$\,}
\newcommand{\Ileak}{$I_{\rm leak}$\,}
\newcommand{\Imon}{$I_{\rm mon}$\,}
\newcommand{\uA}{\,$\mu$A\,}
The high voltage system distributes the bias voltages to the UT silicon sensors. We use CAEN A1511B high-voltage (HV) boards, with each channel providing up to 500 V and up to 1 mA. This is a floating power supply: the reference ground is provided locally at the hybrid ground. As only 288 HV channels are available to serve 968 UT sensors, individual bias channels are assigned primarily to the innermost sensors (module types C and D). 
At increasing distance from the beam axis, sensors are grouped to share a single HV channel, with each channel connected to between two and nine sensors. The bias voltages, \Vbias, are chosen to operate all the sensors slightly over-depleted. The initial bias voltages ranged from 160 to 370 V, based on the depletion voltages determined from capacitance-voltage measurements performed during the initial sensor qualification. When sensors with different depletion voltages are grouped, we make sure that the sensor with the highest depletion voltage is at least fully depleted. In absence of collisions in LHCb, the current drawn by each HV group, \Imon, is equal to the leakage current, \Ileak. However, during high-luminosity $pp$ running, the ionization from particles passing through the detector contributes to \Imon by up to 4.5\uA. This contribution is negligible in heavy-ion collisions~\cite{PhysRevAccelBeams.27.111001}.

\begin{figure}[tb]
\centering
\includegraphics[width=0.96\textwidth]{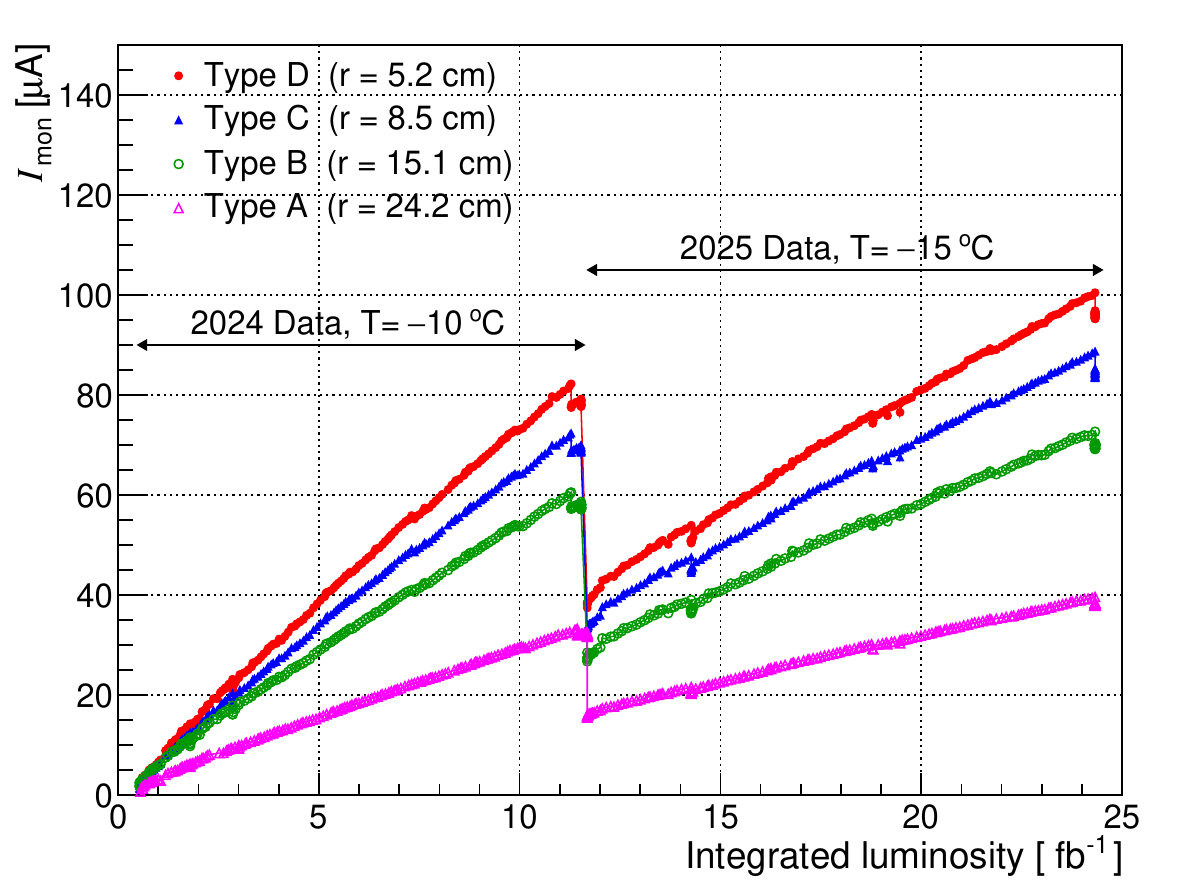}
\caption{\small{Currents (\Imon) in one of each sensor type as a function of the integrated luminosity for the 2024 and 2025 running periods. The small dips in \Imon correspond to heavy-ion data sets, where the contribution to \Imon from ionizing particles is negligible. 
}}
\label{fig:Imon_vs_Lumi}
\end{figure}
The \Imon values are continuously monitored to track current evolution with accumulated irradiation and to detect potential sensor breakdown. The \Imon trends for four individual sensors (one of each type) as a function of integrated luminosity are shown in Fig.~\ref{fig:Imon_vs_Lumi}. The measurements cover the full 2024--2025 running period. During the 2024 year end technical stop (YETS), the UT was at room-temperature for about 4 days, providing some beneficial annealing. To reduce the overall leakage current, the operating temperature was reduced from $-10^{\circ}$C to $-15^{\circ}$C during the 2024 YETS. These two factors account for the drop in \Imon seen in Fig.~\ref{fig:Imon_vs_Lumi} between the 2024 and 2025 running periods. The cooler operating temperature in 2025 also results in a reduction in the CM-subtracted noise (see Fig.~\ref{fig:noise_temp}). 

For safety reasons, strict alarm limits are implemented for \Imon in all HV channels. To account for the increasing leakage current, these limits — along with the HV trip thresholds — are automatically updated after each LHC fill. The trip thresholds are set at least 15\uA above the upper boundary of the corresponding warning range.

\begin{figure}[tb]
\centering
\includegraphics[width=0.48\textwidth]{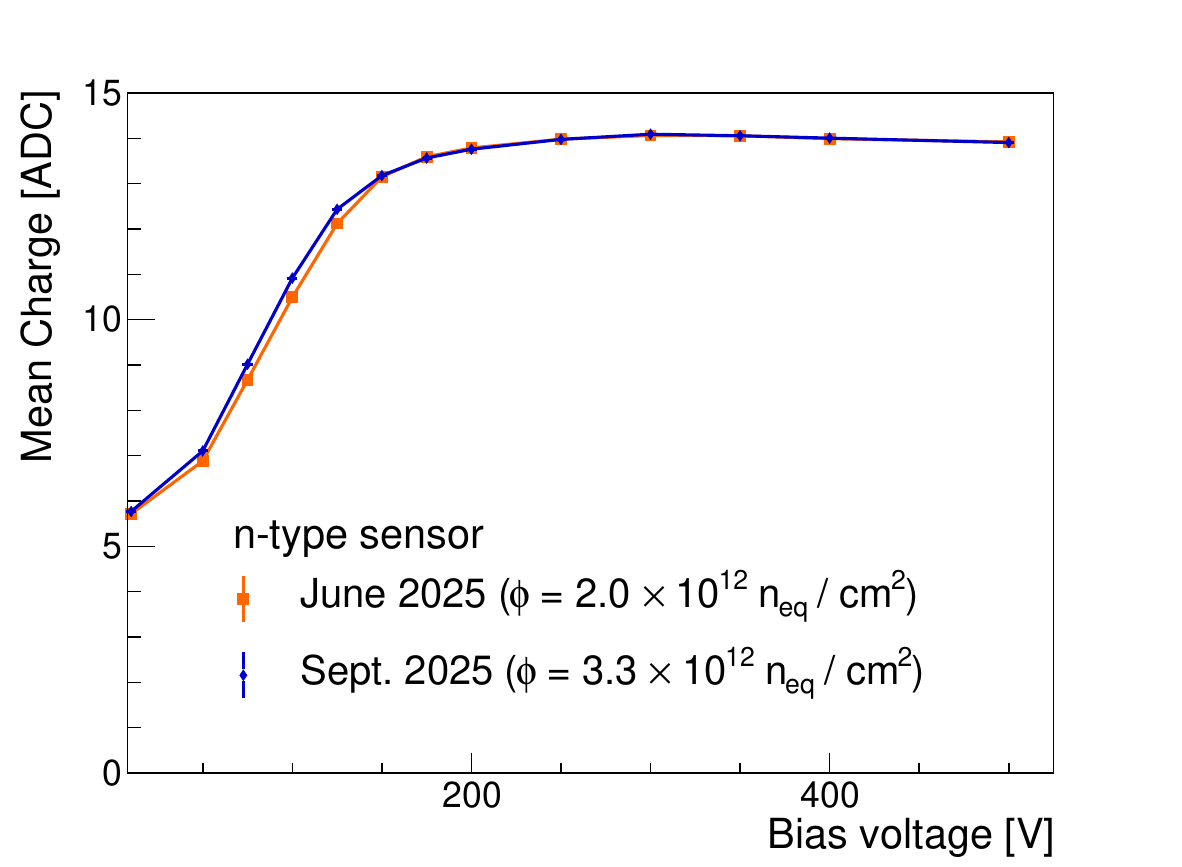}
\includegraphics[width=0.48\textwidth]{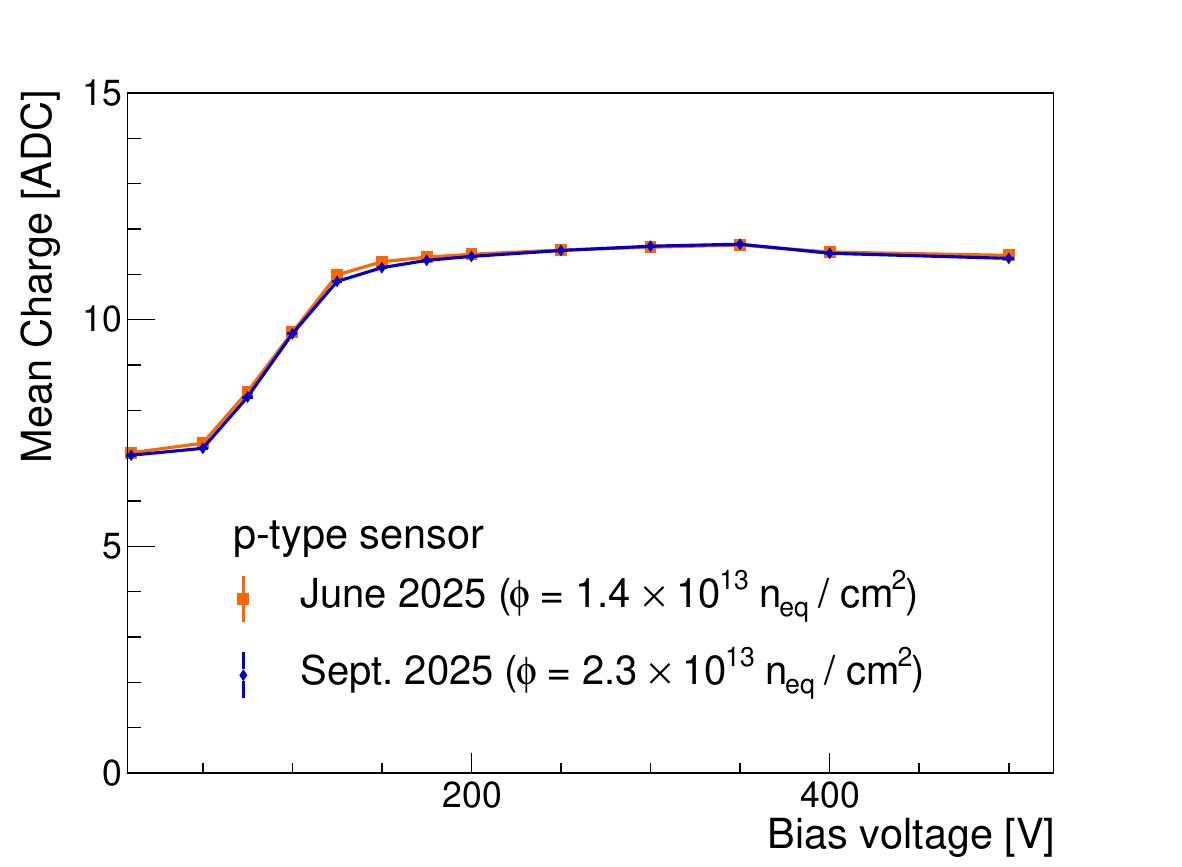}
\includegraphics[width=0.48\textwidth]{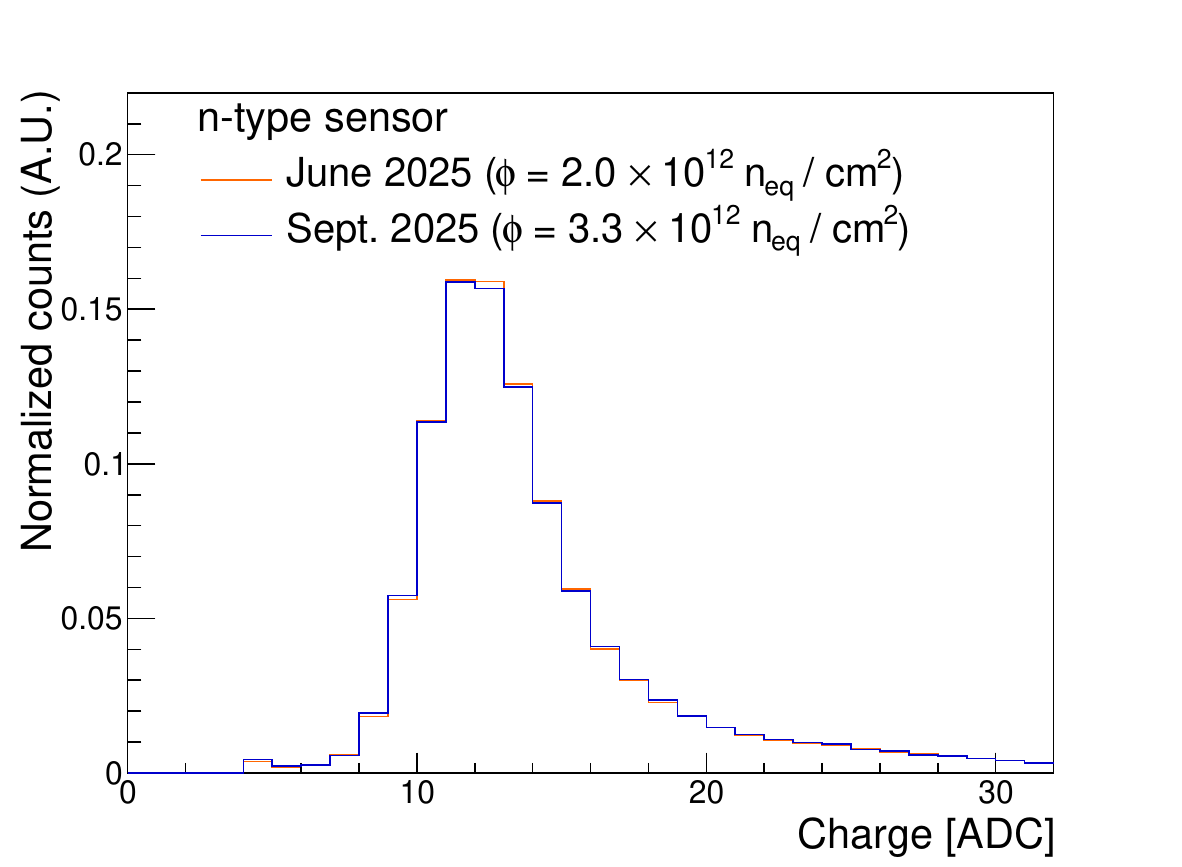}
\includegraphics[width=0.48\textwidth]{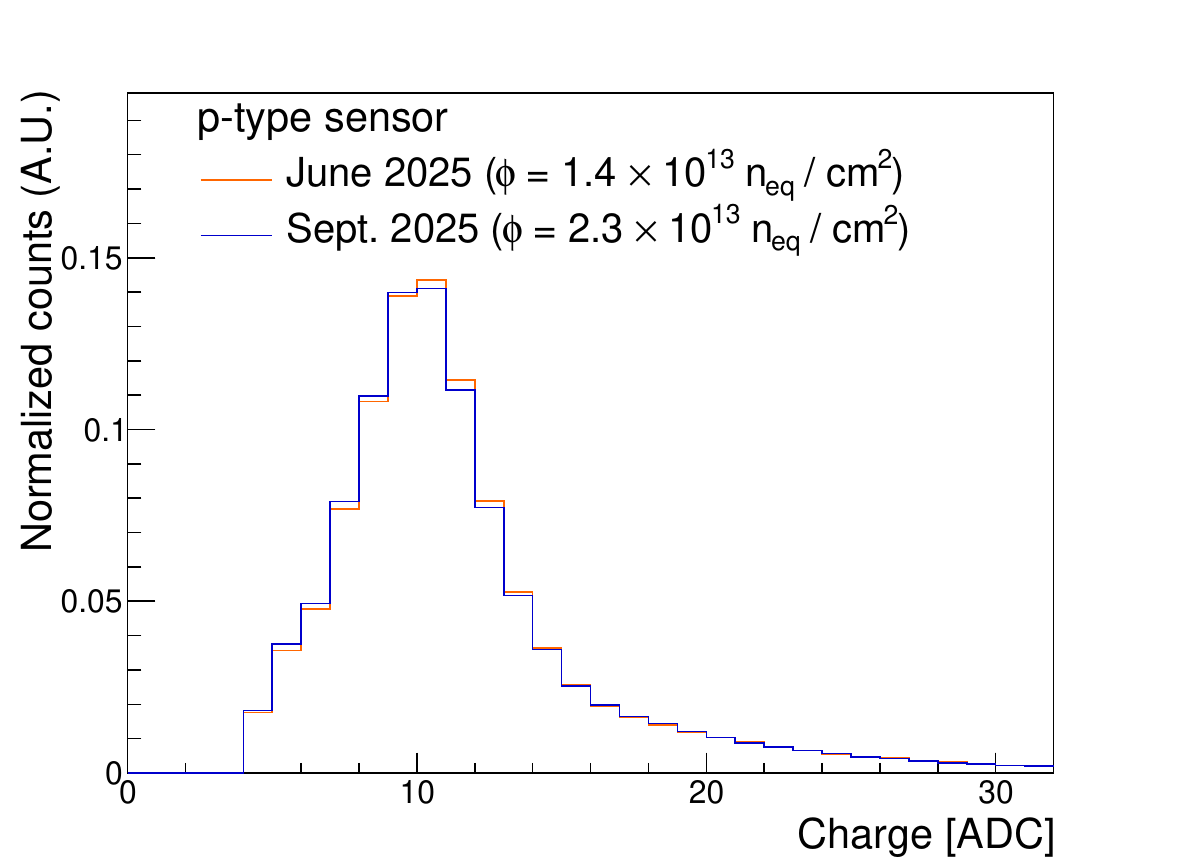}
\caption{\small{Mean charge collected (in ADC) as a function of the applied bias voltage for (top left) an innermost p-type sensor, and (top right) an n-type sensor for data collected in June 2025 and Sept. 2025, corresponding to the estimated fluences ($\phi$) shown in 1\mev n$_{\rm eq}/cm^2$. Below, the distribution of collected charge (in ADC) for the corresponding sensors at 200\,V (250\,V) for p-type (n-type) sensor.
}}
\label{fig:CCEscans}
\end{figure}

Starting at the end of 2024, regular scans of the charge collection versus \Vbias (from 0 to 500 V) were also performed in order to ensure the optimal values for maximal charge collection are used. Figure~\ref{fig:CCEscans}\,(top) shows the mean collected charge (in ADC) for UT hits on long tracks  versus the bias voltage for an n-type and a p-type sensor. Two curves are shown, one for data collected in June 2025 and a second in September 2025, corresponding to integrated luminosities of 9.7\,fb$^{-1}$ and 16.4\,fb$^{-1}$, respectively. The corresponding fluences $\phi$ in 1\mev n$_{\rm eq}/{\rm cm}^2$ at the center of the sensors are estimated using the \fluka simulation~\cite{10.3389/fphy.2021.788253, Fluka_battistoni_2015}. As \Vbias\to 0\,V, the sensor noise nearly doubles. When combined with the ZS threshold of at least 4 ADC, the mean ADC is $\sim$7, even though there is almost no charge collected from particles traversing the detector. As \Vbias increases,  the depletion region grows, leading to an increase in signal and a decrease in noise.  When sensors are fully depleted, the electronic noise reaches its minimal value (see Fig.~\ref{fig:RMSNoisePerASIC2D}) and the collected charge plateaus at about 14 and 11 ADC for the n-type and p-type sensors, respectively. The thicker n-type sensors exhibit a correspondingly higher mean charge.

An algorithm determines the optimal bias voltage for each sensor and HV group. For these particular n-type and p-type sensors, the optimal \Vbias values were found to be $\sim$225\,V and $\sim$193\,V, respectively. The distributions of collected charge for the corresponding sensors are shown in Fig.~\ref{fig:CCEscans}\,(bottom). No significant degradation in charge collection is observed with increasing fluence. This is consistent with expectations based on test beam studies~\cite{UTTestBeam2019}: charge-collection studies in irradiated sensors suggest that a maximum 10\% signal loss for the innermost sensors is anticipated by the end of Run 4, corresponding to an integrated luminosity of 50\,fb$^{-1}$.

\subsection{{Temperature and environmental monitoring}}
\newcommand{\COtwo}{CO\textsubscript{2} }
In the UT, temperature and environmental monitoring covers three areas with different readout methods: electronic board temperatures, front-end hybrid temperatures, and internal UT box conditions (temperature and humidity).

\subsubsection{Temperature monitoring}
All UT temperatures are measured by PT1000 sensors which have been all intercalibrated within 0.1-0.2°C. The PT1000 sensors are read out by two complementary systems: stand-alone ELMB2s (Embedded Local Monitor Boards~\cite{ELMB_Upgrade}), and ADC lines on the GBT-SCAs~\cite{GBT_SCA}). The ELMB2 system provides continuous monitoring of the UT temperatures, regardless of whether or not the detector is powered on or off. In particular, it monitors temperature sensors on:
\begin{itemize}  \setlength\itemsep{0em}
    \item{selected UT modules in the innermost region, closest to the beam-pipe;}
    \item{all sensors at the top and bottom of the staves, corresponding to the \COtwo inlet and outlet of the stave;}
    \item{all PEPI backplanes;}
    \item{all input \COtwo manifolds inside the UT box;}
    \item{the gas inlet inside the UT box, as part of the humidity boards readout (see Sec.~\ref{Humidity monitoring}).}
\end{itemize}
The SCA chips have a temperature resolution of about 0.5°C and, unlike the ELMB2s, they require the LV to be powered on. The SCA chips are installed in the DCBs, LVR boards and TCBs and, in addition to many other parameters, each SCA measures and transmits to WinCC-OA the temperatures of the LVRs, the DCBs and all hybrids.\footnote{Thus, we always have two independent measurements of the most central hybrid temperatures when the DCBs are powered on.}

\subsubsection{{Humidity and dew point measurements}}
\label{Humidity monitoring}
The humidity of the gas inside the UT detector box is carefully monitored to avoid condensation or frost formation on the sensors. It is measured by custom humidity boards, which are installed in the four corners of the box. Each board carries three identical relative humidity (RH) sensors, HiH4000~\cite{HIH-4000}, and a single PT1000 temperature sensor. All sensors are read out by the ELMB2s, along with the supply voltage and the total current drawn by the board. The readings are converted to physical units by the monitoring software, using standard PT1000 and individual HiH4000 factory calibrations, and then further converted to dew points (DP).\footnote{Using the Magnus formula: DP$ = b\cdot H/(a-H)$, where $H = \ln(RH/100)+ a\cdot T/(b+T)$,  $a=$17.62, $b=$243.12\,°C, temperature $T$ is in °C and the relative humidity $RH$ is in \% (see, Ref.~\cite{Sensirion_Humidity}).} 

To prevent condensation or frost, the difference between the coldest part inside the UT box and the average DP is required to be at least 15°C. If this difference drops below 15°C, a warning is issued, and if it falls below 10°C, actions are taken to stabilize the DP, including warming up the detector if necessary. Generally, DP variations have always been managed by increasing the dry nitrogen flushing rate.

\subsection{{Detector safety system}}
The UT detector safety system (DSS), like in other LHCb sub-detectors and other LHC experiments \cite{CERN_DSS}, is part of the hardware infrastructure that protects the experiment from major incidents. It has a very high level of availability and reliability, and while independent of sub-detector ECS/DCS-based safety systems, it is complementary to them. The DSS runs autonomously and is built around industrial PLCs connected to numerous independent sensors, which trigger predefined protective actions before an incident can cause serious damage to experimental equipment. The anticipated incidents include fires in the HV or MARATON racks, water leaks, overheating of the LVR and DCB crates, and failures of the \COtwo cooling and the UT box nitrogen flushing.
The UT DSS contributes to the common LHCb DSS by providing a DSS Alarm–Action Matrix, which defines primary alarms, their latencies, and the automatic actions to be taken if the DSS alarm(s) persist. These actions range from switching off LV and/or HV power (or stopping the \COtwo cooling plant, if needed) to activating fire extinguishers in the electronics rack room. Various monitoring data, and any warnings or alarms are also automatically reported in a dedicated DSS display panel in the control room for further investigations by UT detector experts.

The minimum DSS action is to request intervention by a trained UT person.

\subsection{Online data monitoring }

The UT detectors and recorded data are monitored at multiple stages of the data acquisition and reconstruction chain to identify potential issues as early as possible. The LHCb online monitoring system analyses a fraction of the ZS data sample selected by the HLT1 trigger shortly after it is recorded. This data is used for both monitoring, as well as setting alarms if the integrity of the data being collected is in question.

\begin{figure}[htbp]
\centering
\includegraphics[width=0.48\textwidth]{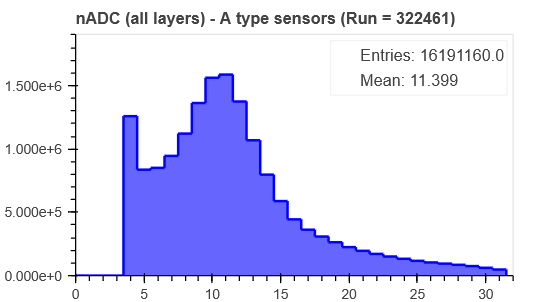}
\includegraphics[width=0.48\textwidth]{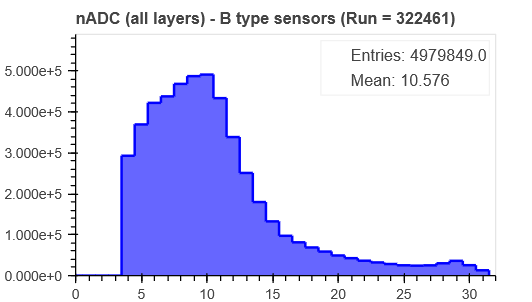}
\includegraphics[width=0.48\textwidth]{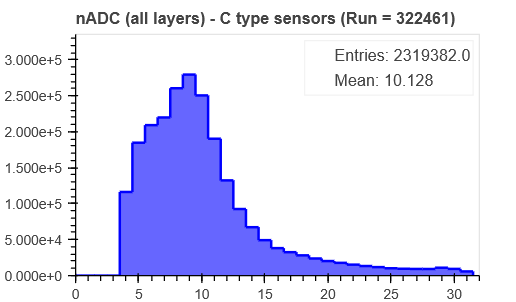}
\includegraphics[width=0.48\textwidth]{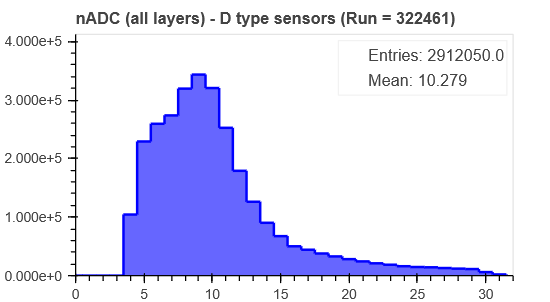}
\caption{\small{Online UT raw-data monitoring display for a typical $pp$ run showing the zero-suppressed (ZS) ADC distribution of the highest-ADC strip in each cluster, for (top left) type A, (top right) type B, (bottom left) type C, and (bottom right) type D.}}
\label{fig:ClusterMax}
\end{figure}

\begin{figure}[htbp]
\centering
\includegraphics[width=0.7\textwidth]{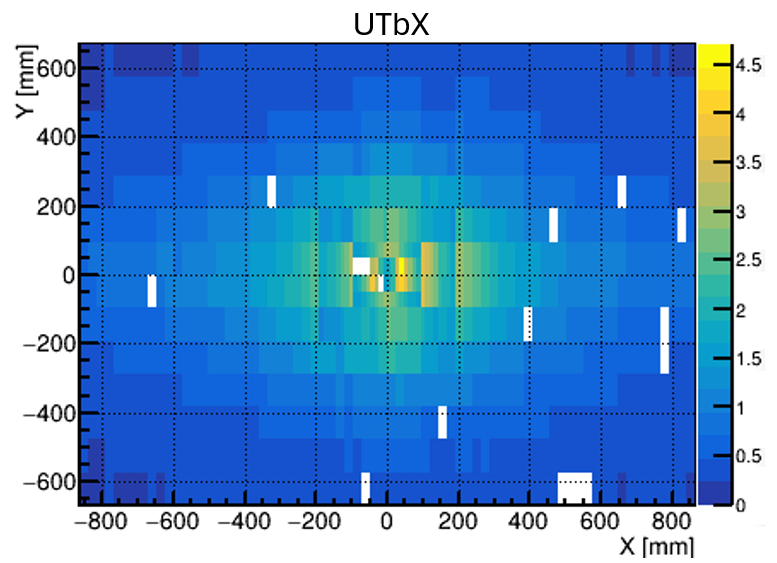}
\caption{\small{Average number of hits per event in an ASIC as a function of position for one UT layer.  Each SALT ASIC serves 128 silicon strips. The white areas correspond to ASICs which were disabled for hardware reasons (4.4\% in 2025 run).}}
\label{fig:RatePerChip}
\end{figure}

Two complementary monitoring approaches were employed. The first relied solely on UT raw data and monitored quantities such as the overall hit multiplicity, per-ASIC occupancy, and pulse-height distributions, providing information on the minimum-ionizing particle peak and background levels (Figs.~\ref{fig:ClusterMax}-\ref{fig:RatePerChip}). For data reconstructed together with the other tracking detectors, the second set monitored the UT tracking efficiency using long tracks. During the 2025 data-taking period, the efficiency of associating three or more UT hits with a long track was approximately 98.7\% (see Section~\ref{sec:UTTrackEff}).

The UT online monitoring was also equipped with the ability to analyse a fraction of the NZS data during electronic calibration runs to provide immediate feedback on the average baseline and noise values down to an individual ASIC or channel. The HLT2, which typically processed events with a short latency, performed full event reconstruction and provided dedicated monitoring of the UT performance using long and downstream tracks.

\subsection{Single-event upset}
As discussed in Section~\ref{sec:radhardasic}, SEU events in either the baseline DAC or pedestal registers are expected to occur at a rate of about 3.5\,mHz for the full UT system. The calibration of the baseline DAC (see Section~\ref{sec:signal_proc}) results in pedestal values of 0 for nearly all channels (see Section~\ref{sec:signal_proc}), with only 4.4\% of pedestal register bits were set to “1”. This can be compared to the baseline DAC registers, where around 50\% of the bits are set to “1”.

All ASICs were initialized with their nominal pedestal and baseline DAC values prior to starting a data-taking run. To mitigate bit-flip errors due to SEU events, these register values were read back from the hardware every 30 seconds and compared with their original settings. Whenever a discrepancy was detected, the entire register block of the affected ASIC was reconfigured. These SEU events were also systematically logged for further analysis.

\begin{figure}[tb]
\centering
\includegraphics[width=0.7\textwidth]{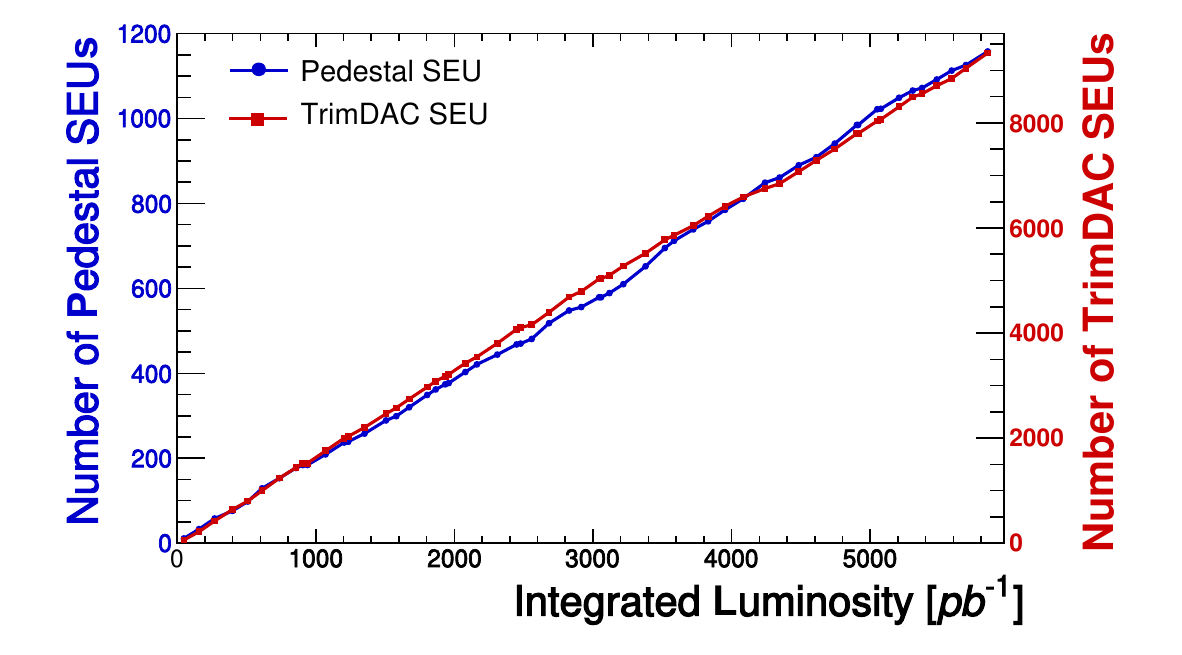}
\caption{\small{Cumulative number of SEU events in the Pedestal registers (in blue) and in the baseline DAC registers (in red) vs the integrated luminosity, for data taken from July 18 to October 18, 2025.}}
\label{fig:seuVsLumi}
\end{figure}

Figure~\ref{fig:seuVsLumi} shows the number of SEU events recorded during the 3-month period from July 18 to October 18, 2025, where a single set of baseline DAC and pedestal values were used. A clear linear correlation is evident between the cumulative number of SEUs and the integrated luminosity. 
A total of 1,149 Pedestal SEUs and 9,255 baseline DAC SEUs were observed, resulting in an overall SEU rate of approximately 3.6\,mHz across both register types. This measured rate is close to the estimated value of 3.5\,mHz, as discussed in Section~\ref{sec:radhardasic}. 

Further analysis showed that the “1-to-0” bit flip transitions were overwhelmingly dominant, in agreement with the irradiation beam tests (see Section~\ref{sec:radhardasic}). During the 3-months of stable data-taking, only 90 “0-to-1” bit-flips occurred, compared to a total of 341,439 “1-to-0” bit-flips. After normalizing for the respective numbers of bits initially set to “1” and “0”, the probability ratio of “0-to-1” to “1-to-0” flips is approximately $1.2\times 10^{-4}$. Overall, the SEU mitigation procedure described above was effective at suppressing the negative impacts of SEUs to a negligible level during data taking.

\subsection{Geometrical alignment}
The UT sensors provide an intrinsic hit-position resolution of approximately 55$\mum$ (Type A) and 26$\mum$ (Types B–D) in the $x$ coordinate. To avoid degrading the tracking performance, the sensor positions must be determined by the alignment procedure with a precision of 10–20$\mum$.

The UT geometry is described in a hierarchical and nested structure from larger structures to smaller as
\begin{itemize}
    \item UT: the entire UT detector
    \item UTSide: A-side or C-side (four layers each) 
    \item UTLayer: 4 half-layers within each UTSide
    \item UTStave: 8 or 9 staves per half-layer
    \item UTFace: The stave are two-sided and each side is referred to as a face
    \item UTModule: the basic element on which the silicon sensor is mounted.
\end{itemize}

Each structure has six alignable degrees of freedom (DOF): $x, y, z,\theta_x, \theta_y, \theta_z$, and is aligned relative to its parent, corresponding to translations and rotations about the three axes with respect to the parent element. The initial module positions were obtained from Smartscope measurements~\cite{smartscope}, while the initial positions of the staves, layers, and half-boxes were determined from photogrammetry and laser survey measurements. The standard LHCb alignment software uses reconstructed long tracks (see Fig~\ref{fig:TrackTypes}) to minimize a global $\chi^2$ of all tracks with respect to the alignment DOFs~\cite{HULSBERGEN2009471}. The software can be configured to specify which structures and DOFs are to be aligned. Prior to aligning the UT, the Vertex Locator (VELO) and Scintillating Fibre Tracker (SciFi) systems~\cite{LHCb-DP-2022-002} were aligned to one another after which the UT alignment was performed. 

At the beginning of the 2024 data run, the UT was first aligned using in a stand-alone procedure, by analyzing the residual distributions between UT hits and tracks formed using only hits in the VELO and SciFi detectors. The detector elements were aligned sequentially, where only certain elements and DOFs were considered. In the first step, the UT layers were aligned in $x$ and $z$. Secondly, the individual staves were aligned allowing for translations in $x$ and rotations around the $z$ axis, keeping all other elements and DOFs fixed. Lastly, the modules were aligned, again, allowing for translations in $x$ and rotations around the $z$ axis.

\begin{figure}[tb]
\centering
\includegraphics[width=0.48\textwidth]{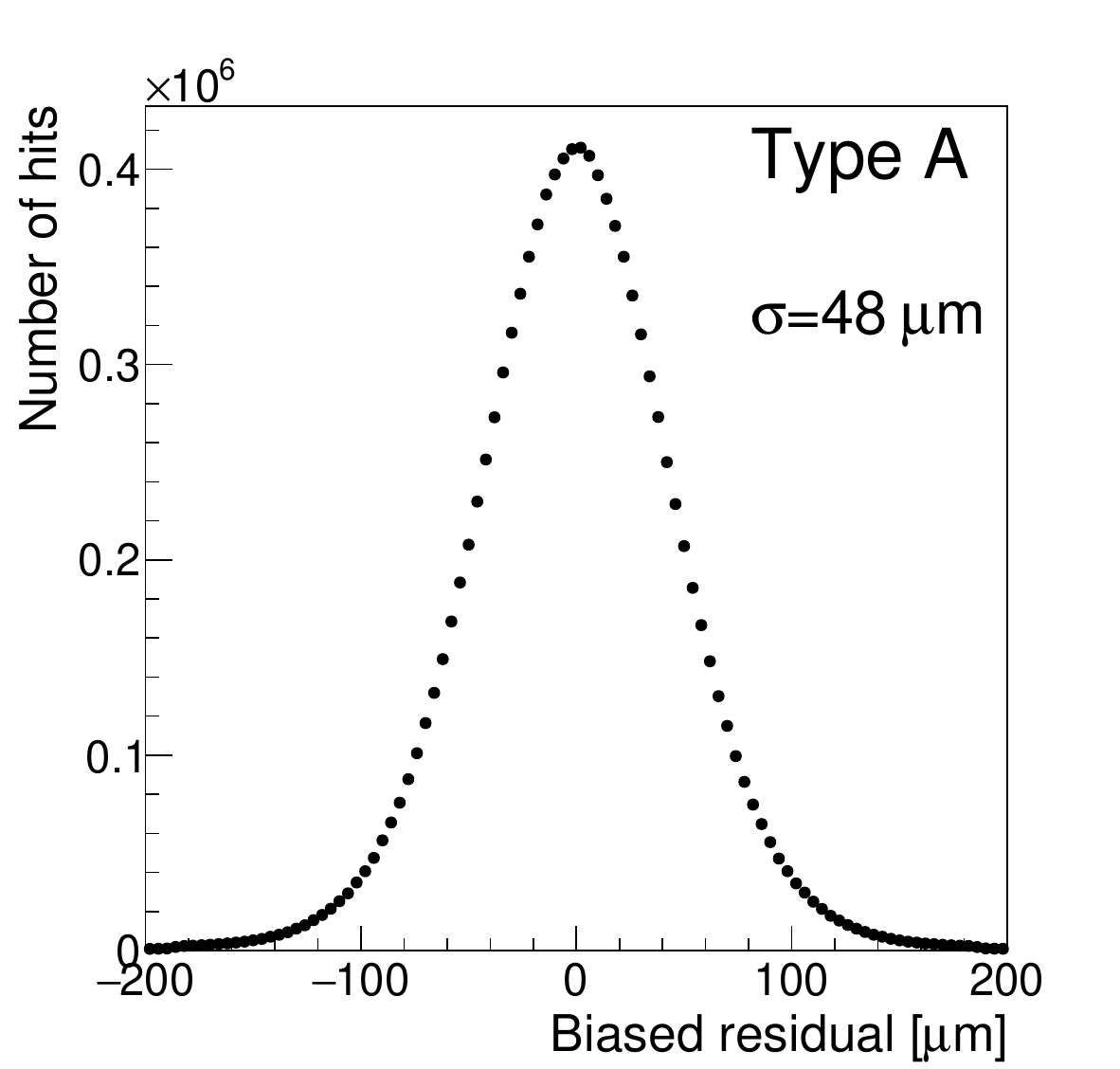}
\includegraphics[width=0.48\textwidth]{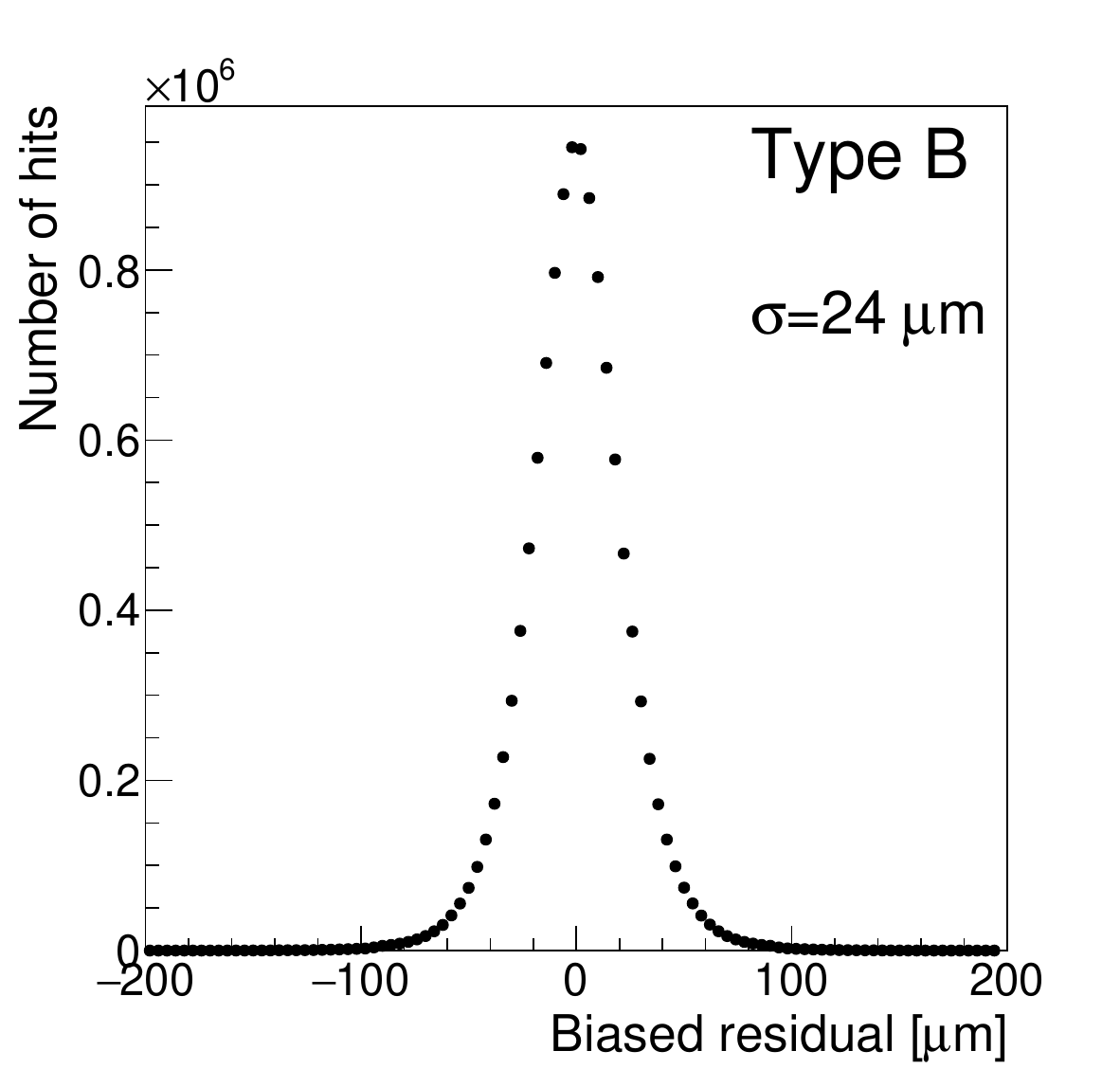}
\includegraphics[width=0.48\textwidth]{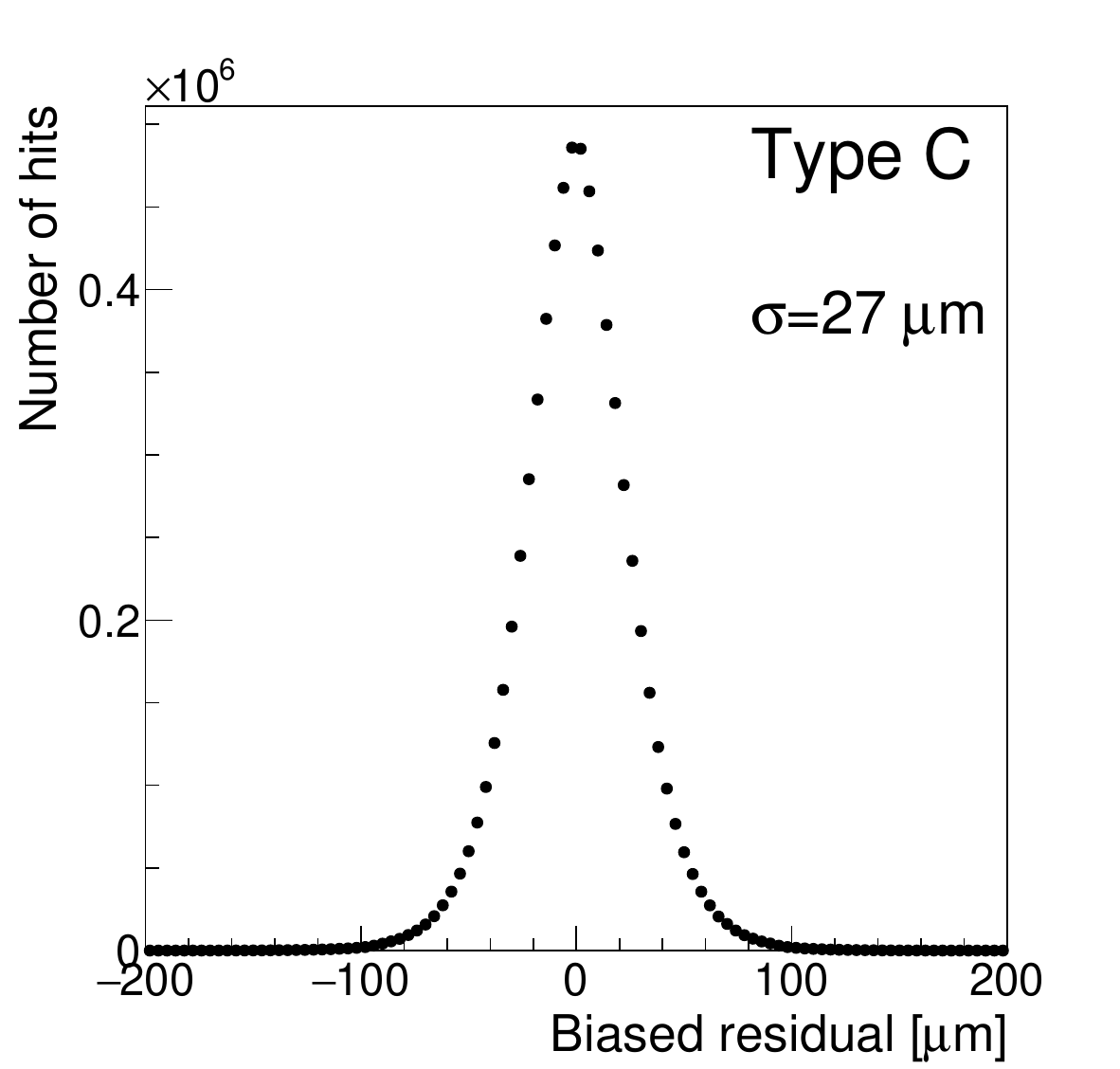}
\includegraphics[width=0.48\textwidth]{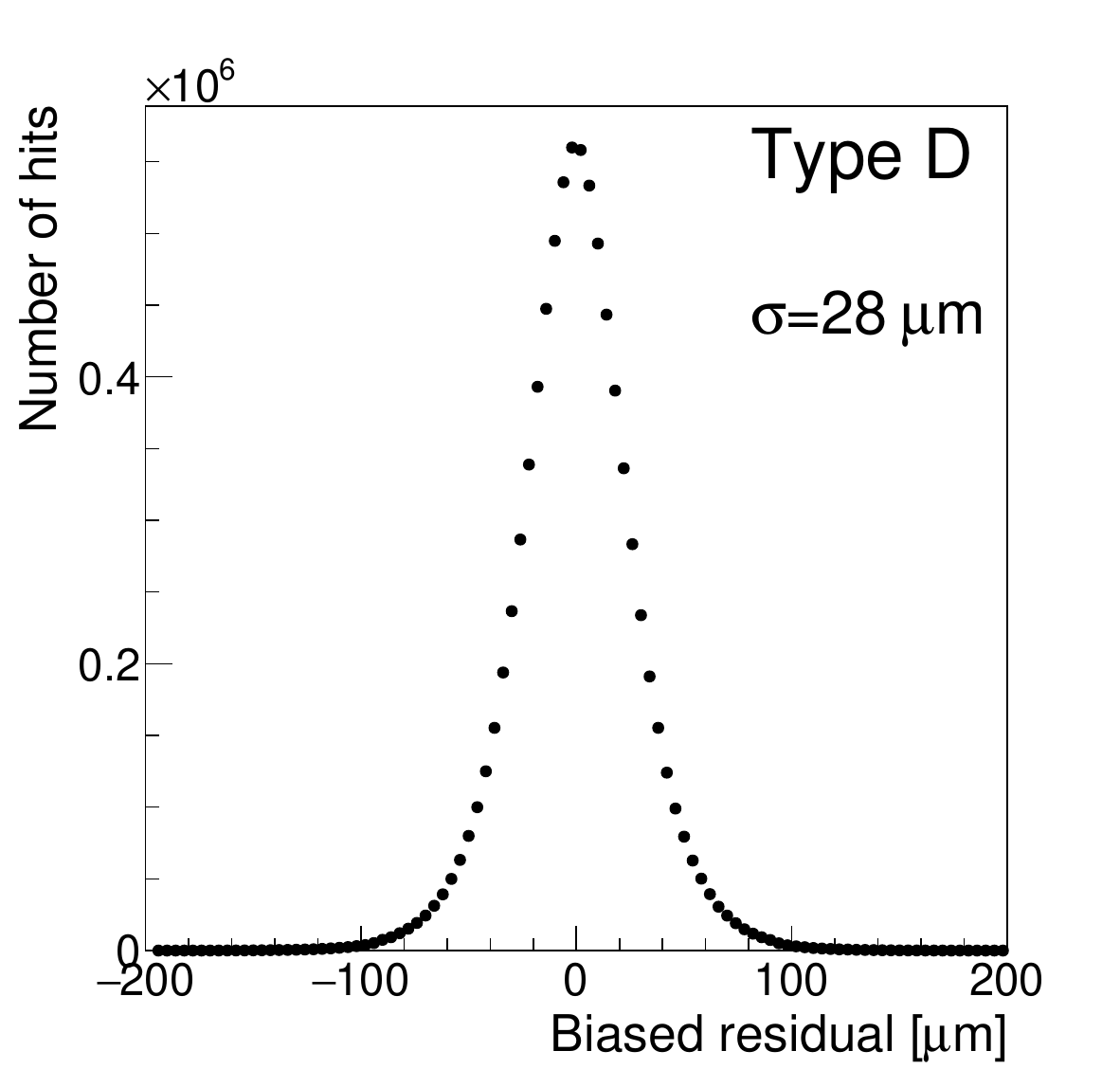}
\caption{\small{Biased residuals between long tracks and UT clusters for each module type for mu=5.3 $pp$ collision data.}}
\label{fig:BiasedResiduals_mu5_3}
\end{figure}

In the final stage, small corrections to the alignment parameters are made through a global fit using a large sample of reconstructed tracks. The global alignment also uses samples of $\jpsi\to\mup\mun$,
$\Dz\to\Km\pip$ and $Z^0\to\mup\mun$ decays, along with an additional $\chisq$ term to aid the alignment in converging to a solution where the reconstructed masses are consistent with the known masses~\cite{HULSBERGEN2009471,LHCb-DP-2026-XXX}. For the UT, only the UTLayer ($x$ and $z$) and UTStave ($x$, $\theta_z$) DOFs are aligned. The biased residual distributions for each module type are shown in Fig.~\ref{fig:BiasedResiduals_mu5_3} for data taken with mu$=$5.3. The widths of these distributions are within 20\% of the values expected from simulation. Small residual misalignments are possible due to the complex coupling between the precise knowledge of the magnetic field through the full tracking system and alignment degrees of freedom. The UT alignment is typically repeated after an LHCb magnet polarity reversal, major updates to the VELO or SciFi alignment, or any long shutdown requiring the UT box to be opened.

\clearpage

\clearpage
\section{UT performance}
A key performance metric of the UT is the efficiency with which hits are associated with long tracks, as the resulting hit information provides strong discrimination against ghost tracks. This section presents hit- and event-level observables that characterise both the operating conditions and the performance of the UT detector.

\subsection{UT occupancy}

\begin{figure}[tb]
\centering
\includegraphics[width=0.48\textwidth]{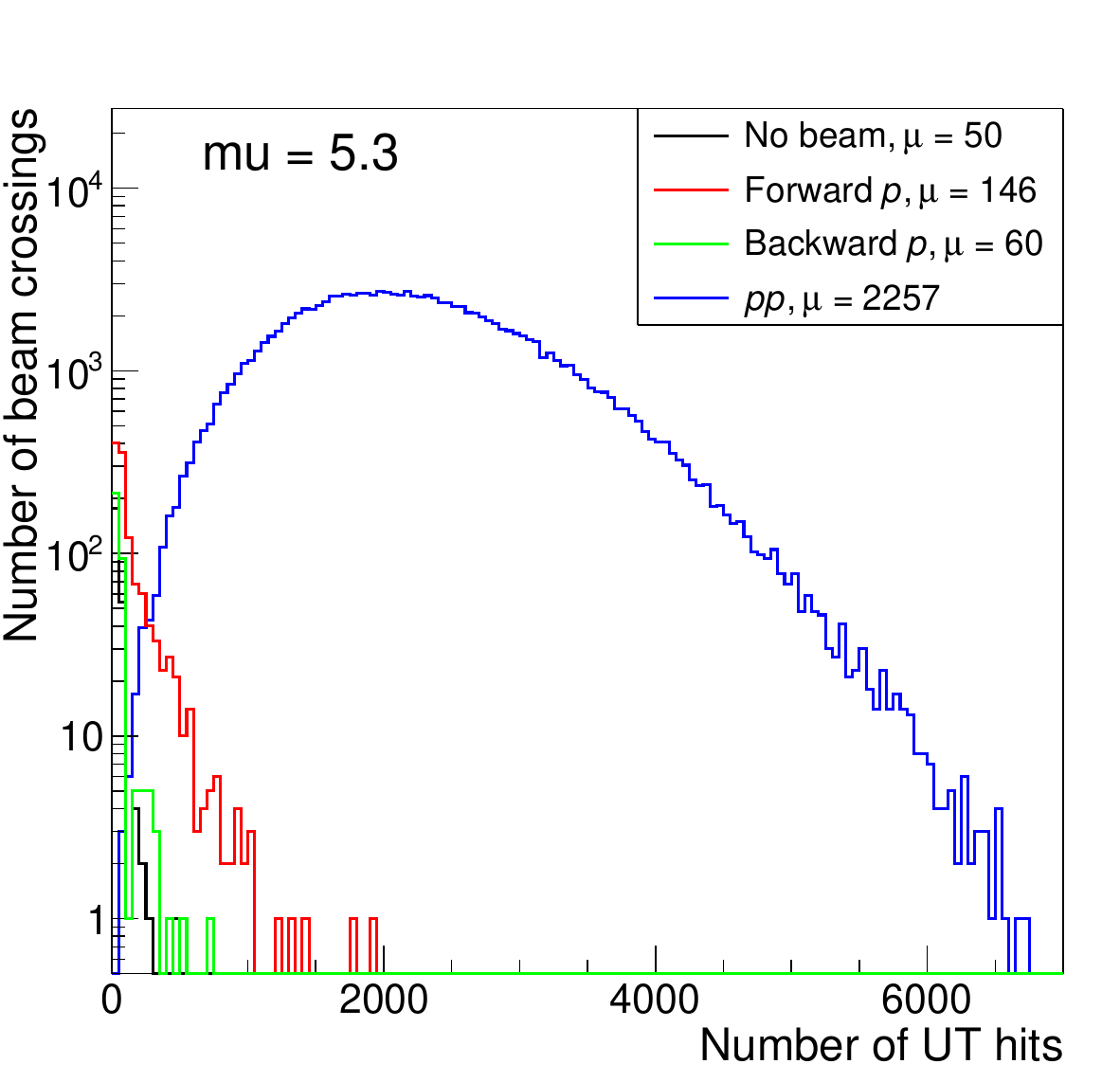}
\includegraphics[width=0.48\textwidth]{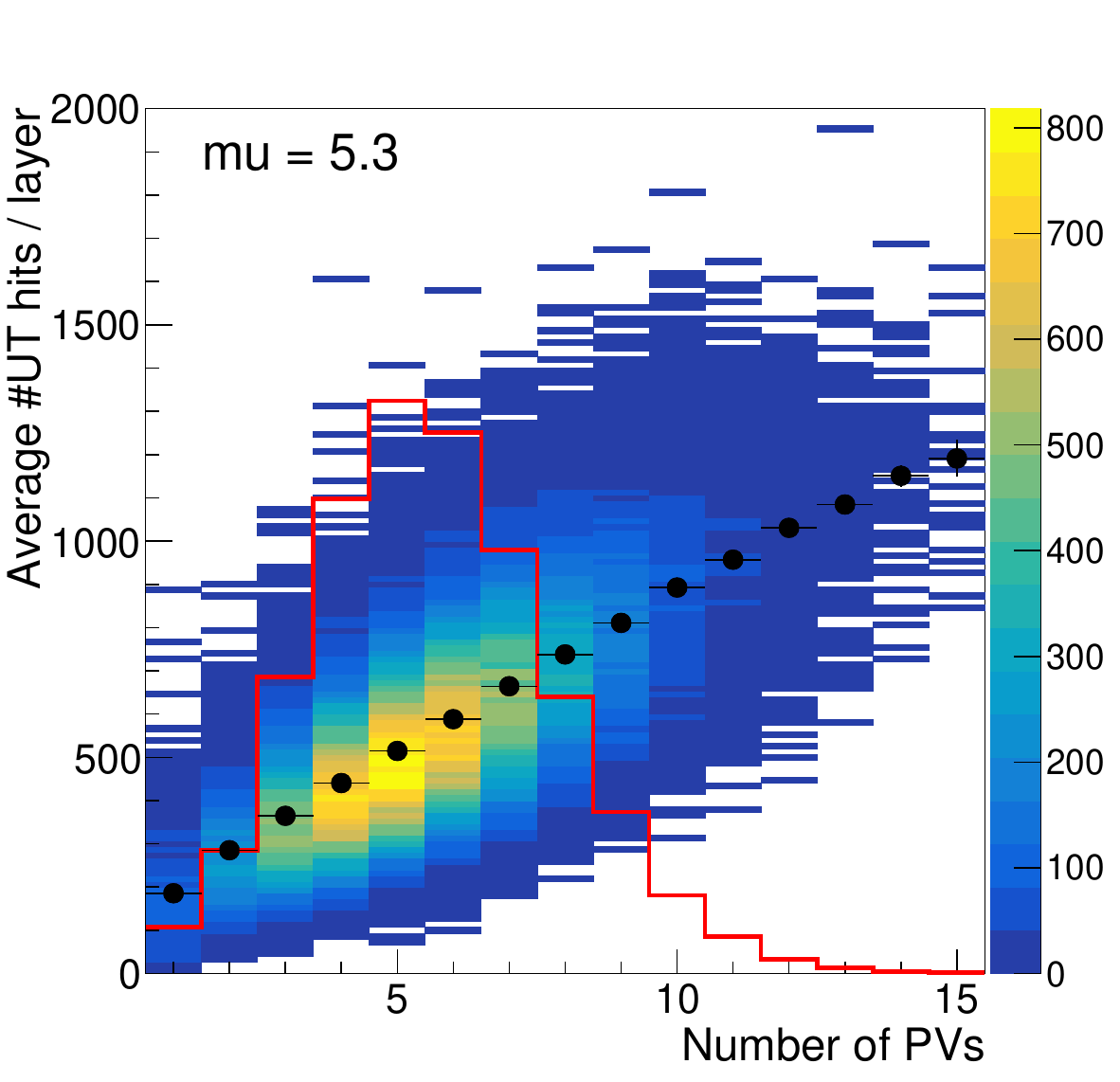}
\caption{\small{(Left) Number of UT hits with different types of LHC bunch crossing types at mu = 5.3. The means of each distribution are indicated in the legend.
(Right) The average number of UT hits per layer versus the number of reconstructed primary vertices (PVs) at mu = 5.3. The distribution of number of PVs is superimposed (red).}}
\label{fig:UTOcc_5_3}
\end{figure}

The UT was designed with sufficient granularity to cope with the large track multiplicities in LHCb Upgrade I environment. To provide context, Fig.~\ref{fig:UTOcc_5_3} shows the number of hits in the UT detector for (left) different types of beam crossings, and (right) as a function of the number of reconstructed $pp$ collisions (PVs) in each crossing. Here, forward (backward) indicates the proton bunch moving in the $+z$ ($-z$) direction in the LHCb coordinate system, while the bunch circulating in the opposite direction is empty. No beam indicates the bunches in both directions are empty at the LHCb interaction region (IR). Colliding bunches at the LHCb IR are denoted as $pp$. For the $pp$ collisions, the number of PVs (nPVs) is required to be larger than zero, whereas for the other cases, no requirement is imposed. In $pp$ collision crossings, there are about 2260 UT hits in the UT, on average, corresponding to an average occupancy of about 0.42\%. In non-$pp$ events, the occupancy is very low, about (0.01--0.02)\%. The average occupancy grows linearly with nPV, as expected.

\begin{figure}[tb]
\centering
\includegraphics[width=0.78\textwidth]{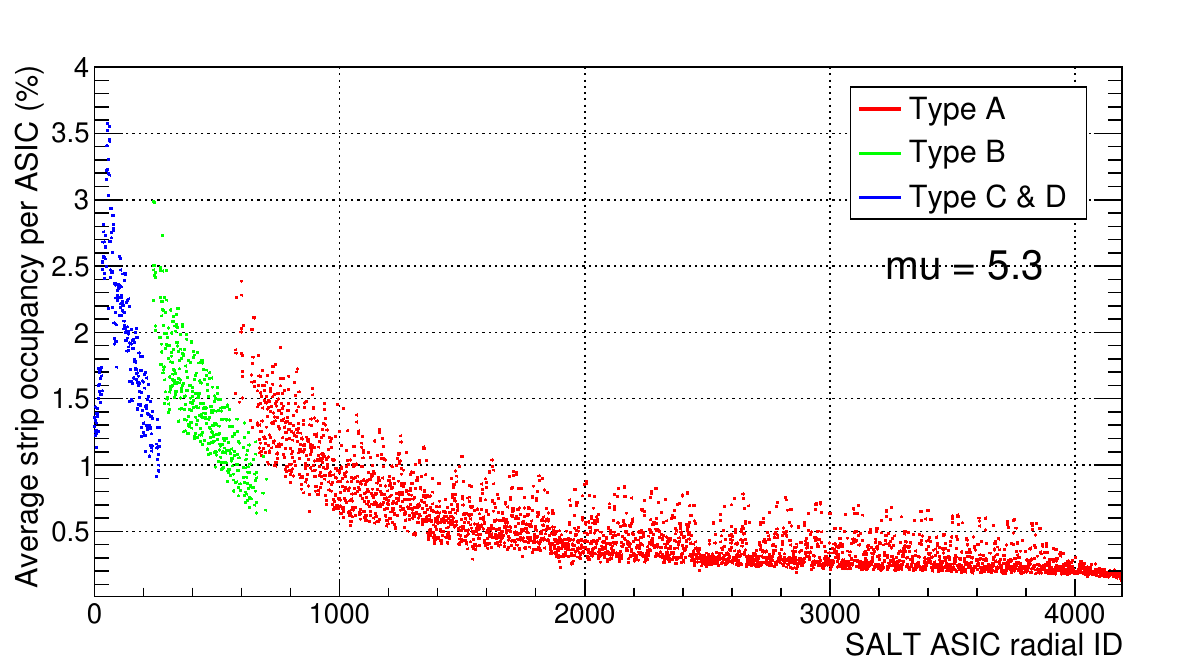}
\caption{\small{Average occupancy per UT strip as a function of the radial distance of the SALT ASIC from the beam line at mu=5.3. The Type A, B and C/D are distinguished by the 3 colours.}}
\label{fig:AverageStripOcc}
\end{figure}

Figure~\ref{fig:AverageStripOcc} shows the average per-strip occupancy versus the SALT ASIC number, where the ASICs are ordered by their distance from the beamline. Here, events are required to pass the HLT1 and have at least 5 reconstructed tracks. The busiest ASICs close to the beam pipe have a per-strip occupancy of about 3.5\%, whereas the outermost ASICs have a per-strip occupancy of (0.2--0.3)\%. The discontinuity from type A to type B is due to the reduction in strip pitch, and the discontinuity between type B and C/D is due to the sensors in the latter having about half of the area of the former. The drop in occupancy in the innermost ASICs $(<50)$ is due to the shorter strip lengths around the quarter-circle cut-outs that accommodate the beam pipe. 

The strip-level occupancies shown in Fig.~\ref{fig:AverageStripOcc} exceed those obtained using only the highest-ADC strip in each UT cluster by about 30\%\footnote{A cluster is 2 or more consecutive strips with a hit, generally due to a single particle passing through the region of more than 1 strip, or due to delta rays.} Moreover, the occupancy is also enhanced by the HLT1 requirements (by about 40\%) as compared to generic beam-beam crossings where HLT1 is not required and there are at least 5 tracks in the event.

\subsection{Cluster charge and size}

\begin{figure}[tb]
\centering
\includegraphics[width=0.3\textwidth]{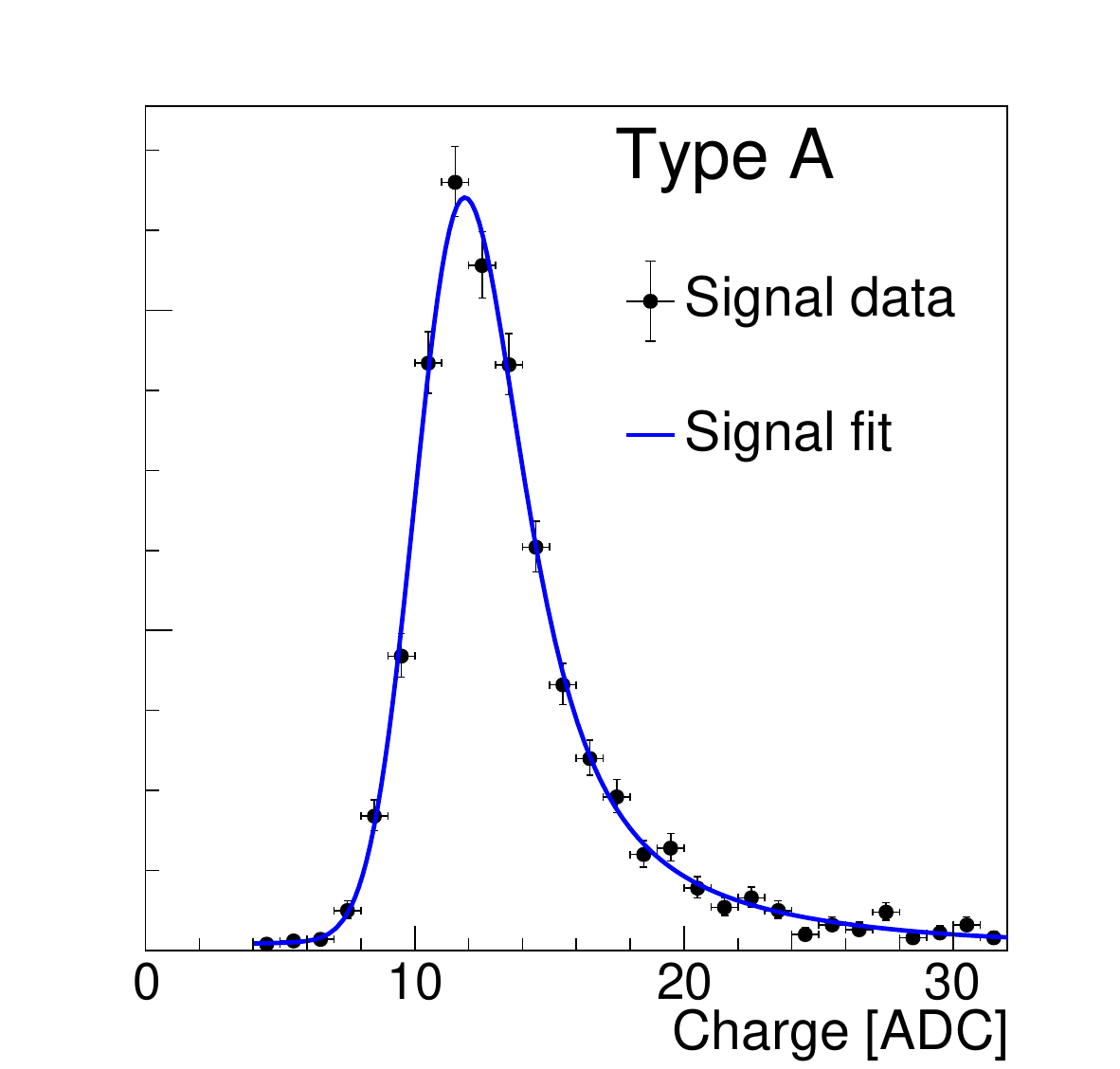}
\includegraphics[width=0.3\textwidth]{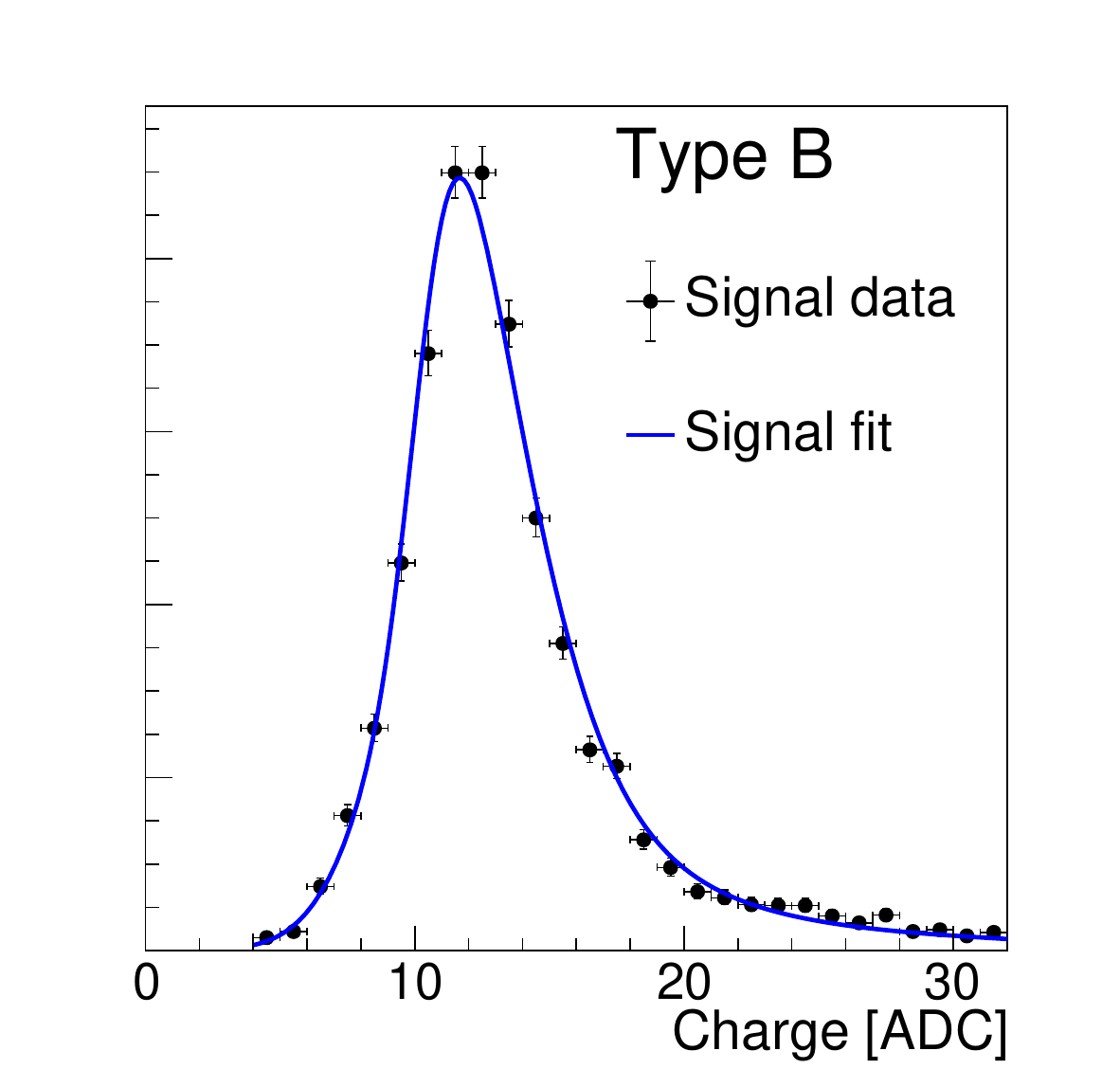}\\
\includegraphics[width=0.3\textwidth]{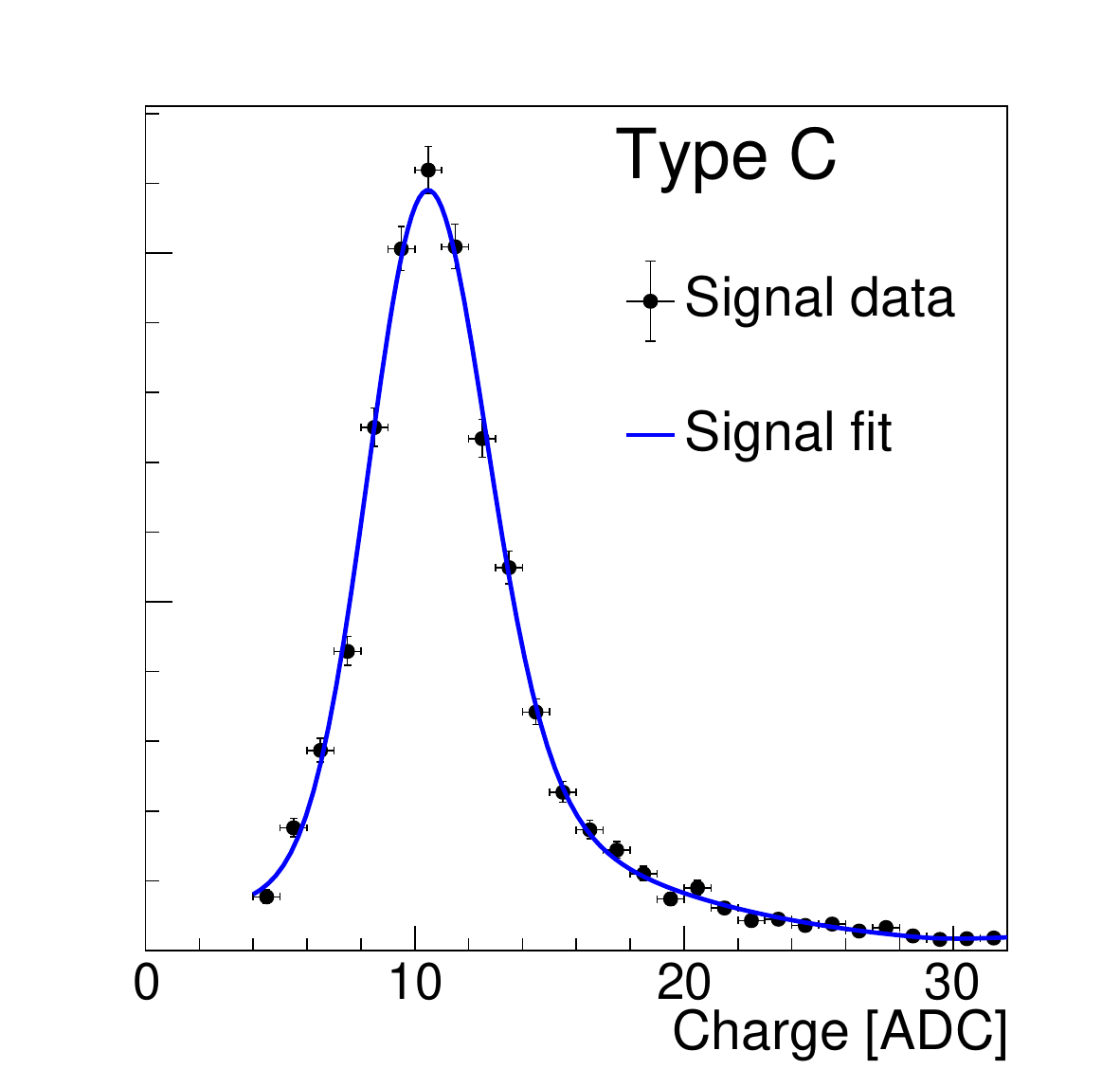}
\includegraphics[width=0.3\textwidth]{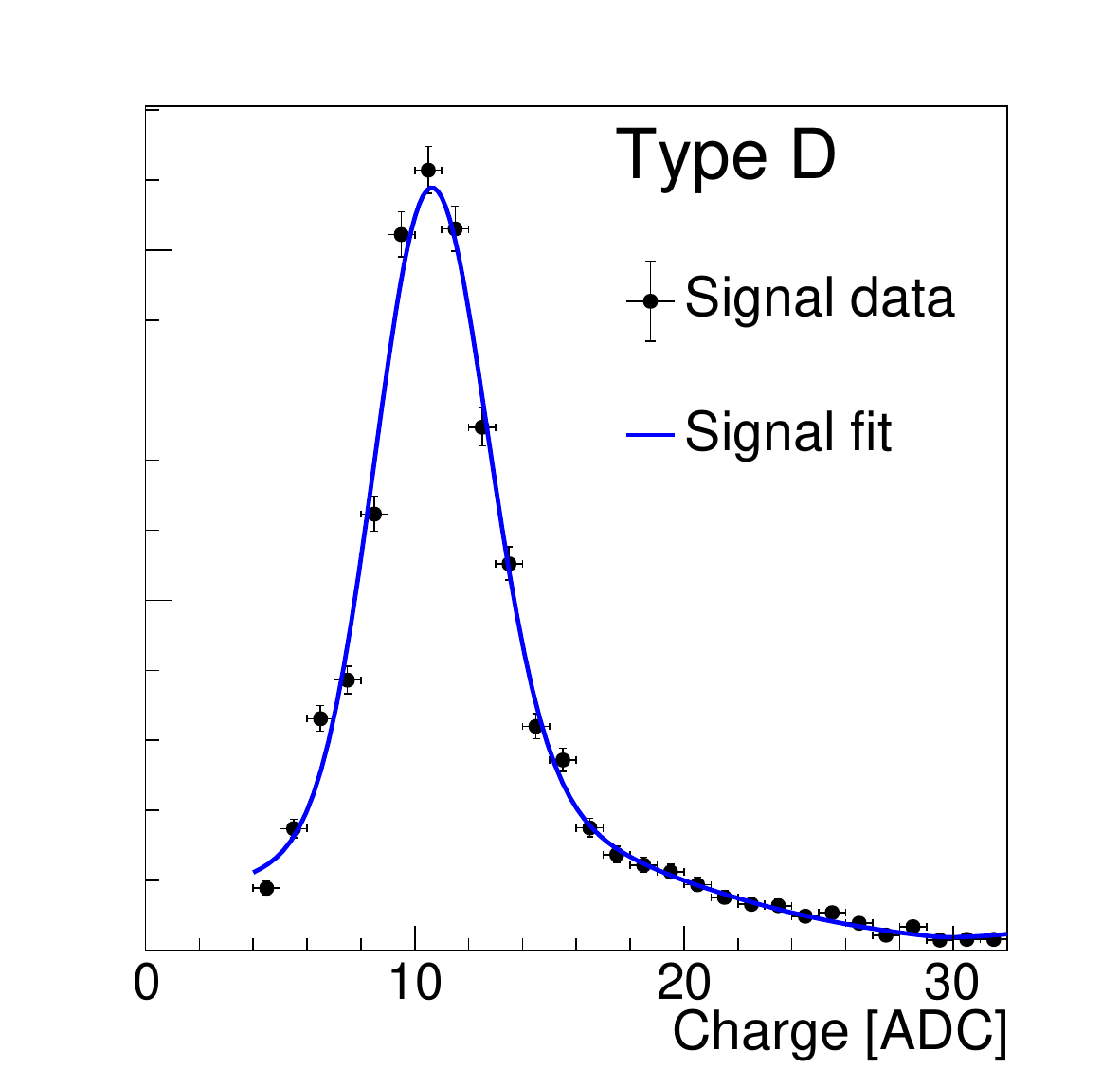}
\caption{\small{Distributions of collected charge (in ADC) for a typical ASIC of each module type for UT clusters associated to reconstructed long tracks at mu = 5.3.}}
\label{fig:ADCDist_mu5_3}
\end{figure}

\begin{figure}[tbh]
\centering
\includegraphics[width=0.6\textwidth]{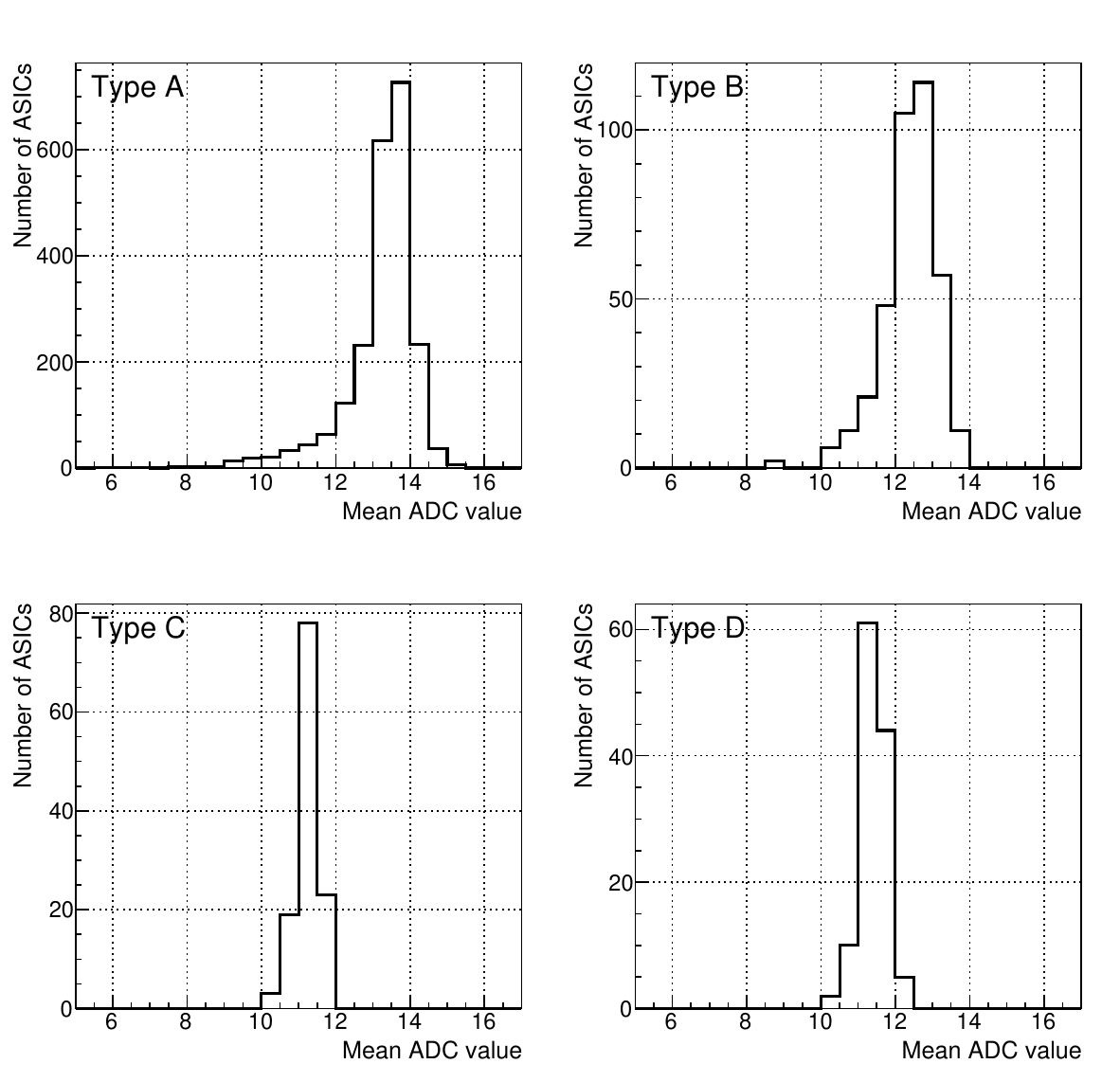}
\caption{\small{Distributions of the mean charge collected per SALT ASIC (in ADC) for each module type.}}
\label{fig:meanCharge}
\end{figure}

Figure~\ref{fig:ADCDist_mu5_3} shows the ADC distribution of clusters associated to long tracks for a typical ASIC in each type of module. Each distribution is fit to a Landau distribution convolved with the sum of two Gaussian functions. The peak value of these type A and B modules is about 12 ADC, while the C and D modules have peak values of about 11 ADC. 
To give a snapshot of the collected charge across the full detector, Fig.~\ref{fig:meanCharge} shows the mean charge for each ASIC with at least 100 hits. The means are about 1--2 ADC larger than the peak values obtained from a fit. There is some spread of values, and the type A sensor ASICs show a tail toward smaller charge collected. Additional studies are being conducted to understand the source of this lower charge collection in some of the ASICs.

\begin{figure}[tbh]
\centering
\includegraphics[width=0.6\textwidth]{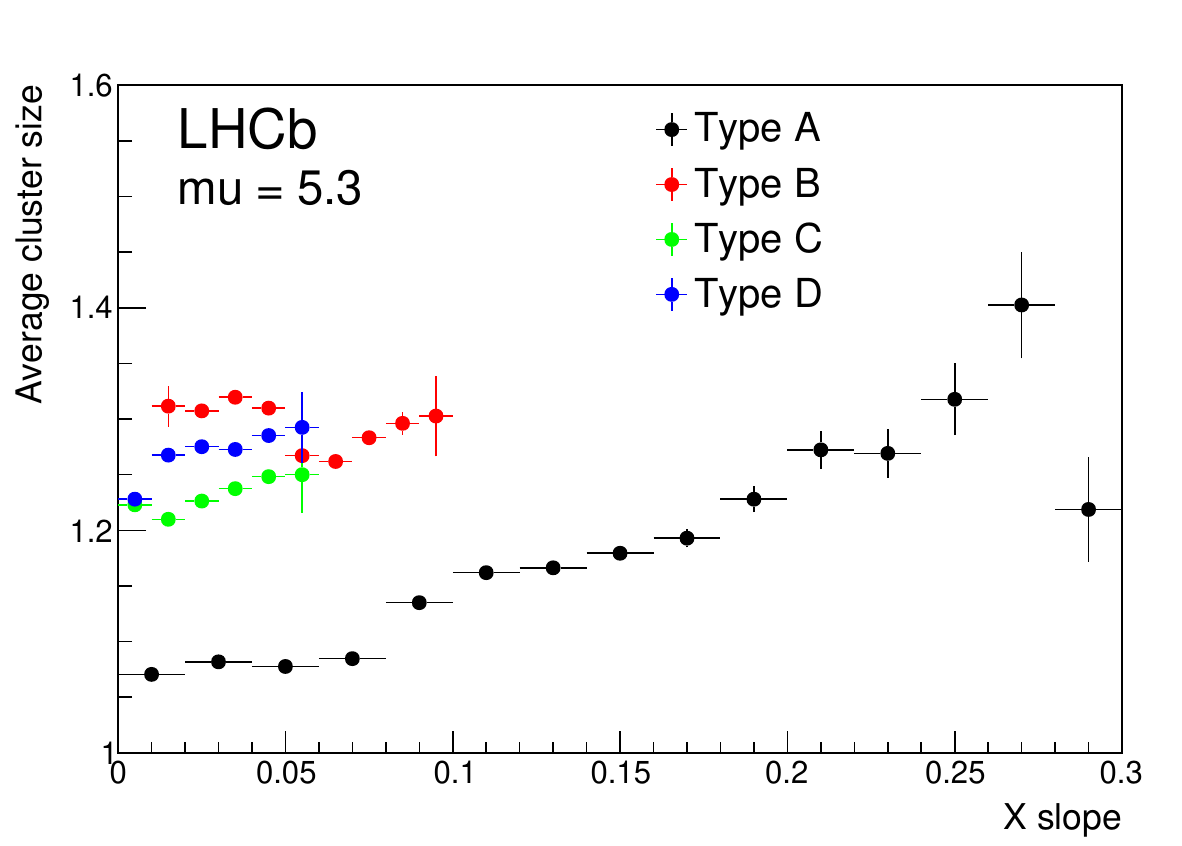}
\caption{\small{Distributions of average UT cluster size as a function of the slope in the $xz$ plane for each module type. The data correspond to mu = 5.3.}}
\label{fig:ClusterSize_5_3}
\end{figure}

The UT strip pitches are $93.5\,\mu$m and $187.5\,\mu$m, and it is expected that most clusters should be formed from a single strip. Tracks with large angles in the $xz$ plane are more likely to deposit significant charge in more than one strip, since the strips are either vertical or have a small stereo angle. Figure~\ref{fig:ClusterSize_5_3} shows the mean cluster size as a function of the track slope in the $xz$ plane. At small angles the type A modules have an average cluster size of about 1.1, whereas the finer-pitch sensors (types B, C, D) have average cluster sizes in the 1.2--1.4 range. The type A modules show the growth of the cluster size as the angle increases, as expected. No such dependence is seen with respect to the track slope in the $yz$ plane.

\subsection{Hit and track efficiency}
\label{sec:UTTrackEff}

\begin{figure}[tbhp]
\centering
\includegraphics[width=0.80\textwidth]{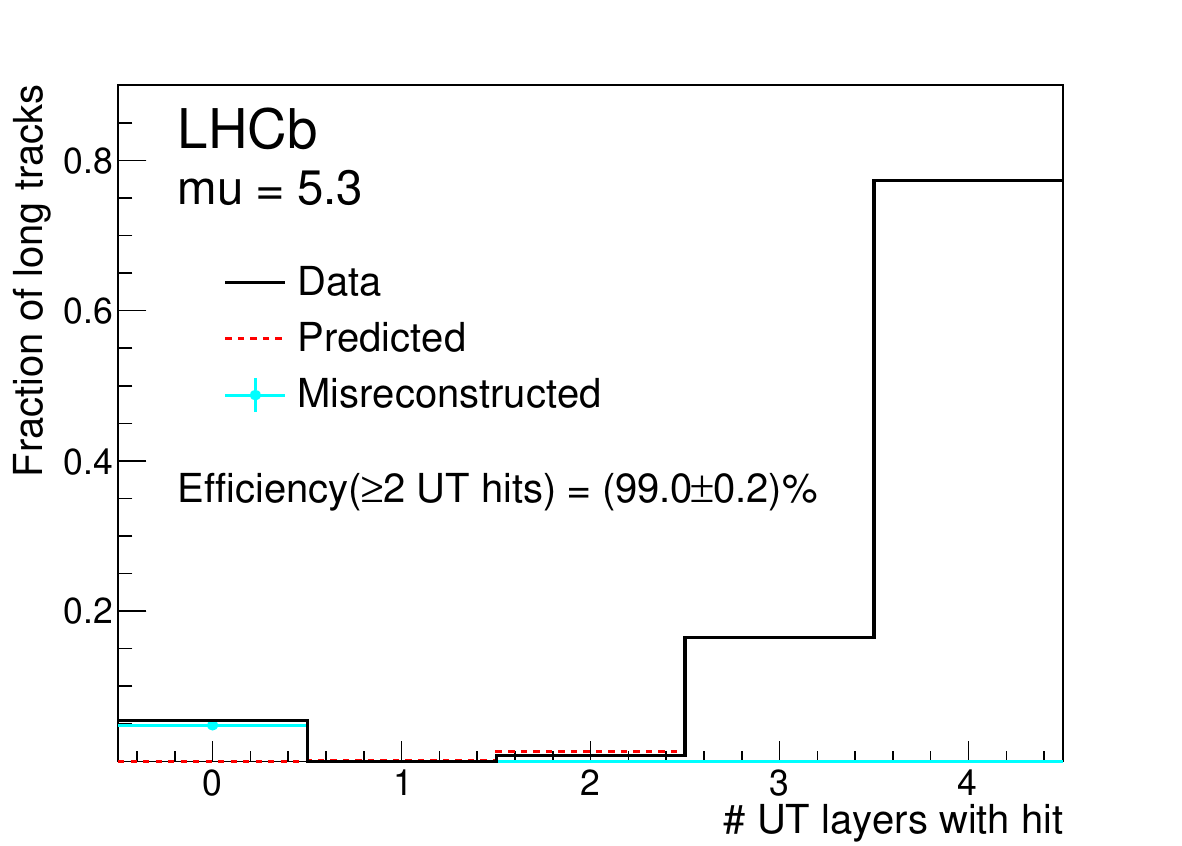}\\
\includegraphics[width=0.40\textwidth]{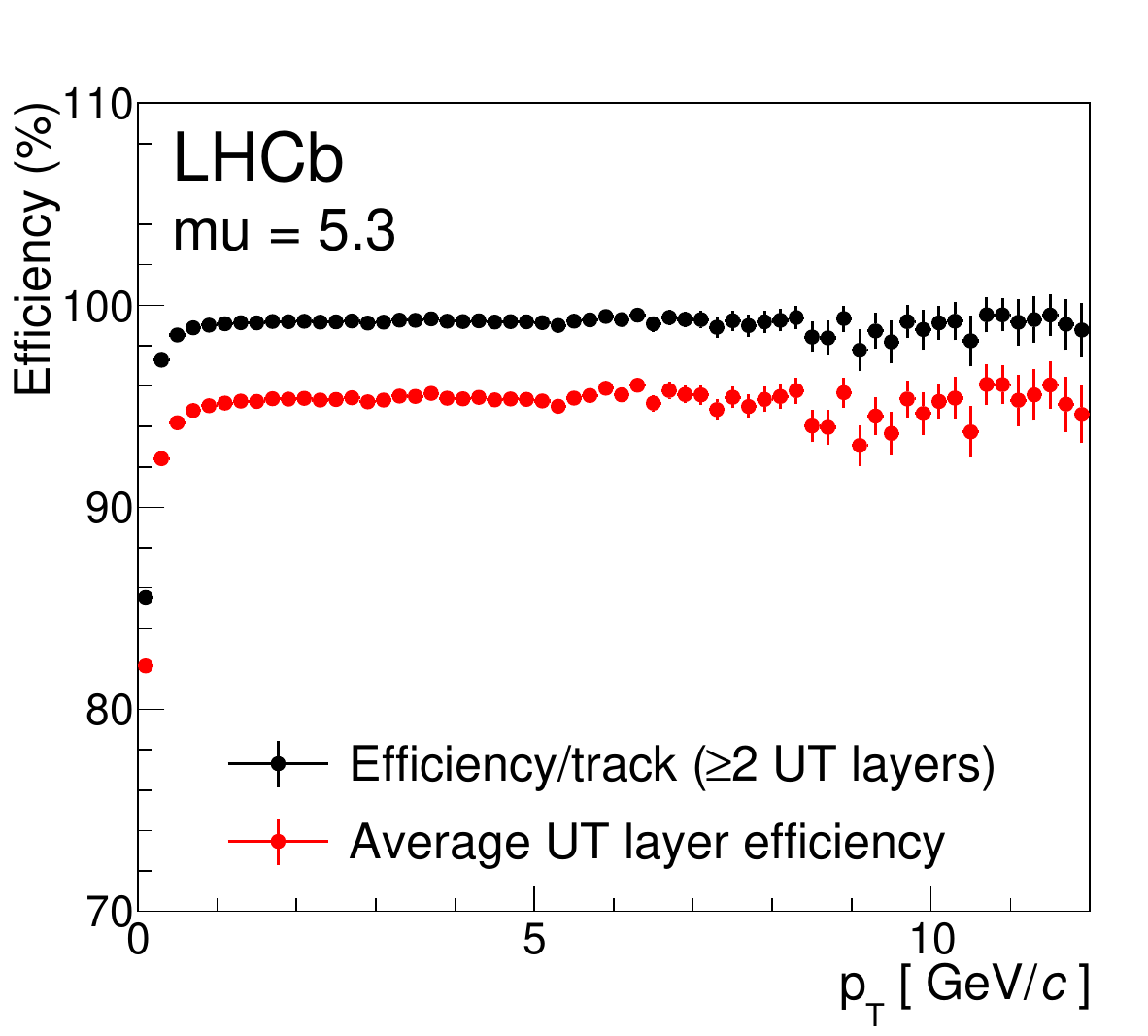}
\includegraphics[width=0.40\textwidth]{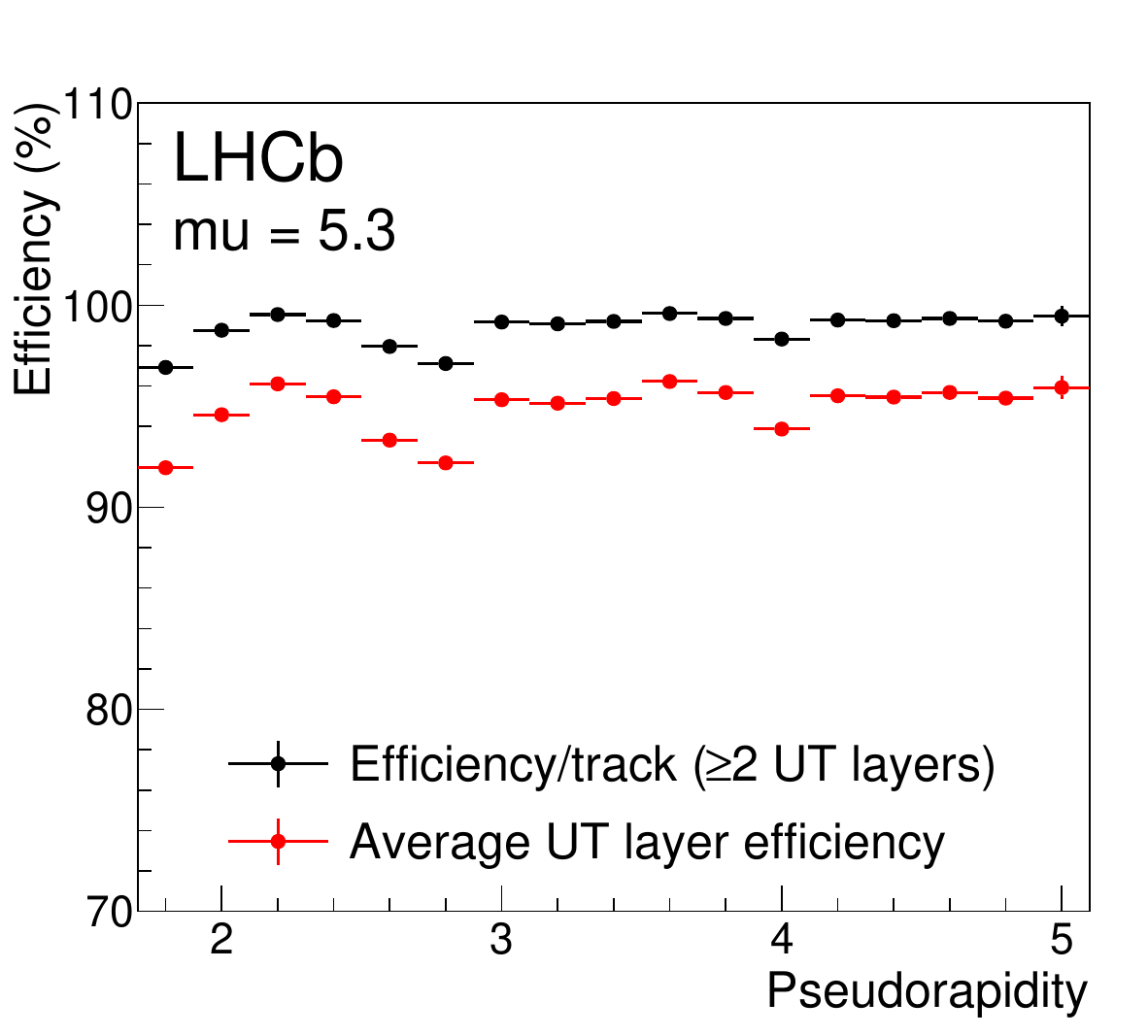}
\caption{\small{(Top) Distribution of the number of UT layers with hits on long tracks at mu = 5.3. (Bottom) Efficiencies of picking up 3 or more UT hits on a long track as a function of (left) $\pt$ and (right) pseudorapidity at mu = 5.3. }}
\label{fig:Eff_mu5_3}
\end{figure}

The most important feature of the detector is the overall efficiency for finding UT hits on the long tracks within the UT acceptance. 
For this study, tracks are required to have $\pt>500\mevc$ and to cross both the first and last UT layers at radial positions greater than 38 mm.
The number of UT layers with hits used on those long tracks is shown in Fig.~\ref{fig:Eff_mu5_3}. The average layer efficiency is estimated from the fraction of four-layer matches relative to three-layer matches, by assuming (i) there are no ghost tracks among the 3 and 4 UT-layer matches, and (ii) each layer has the same average uniform efficiency. With these approximations, the average layer efficiency is estimated to be about 95\%. This is in good agreement with the overall fraction of operational channels in the UT. With an average per-layer efficiency of 95\%, the efficiency of assigning at least 3 UT hits to a long track is found to be $\varepsilon_{\ge3{\rm UT}}=98.6\%$. An additional $\sim$1\% of tracks is expected to have UT hits in only 2 layers, while the data shows about half that value. Based on the assumptions given above for estimating the UT per-layer efficiencies, ${\mathcal{O}}(0.5\%)$ difference in the 2-layer efficiency is not unreasonable. Including these 2 UT-layer matches, the efficiency of associating 2 or more UT hits on a long track within the UT acceptance is estimated to be approximately $\varepsilon_{\ge2{\rm UT}}=(99.0\pm0.2)\%$. The 5.5\% of long tracks in the 0 bin in Fig.~\ref{fig:Eff_mu5_3} include long tracks that have less than 2 UT hits. These tracks are comprised almost entirely of misreconstructed tracks. 

The UT efficiency versus $\pt$ and pseudorapidity ($\eta$) are shown in the bottom half of Fig.~\ref{fig:Eff_mu5_3}. For the $\pt$ dependence, the $\pt$ requirement has been removed. A 10\% drop in efficiency for $\pt<200\mevc$ is seen, which is consistent with increased multiple scattering in the detectors. Beyond about 1\gevc, the efficiency is about 99\% for finding hits in at least 2 UT layers. The efficiency vs $\eta$ shows an 
$\sim2\%$ decrease near $\eta=2.8$, which is due to a set of locally inefficient regions in the UT.

\begin{figure}[hbt]
\centering
\includegraphics[width=0.98\textwidth]{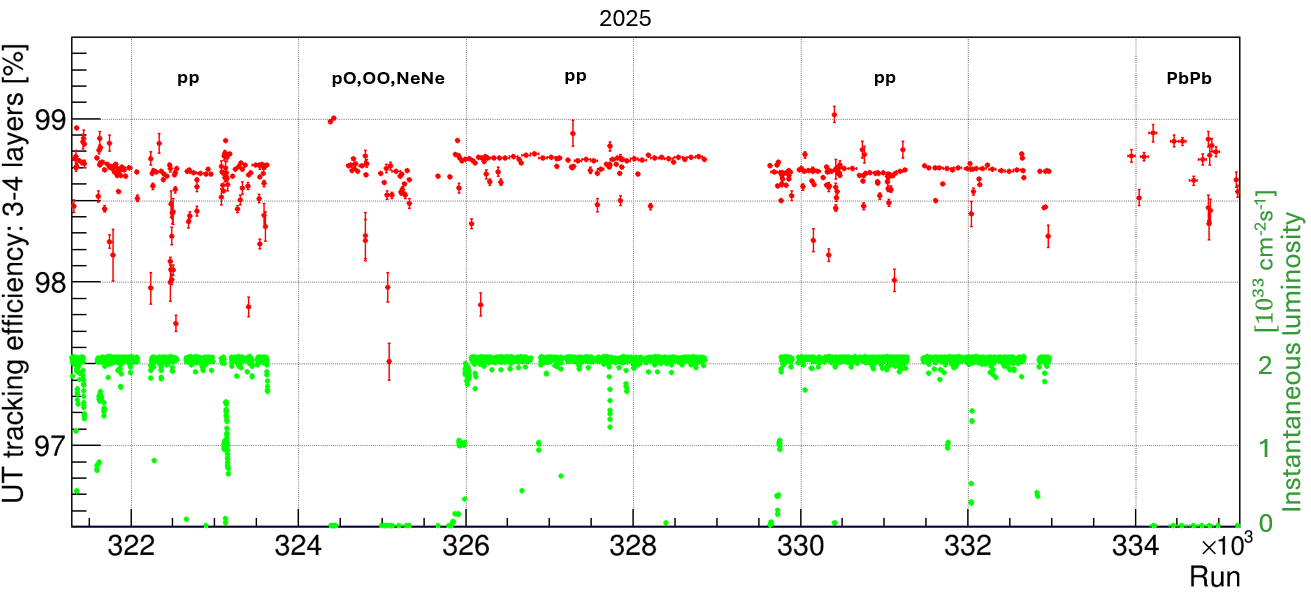}
\caption{\small{ (Red points) Average efficiency of matching 3 or 4 UT layers ($\varepsilon_{\ge3{\rm UT}}$) to a long track as a function of run number for the entire 2025 data taking period. (Green points) Instantaneous luminosity per run. The instantaneous luminosity of light and heavy nuclei collisions is very low, $\sim10^{27}$ cm$^{-2}$ s$^{-1}$.}}
\label{fig:Eff34vsRun2025}
\end{figure}

Uniformly high efficiency was maintained throughout the 2025 running period. Figure~\ref{fig:Eff34vsRun2025} shows the values of $\varepsilon_{\ge3{\rm UT}}$ as a function of the run number, where the efficiency is obtained using the method described in Section~\ref{sec:UTTrackEff}. The instantaneous luminosity for each run is also shown. The results show that the UT maintained an efficiency of about 98.7\% throughout the run, independent of the types of collisions. Small dips in efficiency for isolated runs are due to temporary readout board desynchronizations.

\begin{figure}[tb]
\centering
\includegraphics[width=0.68\textwidth]{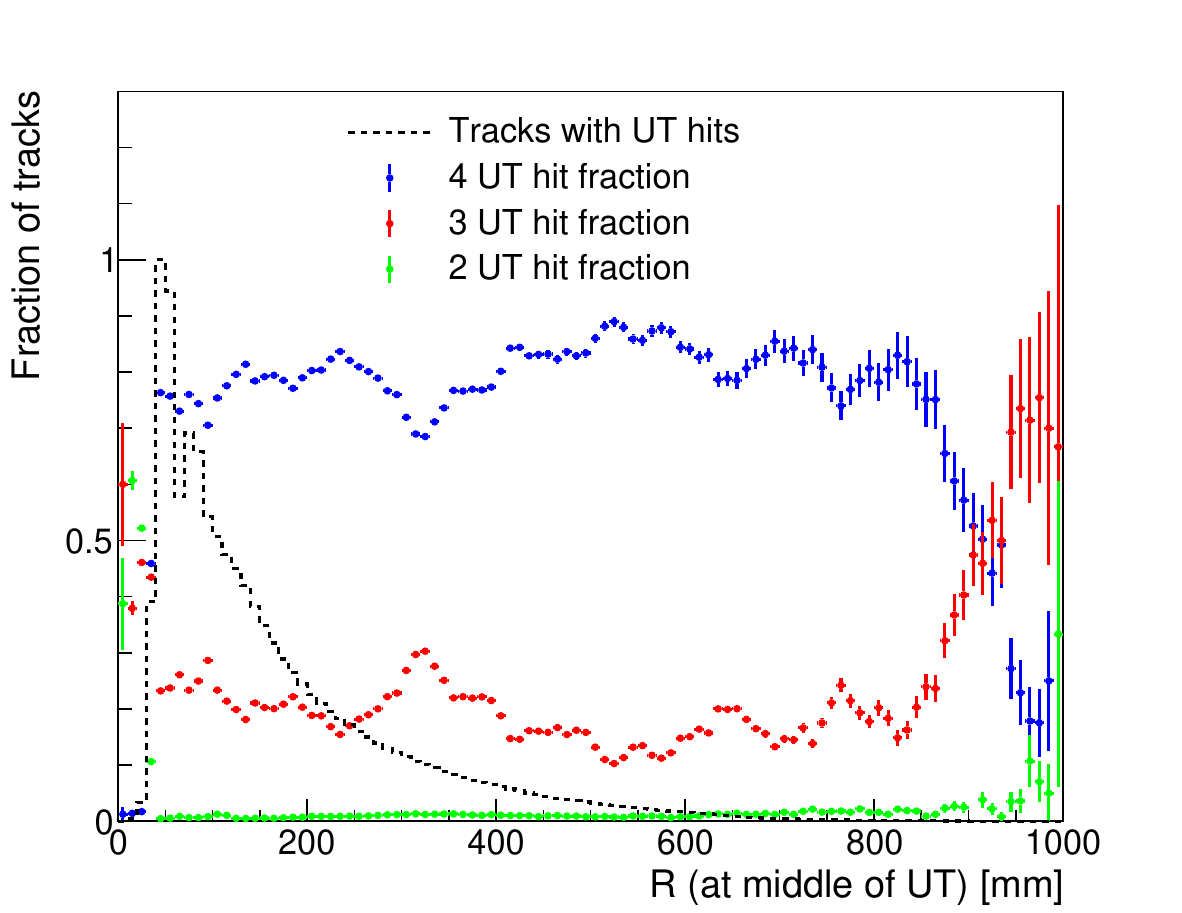}
\caption{\small{Fractions of long tracks with hits in 2, 3 and 4 UT layers as a function of the radius evaluated at the center of the UT detector ($z\simeq2.49$~m). The radial distribution of long tracks with any number of UT hits is superimposed. 
}}
\label{fig:RadialDistribution}
\end{figure}

To study the composition of long tracks according to the number of associated UT layers, Fig.~\ref{fig:RadialDistribution} shows the fractions of long tracks with two, three, and four UT-layer matches as a function of the track radial position, evaluated at the centre of the UT along the $z$ axis ($z\simeq2.49$~m). The radial distribution of all long tracks with at least one UT hit is superimposed, showing the expected increase in track density towards smaller radii. At larger radii, the fraction of long tracks with three UT-layer matches increases, due to the combined effects of the UT acceptance and multiple scattering. Dips in the four-hit fraction at some radius values is due to localized inefficiencies, and they are generally compensated by an increase in the three-hit fraction. Long tracks with three UT-layer matches account for approximately one quarter of all long tracks, while those with only two UT-layer matches represent less than 1\% of the total. Given the low UT occupancy, accidental hit associations are unlikely; however, their relative contribution is expected to be larger for two-layer matches than for three- or four-layer matches. Therefore, the availability of four UT layers is important to maintain high track purity.

\subsection{Performance evaluation using signal decays}
The UT is used extensively in the formation of long tracks, downstream tracks and upstream tracks. The performance is evaluated using $\Bz\to\Dm\pip$ for long tracks and $\Bz\to\jpsi\KS$ decays for downstream tracks.

\subsubsection{Long tracks}
To estimate the per-track efficiency of requiring UT hits on long tracks, the mode $\Bz\to\Dm\pip$, with $\Dm\to\Kp\pim\pim$ is used. Apart from the selections applied in the HLT~\cite{LHCb-TDR-021,LHCb-DP-2026-XXX}, the final state particles are required to be within the UT acceptance by requiring pseudorapidity in the range [2.0, 4.8].\footnote{Some tracks at the largest pseudorapidity could bend into the beam pipe region and be physically outside the UT acceptance.} In addition the $\Dm$ candidate mass is required to be within the range $1844-1894$\mevcc, corresponding to about
3 times the mass resolution centred on the known $\Dm$ mass\cite{PDG2026}.

\begin{figure}[tb]
\centering
\includegraphics[width=0.48\textwidth]{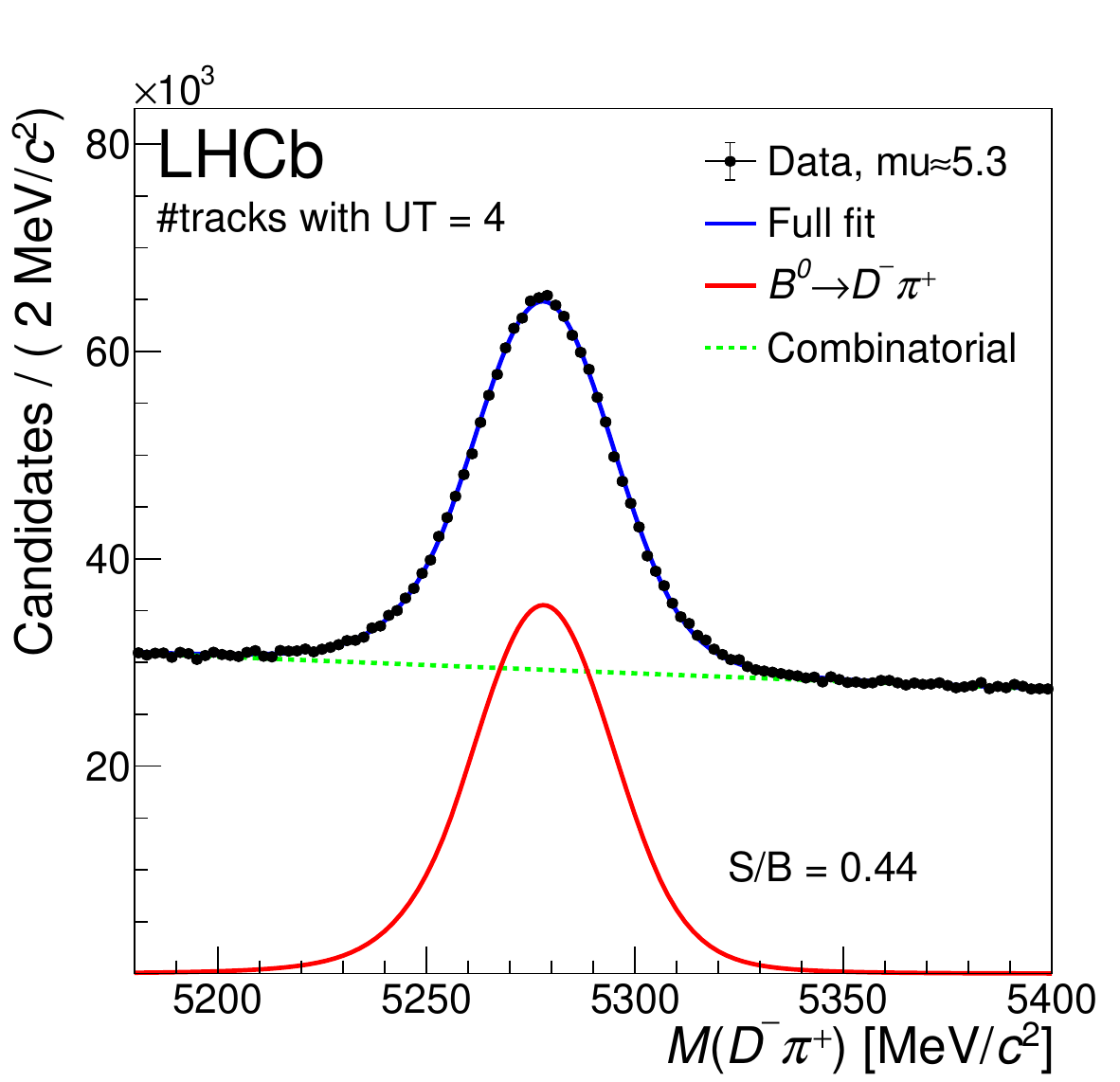}
\includegraphics[width=0.48\textwidth]{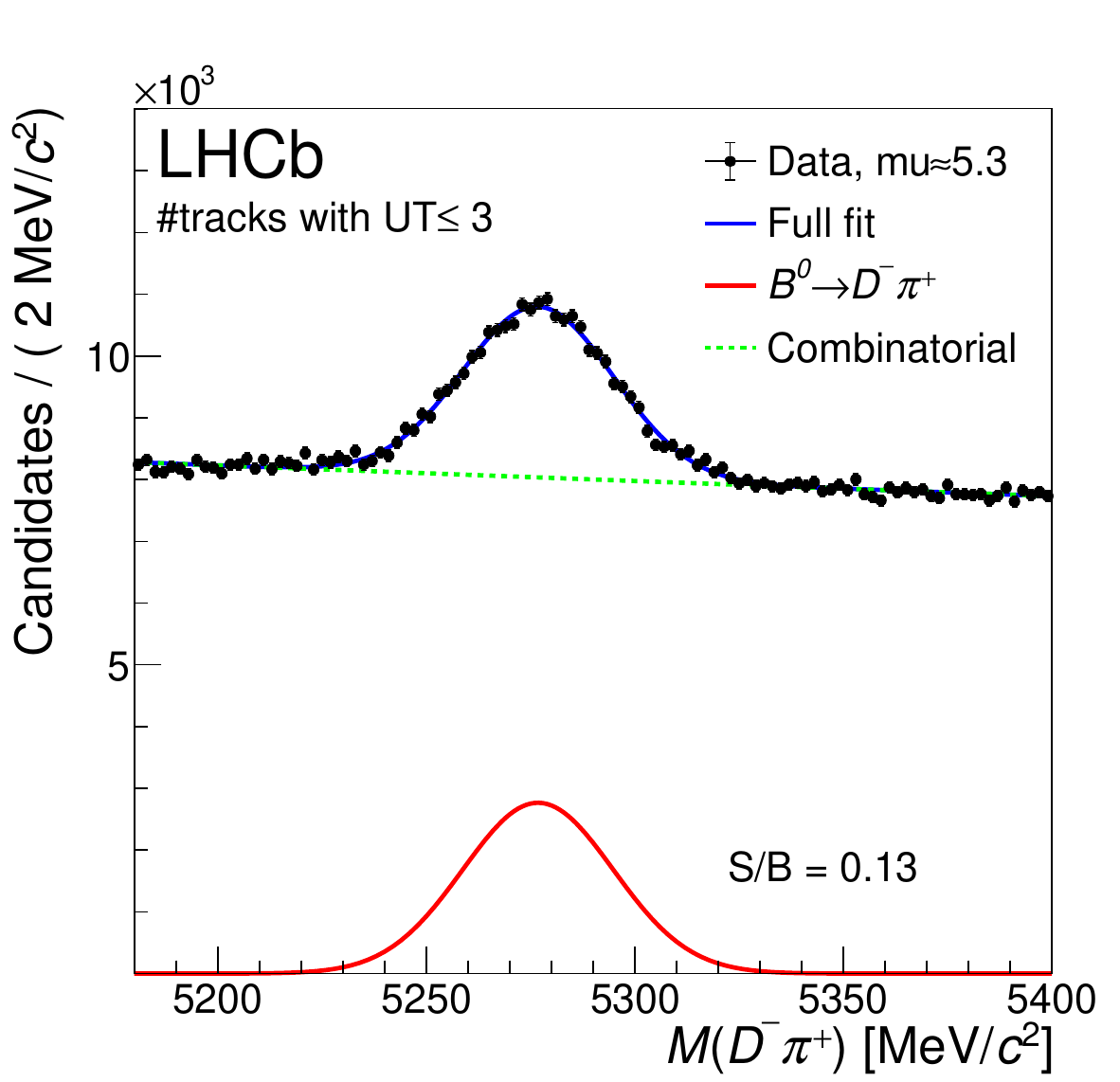}
\caption{\small{Mass spectra for $\Bz\to\Dm\pip$ candidates in 2025 data where (left) all four final state particles have UT hits, and (right) at least one of the final state particles does not have UT hits.}}
\label{fig:B2DPi}
\end{figure}

Figure~\ref{fig:B2DPi} shows the $\Bz$ candidate mass when (left) all four decay products have UT hits included on their tracks, and (right) at least one of the four tracks does not have UT hits. It should be emphasized that these spectra have minimal selections applied, beyond the requirements imposed in HLT. The signal-to-background ratio (S/B) is about 3.5 times better when requiring all four decay products to have UT hits. From the ratio of fitted signal yields in these two distributions, the average efficiency for a track to have UT hits is 98.1\%. This number is about 1\% lower than the 99\% value shown in Section~\ref{sec:UTTrackEff}. This is mainly due to the larger multiple scattering in this track sample due to a less stringent $\pt$ requirement $(\pt>250\mevc)$ as compared to the selection in Section~\ref{sec:UTTrackEff}, which requires $\pt>500\mevc$. The general conclusion is that for tracks within the UT acceptance, the efficiency of assigning 2 or more UT hits to the track is very high, and the S/B is substantially better when the decay products have UT hits attached to their associated long tracks.

\subsubsection{Downstream tracks}
To study the downstream tracking performance, the decay $\Bz\to\jpsi\KS$ with ${\jpsi\to\mup\mun}$ and $\KS\to\pip\pim$ is analysed. At the level of the trigger, events are only required to have a $\jpsi\to\mup\mun$ decay vertex that is significantly detached from all PVs in the event. The long-lived $\KS$ decay is reconstructed using either pairs of long tracks (LL) or downstream tracks (DD), while the muons are reconstructed as long tracks.
Apart from the trigger selections, the reconstructed mass of the $\KS$ candidates is required to be within $\pm25\mevcc$ of the known mass~\cite{PDG2026}, corresponding to about three times the mass resolution\footnote{
The mass resolution of $\KS\to\pip\pim$decays is comparable between Run 2 and Run 3.}. Other selections are applied to ensure the $\Bz$ candidate momentum vector points back to a PV in the event, and that the $\KS$ meson is inconsistent with being a $\Lz\to p\pim$ decay~\cite{Podolanski01011954}.

\begin{figure}[tb]
    \centering
    \includegraphics[width=0.48\textwidth]{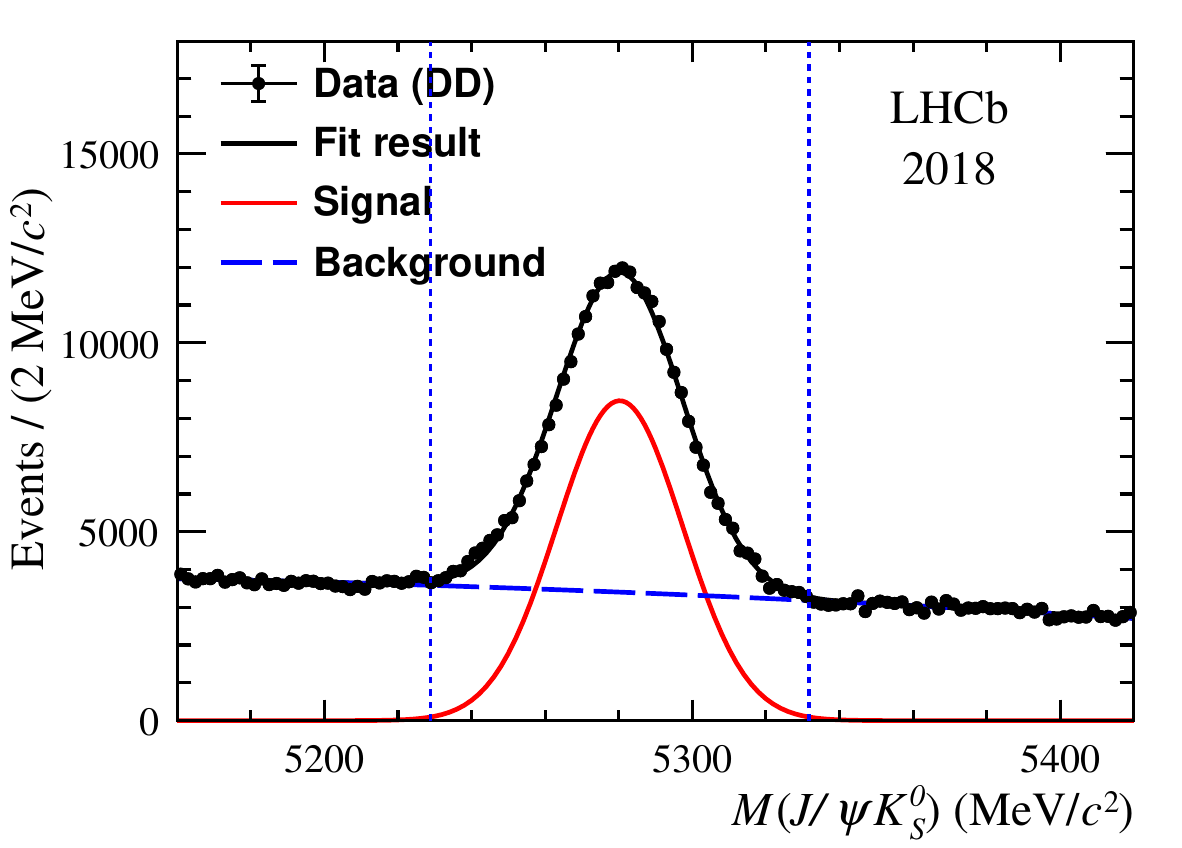}
    \includegraphics[width=0.48\textwidth]{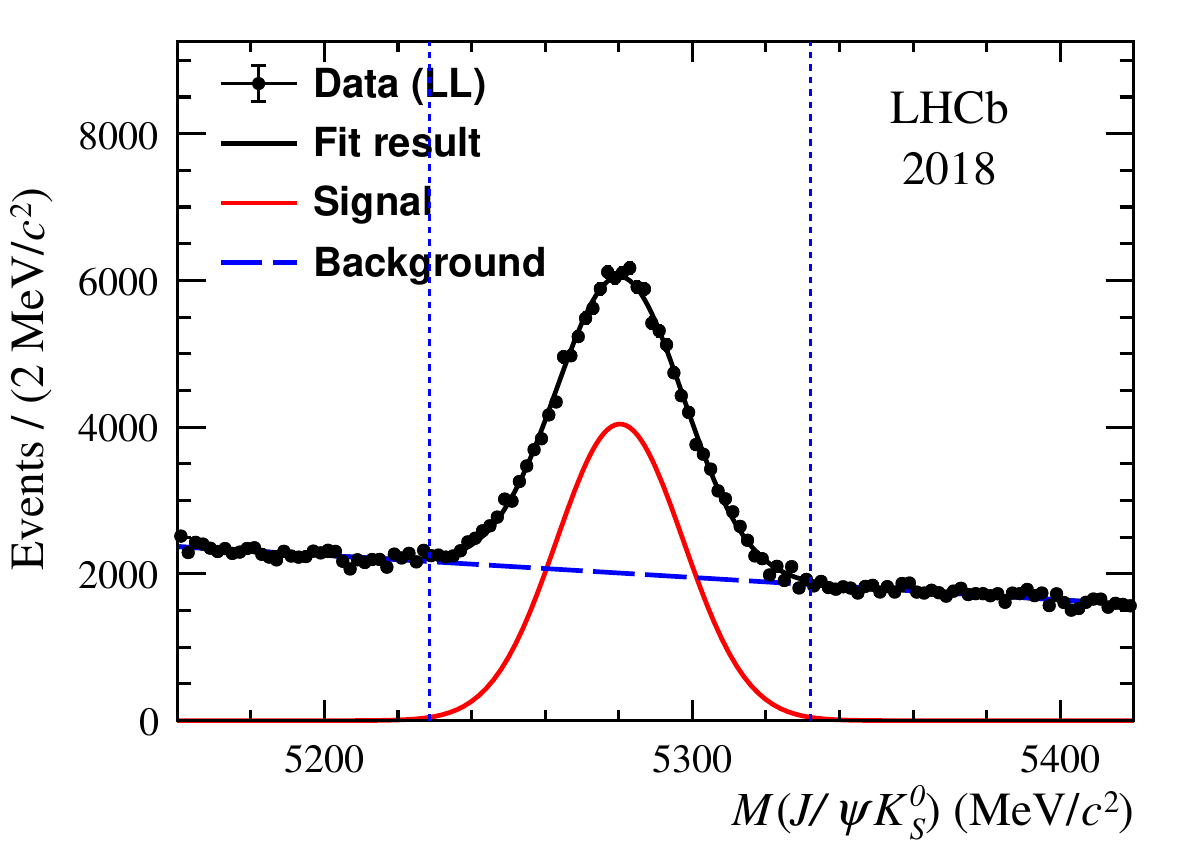}
    \includegraphics[width=0.48\textwidth]{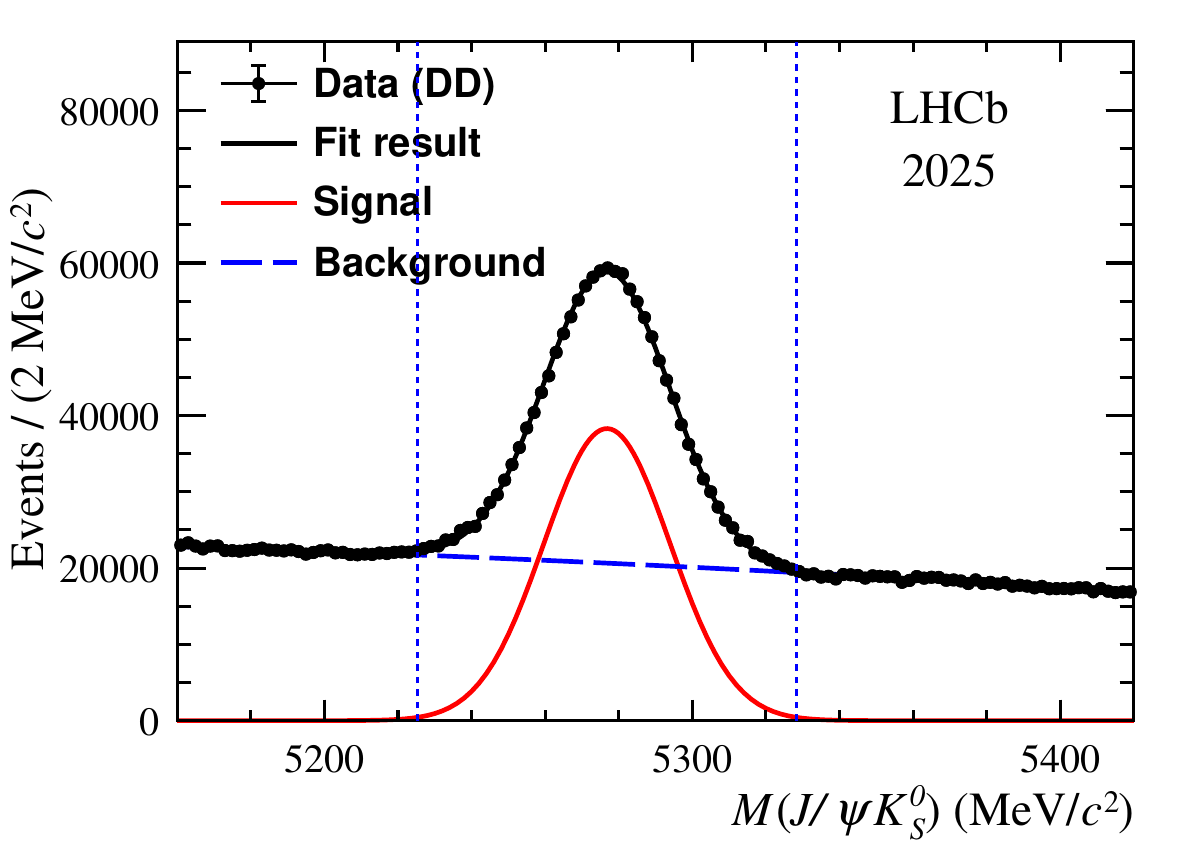}
    \includegraphics[width=0.48\textwidth]{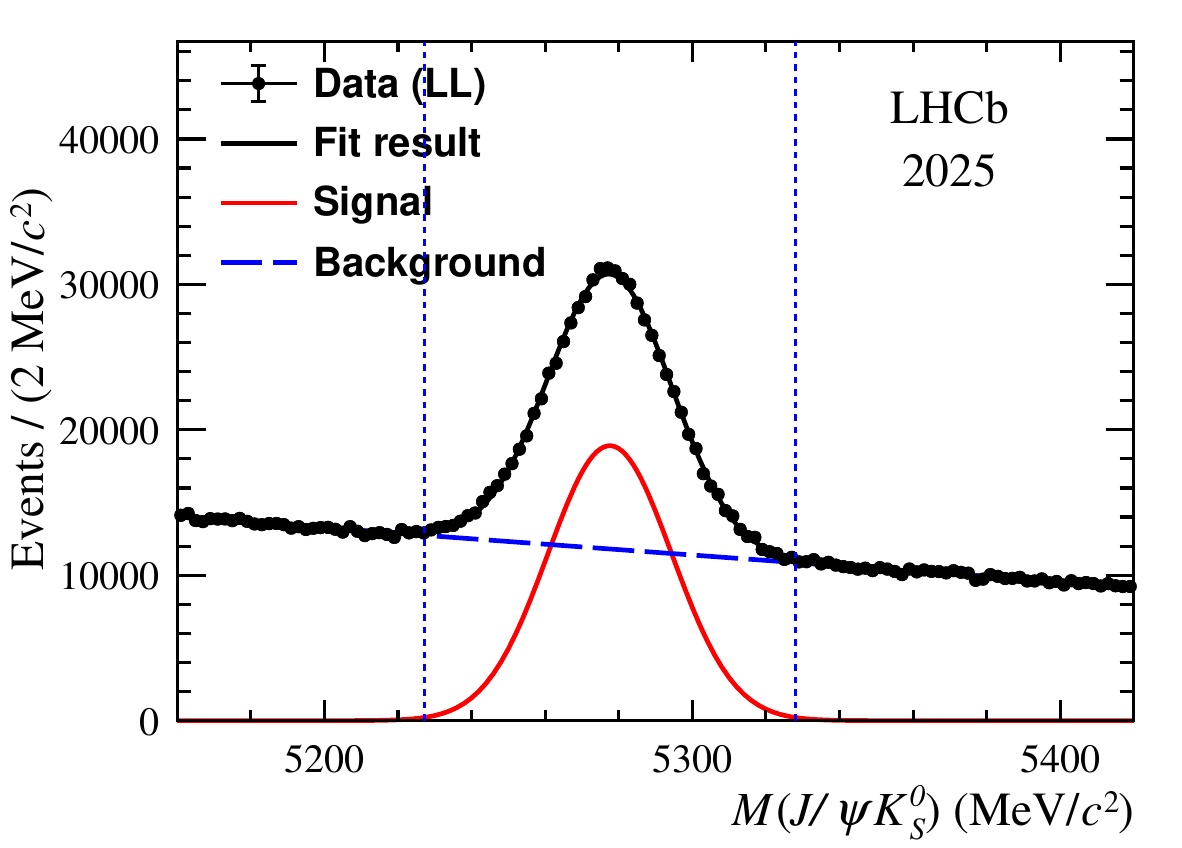}
    \caption{\small{Mass spectra for $\Bz\to\jpsi\KS$ candidates in (top) 2018 data and (bottom) 2025 data, where $\KS$ is reconstructed by pairs of (left) downstream tracks (DD) or (right) long tracks (LL).}}
    \label{fig:YieldsLumi_LmdKS}
\end{figure}

\begin{figure}[tb]
    \centering
    \includegraphics[width=0.48\textwidth]{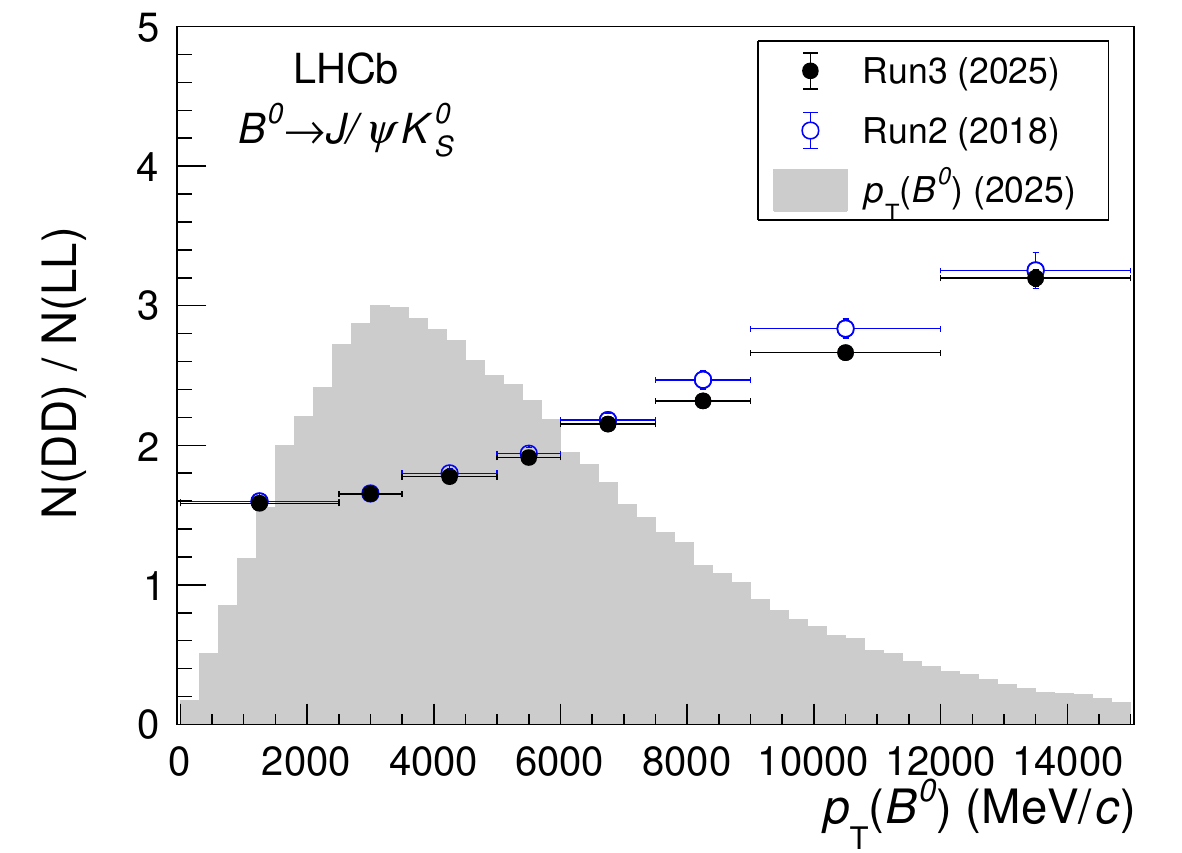}
    \includegraphics[width=0.48\textwidth]{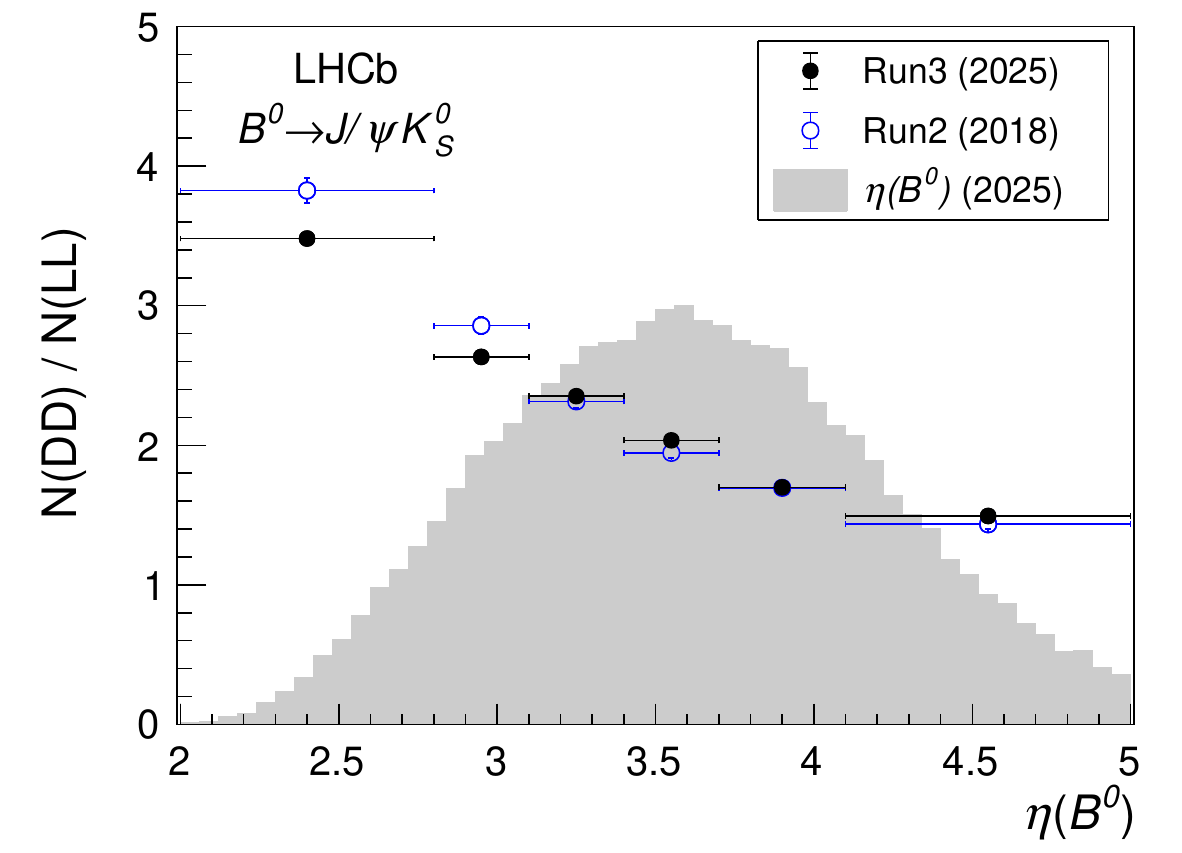}
    \caption{\small{Comparison of the yield ratios N(DD)/N(LL) between 2018 and 2025 data samples in different (left) transverse momentum and (right) pseudorapidity ranges, along with the corresponding distributions in data. The horizontal errors indicate the bin boundaries. Since the distributions of transverse momentum and pseudorapidity in 2018 and 2025 are similar, only those from 2025 are shown.}}
    \label{fig:YieldsLumi_LmdKS_Dist}
\end{figure}

To quantify the overall performance of the downstream tracking in Run 3 compared to Run 2, the $\Bz\to\jpsi\KS$ signal yields for DD and LL candidates are obtained for samples of data collected in Run 2\,(2018) and Run 3\,(2025), as shown in Fig.~\ref{fig:YieldsLumi_LmdKS}. Fits are also performed in bins of $\pt$ and $\eta$, and the corresponding values fractions N(DD)/N(LL) are shown in the Fig.~\ref{fig:YieldsLumi_LmdKS_Dist}. The gray distributions show the $\pt$ and $\eta$ spectra of $\Bz\to\jpsi\KS$ decays\footnote{The background is subtracted using the \sPlot~\cite{Pivk:2004ty} method.}. The relative yields between the DD and LL samples in Run 3 are very close to those in Run 2 over the full phase space. The fraction N(DD)/N(LL) of $\Bz\to\jpsi\KS$ decays averaged over all $\pt$ and $\eta$ is 2.1 for both the Run 2 and Run 3 samples. These studies demonstrate that the relative tracking efficiency between downstream and long tracks has been maintained in Run 3 relative to Run 2, while operating at five times larger instantaneous luminosity. 

\section{Summary}
The UT detector is an integral part of the Upgrade I tracking system of LHCb. Key aspects related to the calibration, commissioning, monitoring and performance of the UT detector are described. With about 20 fb$^{-1}$ collected to date, there is minimal loss in charge collection efficiency. The detector has about 96\% of its channels operational, and provides an efficiency of about 99\% for assigning at least 2 UT hits to long tracks. The presence of UT hits is shown to be a powerful discriminant against misreconstructed tracks, as anticipated when the UT was designed.  The inclusion of the UT in the HLT1 pattern recognition proved essential for LHCb to reach its design luminosity $2.0\times10^{33}$ cm$^{-2}$s$^{-1}$. Without UT information, the rapid increase in ghost rate and combinatorics at high pile-up prevented efficient operation. Including UT hits in the pattern recognition significantly reduced these effects, enabling stable operation at mu = 4 and above. The UT has also enabled LHCb to fully exploit the fully software-based trigger with respect to downstream tracking, resulting in signal yields (per fb${-1}$) of decays with $\KS$ and $\Lz$ baryons in the final state that are about 3 times larger than those obtained in Run 2. 

\section*{Acknowledgements}
%
%
\noindent We express our gratitude to our colleagues in the CERN accelerator departments for the excellent performance of the LHC and for support of numerous test beam campaigns to optimize the UT sensor design. We are grateful for the support from CERN and from the national agencies, including: the National Science Foundation of USA awards PHY-1433122, PHY-1433169 for the construction of the UT, grant PHY-2512443 for UT operations, and the base grants that support the scientists to exploit the science enabled by the UT, including PHY-1507572, PHY-1803004, PHY-2102879, PHY-2411639 [Syracuse University]; PHY-1307526, PHY-1601128, PHY-1907466, PHY-2012793, PHY-2309144 [University of Maryland]; and PHY-2309922 [University of Michigan]. We also acknowledge the support to the University of Milano through the European Union NextGenerationEU Project Title 43 202277EWLW grant and the ERC Consolidator Grant SELDOM G. A. 771642, and grants 20FL20\_173580, 20FL20\_201480, 20FL20\_216604, 20FL21\_232669 from the Swiss National Science Foundation. We are also grateful for the support from the National Natural Science Foundation of China under project numbers 12188102, 11961141015, W2443008, and the National Key Programme for S\&T Research and Development (grant number 2023YFA1606300). We are also acknowledge the funding provided by the Ministry of Science and Higher Education of Poland grant number MEiN DIR/WK/2017/2020/04-1, and BMFTR [Germany]. Lastly, we are also grateful for the contributions of numerous administrative and technical staff at the various institutes, including C. Brown and D. Pratt [Syracuse], Robert Bard, Jack Touart, and Larry Lutz [Maryland], and Andrew Volz [Cincinnati]. 
 



\clearpage\newpage
\addcontentsline{toc}{section}{References}
\bibliographystyle{LHCb}
\bibliography{main,standard,LHCb-PAPER,LHCb-CONF,LHCb-DP,LHCb-TDR}
 
\end{document}